      [1999/12/01 v1.4c Il Nuovo Cimento]
\documentclass[varenna]{cimento}

\usepackage{amssymb}
\usepackage{amsmath}
\usepackage{citesort}
\usepackage{latexsym}
\usepackage{graphicx}  

\newcommand{\mc}{\multicolumn}
\newcommand{\lsim}{\mathrel{\mathop{\kern 0pt \rlap
  {\raise.2ex\hbox{$<$}}}
  \lower.9ex\hbox{\kern-.190em $\sim$}}}
\newcommand{\gsim}{\mathrel{\mathop{\kern 0pt \rlap
  {\raise.2ex\hbox{$>$}}}
  \lower.9ex\hbox{\kern-.190em $\sim$}}}
\def\dsla{\rlap{/}\partial}
\def\non{\rlap{\small /}\in}
\newcommand{\beq}{\begin{equation}}
\newcommand{\eeq}{\end{equation}}
\newcommand{\bea}{\begin{eqnarray}}
\newcommand{\eea}{\end{eqnarray}}

\def\Lam{\Lambda}

\title{Weak decay of hypernuclei}
\author{W.~M.~Alberico \atque G.~Garbarino}
\institute{Dipartimento di Fisica Teorica, Universit\`a di Torino
and INFN, Sezione di Torino, \\I--10125 Torino, Italy}
\PACSes{\PACSit{21.80.+a}{By the way, which PACS is it, the 00.00? GOK.}
\PACSit{13.75.Ev}{which PACS is it}
\PACSit{13.75.Ev}{which PACS is it}}

\begin{document}

\maketitle


\section{Introduction}

The focus of these lectures is on the weak decay modes of hypernuclei,
 with special attention to  $\Lambda$--hypernuclei. The subject involves 
many fields of modern theoretical and experimental physics, from nuclear 
structure to the fundamental constituents of matter and their interactions. The 
peculiar behaviour of matter containing strange quarks has raised in 
recent years many interesting problems, one of which being the physics
of hypernuclei.

Hypernuclear physics was born in 1952, when the first hypernucleus was 
observed through its decays \cite{Da53}. Since then, it has been characterized by more 
and more new challenging questions and answers. The interest was further
raised by the great advances made in the last 15--20 years.
Moreover, the existence of hypernuclei  gives a new dimension 
to the traditional world of nuclei (states with new symmetries, 
selection rules, etc). They represent the first kind of 
\emph{flavoured nuclei}, in the direction of other exotic systems 
(charmed nuclei and so on).

Hyperons (${\Lambda}$, ${\Sigma}$, $\Xi$, $\Omega$)
have lifetimes of the order of $10^{-10}$~sec (apart from the ${\Sigma}^0$, 
which decays into $\Lambda \gamma$). They decay weakly, with a mean free path 
$\lambda \approx c\tau ={\mathcal O}(10$ cm$)$. 
A hypernucleus is a bound system of neutrons, 
protons and one or more hyperons. We will denote with $^{A+1}_YZ$ a 
hypernucleus with $Z$ protons, $A-Z$ neutrons and a hyperon $Y$. 
In order to describe the structure of these \emph{strange nuclei} it
is crucial the knowledge of the elementary hyperon--nucleon ($YN$) and 
hyperon--hyperon ($YY$) interactions. Hyperon masses differ remarkably from 
the nucleonic mass, hence the flavour $SU(3)$ symmetry
is broken. The amount of this breaking is a fundamental question in order to 
understand the baryon--baryon interaction in the strange sector.

Nowadays, the knowledge of hypernuclear phenomena is rather good, but some
open problems still remain. The study of this field may help in 
understanding some important questions, related to:
\begin{enumerate}
\item some aspects of the baryon--baryon weak interactions;
\item the $YN$ and $YY$ strong interactions in the $J^P=1/2^+$ baryon octet;
\item the possible existence of di--baryon particles;
\item the renormalization of hyperon and meson properties in the 
nuclear medium;
\item the nuclear structure: for instance, the long standing question 
of the origin of the spin--orbit interaction and other aspects of the 
many--body nuclear dynamics;
\item the role played by quark degrees of freedom, flavour symmetry 
and chiral models in nuclear and hypernuclear phenomena.
\end{enumerate}
Many of these aspects can be discussed and understood by investigating
the hypernuclear weak decays. Important related arguments, which here 
will not be considered or only briefly mentioned, can be found in the 
lectures by A.~Gal~\cite{Gal-Varenna} and T.~Nagae~\cite{Nagae-Varenna},
while the experimental viewpoint on the same subject is presented by 
H.~Outa~\cite{Outa-Varenna}.

In these lectures the various weak decay modes of $\Lambda$--hypernuclei
are described: indeed in a nucleus the $\Lambda$ can decay by emitting 
a nucleon and a pion (\emph{mesonic mode}) as it happens in free space, but 
its (weak) interaction with the nucleons opens new channels, customarily
indicated as \emph{non--mesonic} decay modes. These are the dominant decay 
channels of medium--heavy nuclei, where, on the contrary, the mesonic decay
is disfavoured by Pauli blocking effect on the outgoing nucleon. 
In particular, one can distinguish between one--body and two--body induced 
decays, according whether the hyperon interacts with a single nucleon or
with a pair of correlated nucleons.

An interesting rule for the amount of isospin violation ($\Delta I=1/2$) 
is strongly suggested by the mesonic decay of free $\Lambda$'s, whose
branching ratios are almost in the proportion 2 to 1, according 
whether a $\pi^- p$
or $\pi^0 n$ are emitted. This totally empirical rule has been generally
adopted in most of the models proposed for the evaluation of the 
$\Lambda$--hypernuclei decay widths: some of the expected consequences, 
however, seem to require additional work and investigation. 
Indeed, the total non--mesonic ($\Gamma_{\rm NM}=\Gamma_n+\Gamma_p\, (+\Gamma_2)$)
and mesonic ($\Gamma_{\rm M}=\Gamma_{\pi^0}+\Gamma_{\pi^-}$) 
decay rates are well explained by several
calculations; however, for many years the main open problem in the decay of
$\Lam$--hypernuclei has been the discrepancy between theoretical and 
experimental values of the ratio $\Gamma_n/\Gamma_p$. This topic will be discussed 
at length here, together with the most recent indications toward a solution
of the puzzle.

Another interesting and open question concerns the asymmetric non--mesonic 
decay of polarized hypernuclei: strong inconsistencies appear already among 
data. Also in this case, as for the $\Gamma_n/\Gamma_p$ puzzle, one can expect
important progress from the present and future improved experiments, 
which will provide a guidance for a deeper theoretical understanding of
hypernuclear dynamics and decay mechanisms.

For a comprehensive review on the subject of these lectures we refer the
reader to Ref.~\cite{Al02} and references therein.


\section{Weak decay modes of $\Lambda$--hypernuclei}

In the production of hypernuclei, the populated state may be highly excited, 
above one or more threshold energies for particle decays. These states are 
unstable with respect to the emission of the hyperon, of photons and nucleons. 
The spectroscopic studies of strong and electromagnetic de--excitations give
information on the hypernuclear structure which are complementary
to those we can extract from excitation functions and angular distributions
studies. Once the hypernucleus is stable with respect to
electromagnetic and strong processes, it is in the ground state, with
the hyperon in the $1s$ level, and can only 
decay via a strangeness--changing weak interaction, through the 
disappearance of the hyperon.

\subsection{Mesonic decay} 

The mesonic mode is the main decay channel of a $\Lambda$ in free space:
\beq
\begin{array}{l l l l l}
\Lambda &\rightarrow & \pi^- p & & ({\rm B.R.}= 63.9 \times 10^{-2})
 \nonumber \\
&   & \pi^0 n & & ({\rm B.R.}= 35.8 \times 10^{-2})
 \nonumber
\end{array}
\label{lambdafreedec}
\eeq
with a lifetime $\tau^{\rm free}_{\Lambda}\equiv 
\hbar/\Gamma^{\rm free}_{\Lambda}= 2.632\times 10^{-10}$ sec.

Semi--leptonic and weak radiative $\Lambda$ decay modes have negligible 
branching ratios:
\begin{equation}
\begin{array}{l l l l l}
\Lambda &\rightarrow & n\gamma        & & ({\rm B.R.}= 1.75\times 10^{-3}) 
\nonumber \\
     & &   p\pi^-\gamma               & & ({\rm B.R.}= 8.4\times 10^{-4})  
\nonumber \\
     & &   pe^-\overline{\nu}_e       & & ({\rm B.R.}= 8.32\times 10^{-4})  
\nonumber \\
     & &   p\mu^-\overline{\nu}_{\mu} & & ({\rm B.R.}= 1.57\times 10^{-4})   
\nonumber  
\end{array}
\end{equation} 
and will not be considered here.

The $\Lambda$ hyperon is an isospin singlet ($I_{\Lambda}=0$), while 
the $\pi N$ system can be either in $I=1/2$ or in $I=3/2$ isospin states. 
The customary angular momentum coupling implies:
\bea
|\pi^- p\rangle &&=\sqrt{\frac{1}{3}}\left|\frac{3}{2}, 
-\frac{1}{2}\right\rangle
- \sqrt{\frac{2}{3}}\left|\frac{1}{2}, -\frac{1}{2}\right\rangle ,
\nonumber\\
|\pi^0 n\rangle &&=\sqrt{\frac{2}{3}}\left|\frac{3}{2}, 
-\frac{1}{2}\right\rangle
+ \sqrt{\frac{1}{3}}\left|\frac{1}{2}, -\frac{1}{2}\right\rangle .
\nonumber
\eea
Hence the ratio of amplitudes for $\Delta I=1/2$ transitions yields:
\bea
\frac{\Gamma^{\rm free}_{\Lam\to\pi^- p}}{\Gamma^{\rm free}_{\Lam\to\pi^0 n}}
\simeq \frac{\left|\langle \pi^- p|T_{1/2, -1/2}|\Lam\rangle\right|^2}
{\left|\langle \pi^0 n|T_{1/2, -1/2}|\Lam\rangle\right|^2}=
\left|\frac{\sqrt{2/3}}{\sqrt{1/3}}\right|^2=2 ,
\nonumber
\eea
while a $\Delta I=3/2$ process should give:
\bea
\frac{\Gamma^{\rm free}_{\Lam\to\pi^- p}}{\Gamma^{\rm free}_{\Lam\to\pi^0 n}}
\simeq \frac{\left|\langle \pi^- p|T_{3/2, -1/2}|\Lam\rangle\right|^2}
{\left|\langle \pi^0 n|T_{3/2, -1/2}|\Lam\rangle\right|^2}=
\left|\frac{\sqrt{1/3}}{\sqrt{2/3}}\right|^2=\frac{1}{2} .
\nonumber
\eea
Experimentally the above ratio turns out to be:
\bea
\left\{\frac{\Gamma^{\rm free}_{\Lam\to\pi^- p}}
{\Gamma^{\rm free}_{\Lam\to\pi^0 n}}\right\}^{\rm Exp}
\simeq  1.78,
\nonumber
\eea
which is very close to 2 and strongly suggests the $\Delta I=1/2$ rule 
on the isospin change. From the above considerations and from analyses of the 
$\Lambda$ polarization observables it follows that the measured ratio 
between $\Delta I=1/2$ and 
$\Delta I=3/2$ transition amplitudes is very large:
\[ \left| \frac{A_{1/2}}{A_{3/2}}\right| \simeq 30.\]
The $\Delta I=1/2$ rule is based on experimental observations
but its dynamical origin is not yet understood on theoretical 
grounds.
It is also valid for the decay of the $\Sigma$ hyperon and for pionic kaon 
decays (namely in non--leptonic strangeness--changing processes). 
Actually, this rule is slightly violated in the $\Lambda$ free decay, and 
it is not clear whether it is a universal characteristic 
of all non--leptonic processes with $\Delta S\neq 0$. The $\Lambda$ free
decay in the Standard Model can occur through both $\Delta I=1/2$ 
and $\Delta I=3/2$ transitions, with comparable strengths: an $s$ quark 
converts into a $u$ quark through the exchange of a $W$ boson. Moreover, 
the effective 4--quark weak interaction derived from the Standard Model 
including perturbative QCD corrections 
gives too small $|A_{1/2}/A_{3/2}|$ ratios ($\simeq 3\div 4$, as calculated at
the hadronic scale of about $1$ GeV by using renormalization 
group techniques \cite{Gi79}). Therefore, non--perturbative
QCD effects at low energy (such as hadron structure and reaction mechanism), 
which are more difficult to handle, and/or final state interactions could 
be responsible for the enhancement of the $\Delta I=1/2$ amplitude and/or 
the suppression of the $\Delta I=3/2$ amplitude\footnote{See, for example, 
a recent work based on the Instanton Liquid 
Model~\cite{Faccioli}}.

The $Q$--value for free--$\Lam$ mesonic decay at rest is 
$Q_{\Lambda}\simeq m_{\Lambda}-m_N-m_{\pi}\simeq 40$ MeV. Then, 
taking into account energy--momentum conservation,  
$m_{\Lambda} \simeq
\sqrt{\vec p\,^2+m_{\pi}^2}+\sqrt{\vec p\,^2+m_N^2}$ in the
center--of--mass system and the momentum of the final nucleon turns out to
be $p\simeq 100$ MeV. Inside a hypernucleus, the binding energies of the 
recoil nucleon ($B_N\simeq -8$~MeV) and of the $\Lambda$ 
($B_{\Lam}\ge -27$~MeV) tend to further decrease $Q_{\Lambda}$ 
[$Q_{\Lambda,\mathrm{bound}}= Q_{\Lambda}+B_{\Lam}-B_N$]
and hence $p$.

As a consequence, in nuclei the $\Lam$ mesonic decay is disfavoured by the 
Pauli principle, particularly in heavy systems. It is strictly forbidden in 
normal infinite nuclear matter (where the Fermi momentum is 
$k_F^0\simeq$ 270 MeV), while in finite nuclei it can occur because of 
three important effects: 
\begin{enumerate}
\item In nuclei the hyperon 
has a momentum distribution (being confined in a limited spatial region) 
that allows larger momenta to be available to the final nucleon; 
\item The final pion feels an attraction by the medium such that for fixed
momentum $\vec q$ it has an energy smaller than the free one 
[$\omega(\vec q)<\sqrt{\vec q\,^2+m_{\pi}^2}$], and consequently, due to
energy conservation, the final nucleon again has more chance to come out above 
the Fermi surface. Indeed it has been shown \cite{Os93,Mo94} that the pion 
distortion increases the mesonic width by one or two 
orders of magnitude for very heavy hypernuclei ($A\simeq 200$) with
respect to the value obtained without the medium distortion;
\item At the nuclear surface the local Fermi momentum is considerably
smaller than $k_F^0$, and the Pauli blocking is less effective in 
forbidding the decay. 
\end{enumerate}
In any case  the mesonic width rapidly decreases as 
the nuclear mass number $A$ of the hypernucleus increases. 

The mesonic channel also gives information on the pion--nucleus optical 
potential since  $\Gamma_{\rm M}=\Gamma_{\pi^-}+\Gamma_{\pi^0}$ is very 
sensitive to the pion self--energy in the medium: the latter is enhanced 
by the attractive $P$--wave $\pi$--nucleus interaction and reduced by the 
repulsive $S$--wave one. 
Evidence for a central repulsion in the $\Lambda$--nucleus mean potential 
was obtained from the mesonic decays of $s$--shell hypernuclei~\cite{Ak97,Ou98}.

\subsection{Non--mesonic decay} 

\begin{figure}
\begin{center}
\includegraphics[width=.8\textwidth]{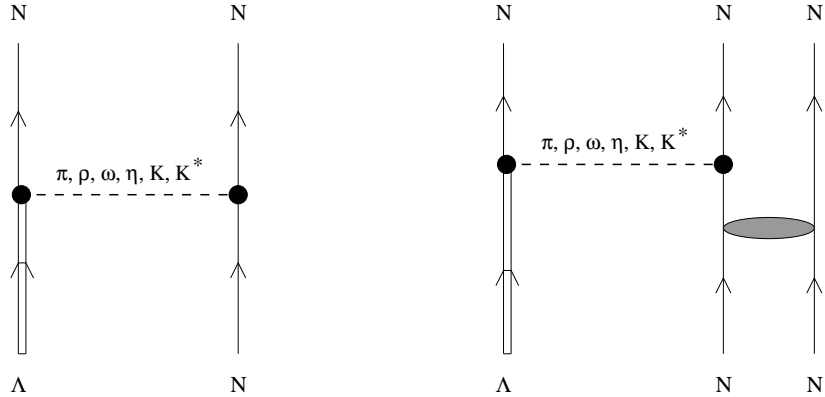}
\caption{One--nucleon (a) and two--nucleon (b) induced $\Lambda$ decay in 
nuclei.}
\label{nm12}
\end{center}
\end{figure}

In hypernuclei the weak decay can occur through processes which involve
a weak interaction of the $\Lam$ with one or more nucleons. Sticking to
the  weak hadronic vertex $\Lambda \rightarrow \pi N$, when the emitted 
pion is virtual, then it will be absorbed by the nuclear medium, resulting in 
a non--mesonic process of the following type:
\begin{eqnarray}
\label{gn}
\Lambda n & \rightarrow & nn \hspace{4mm} \left(\Gamma_n\right), \\
\label{gp}
\Lambda p & \rightarrow & np \hspace{4mm} \left(\Gamma_p\right) , \\
\label{g2}
\Lambda NN & \rightarrow & nNN \hspace{4mm} \left(\Gamma_2\right) .
\end{eqnarray}
The total weak decay rate of a $\Lambda$--hypernucleus is then:
\begin{equation}
\Gamma_{\rm T}=\Gamma_{\rm M}+\Gamma_{\rm NM} , \nonumber
\end{equation}
where:
\begin{equation}
\Gamma_{\rm M}=\Gamma_{\pi^-}+\Gamma_{\pi^0} , \hspace{0.15 in}
\Gamma_{\rm NM}=\Gamma_1+\Gamma_2 , \hspace{0.15 in}
\Gamma_1=\Gamma_n+\Gamma_p , \nonumber
\end{equation}
and the lifetime is $\tau=\hbar/\Gamma_{\rm T}$.
The channel (\ref{g2}) can be interpreted by assuming that the pion 
is absorbed by a pair of nucleons, 
correlated by the strong interaction. Obviously, the non--mesonic
processes can also be mediated by the exchange of more massive mesons than
the pion (see figure \ref{nm12}). 

The non--mesonic mode is \emph{only possible in nuclei} and,
nowadays, the systematic study of the hypernuclear decay is the only 
practical way to get information on the weak process
${\Lambda}N \rightarrow NN$ (which provides the first extension of the weak 
$\Delta S=0$ $NN\rightarrow NN$ interaction to strange baryons), 
especially on its parity--conserving part, which
is masked by the strong interaction in the weak $NN\rightarrow NN$ reaction.

The final nucleons in the non--mesonic processes emerge with large
momenta: disregarding the $\Lambda$ and nucleon binding energies
and assuming the available energy $Q=m_{\Lambda}-m_N\simeq 176$ MeV 
to be equally splitted among the final nucleons, it turns out that
$p_N\simeq 420$ MeV for the one--nucleon induced channels 
[Eqs.~(\ref{gn}), (\ref{gp})] and 
$p_N\simeq 340$ MeV in the case of the two--nucleon induced mechanism 
[Eq.~(\ref{g2})]. Therefore, the non--mesonic decay mode is not forbidden 
by the Pauli principle: on the contrary, the final nucleons have 
great probability to escape from the nucleus.
The non--mesonic mechanism dominates over the mesonic mode for all but 
the $s$--shell hypernuclei. Only for very light systems the two
decay modes are competitive.

Since the non--mesonic channel is characterized by large momentum transfer, 
 the details of the hypernuclear structure do not 
have a substantial influence (then providing useful information 
directly on the hadronic weak interaction).
On the other hand, the $NN$ and $\Lambda N$ short 
range correlations turn out to be very important. 

It is interesting to observe that there is an anticorrelation between 
mesonic and non--mesonic decay modes such that the experimental
lifetime is quite stable from light to heavy hypernuclei \cite{Co90,Os98}, apart
from some fluctuation in light systems because of shell structure effects: 
${\tau}_{\Lambda}=(0.5\div 1)\,{\tau}^{\rm free}_{\Lambda}$.
Since the mesonic width is less than 1\% of the total width
for $A>100$, the above consideration implies that the non--mesonic rate is rather
constant in the region of heavy hypernuclei. 

\begin{figure}
\begin{center}
\includegraphics[width=.75\textwidth]{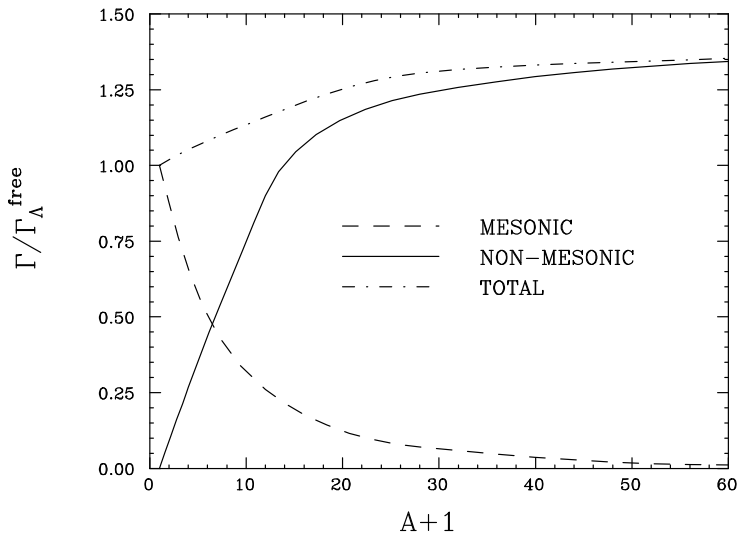}
\caption{Qualitative behaviour of mesonic, non--mesonic and total decay widths
as a function of the hypernuclear baryonic number $A+1$.}
\label{qual}
\end{center}
\end{figure}

This can be simply understood from the following consideration.
If one naively assumes a zero range approximation for the non--mesonic 
weak interaction $\Lambda N\rightarrow nN$, then $\Gamma_1$ is proportional 
to the overlap between the $\Lambda$ wave function and the nuclear density:
\begin{equation}
\Gamma_1(A)\propto \int d{\vec r}\,|\psi_{\Lambda}(\vec r)|^2\rho_A(\vec r) , 
\nonumber
\end{equation}
where the $\Lambda$ wave function $\psi_{\Lambda}$ (nuclear density $\rho_A$) 
is normalized to unity (to the nuclear mass number $A$). This overlap integral 
increases with the mass number and reaches a constant value: 
by using, e.g., $\Lambda$ harmonic oscillator wave functions 
(with frequency $\omega$ adjusted to the experimental hyperon levels in 
hypernuclei) and Fermi distributions for the nuclear densities, we find 
$\Gamma_1(^{12}_{\Lambda}{\rm C})/\Gamma_1(^{208}_{\Lambda}{\rm Pb})\simeq 0.56$,
while $\Gamma_1$ is $90$ \% of the saturation value for $A\simeq 65$. 
In figure \ref{qual} the qualitative behaviour of mesonic, non--mesonic and 
total widths as a function of the nuclear mass number $A$ is shown.

For $A\leq 11$ the experimental data are quite well fitted by 
$\Gamma_{\rm NM}/\Gamma^{\rm free}_{\Lambda}\simeq 0.1A$:
$\Gamma_1$  is proportional to the number of $\Lambda N$ pairs, 
$A$, as it is expected from the above simple description, where we neglect 
the contribution of $\Gamma_2$. However, $\Gamma_1$ saturates  when the radius
of the hypernucleus becomes sensitively larger than the range of the 
$\Lambda N\to nN$ interaction. 
For a more quantitative explanation it will be important to collect data 
with good precision. Yet, from the available data one can roughly say 
that the long distance component of the $\Lambda N\rightarrow nN$ transition 
has a range of about 1.5 fm and corresponds, as we expect, to the 
one--pion--exchange component of the interaction.

\subsection{The $\Gamma_n/\Gamma_p$ puzzle} 

Nowadays, an important question concerning the weak decay rates is the
longstanding disagreement between theoretical estimates and experimental
determinations of the ratio ${\Gamma}_n/{\Gamma}_p$
between the neutron-- and the proton--induced decay widths.

This problem will be extensively discussed in Section \ref{ratio}. However, 
it is worth recalling here that, up to short time ago, all theoretical
calculations appeared to strongly underestimate the available central data 
measured in several hypernuclei:
\begin{equation}
\left\{\frac{{\Gamma}_n}{{\Gamma}_p}\right\}^{\rm Th}\ll
\left\{\frac{{\Gamma}_n}{{\Gamma}_p}\right\}^{\rm Exp} ,
\hspace{0.2in}
0.5\lsim \left\{\frac{{\Gamma}_n}{{\Gamma}_p}\right\}^{\rm Exp}\lsim 2 .
\nonumber
\end{equation}

Until recently the data were quite limited and not precise because of the 
difficulty in detecting the products of the non--mesonic decays, especially 
the neutrons. 
Moreover, the experimental energy resolution for the detection of the outgoing
nucleons does not allow to identify the final state of the 
residual nuclei in the processes 
$^A_{\Lambda}{\rm Z}\rightarrow {^{A-2}{\rm Z}} + nn$ and
$^A_{\Lambda}{\rm Z}\rightarrow {^{A-2}({\rm Z-1})} + np$. As a consequence,
the measurements supply decay rates averaged over several nuclear final states.

In the one--pion--exchange (OPE) approximation, by assuming the 
$\Delta I=1/2$ rule in the $\Lambda \rightarrow \pi^-p$ and 
$\Lambda \rightarrow \pi^0n$ 
free couplings, the calculations (which will be reported later)
give small ratios, in the range $0.05\div 0.20$
for all the considered systems.
However, as we shall see in Section \ref{exth}, the OPE model 
with $\Delta I=1/2$ couplings has been able to reproduce the one--body 
stimulated non--mesonic rates $\Gamma_{1}=\Gamma_n+\Gamma_p$ for light 
and medium hypernuclei. Hence, the 
problem seems to consist in overestimating the proton--induced  
and underestimating the neutron--induced transition rates.

In order to solve this puzzle (namely to explain
both $\Gamma_n+\Gamma_p$ and $\Gamma_n/\Gamma_p$), many attempts have
been made up to now, mainly without success. We recall the 
inclusion in the ${\Lambda}N\rightarrow nN$
transition potential of mesons heavier than the pion (also including the 
exchange of correlated or uncorrelated two--pions) 
\cite{Du96,Pa97,It02,Pa01,Os01}, the inclusion of interaction terms that
explicitly violate the ${\Delta}I=1/2$ rule 
\cite{Pa98} and the description of the 
short range baryon--baryon interaction in terms of quark degrees of freedom
\cite{Ch83,Ok99}, which automatically introduces $\Delta I=3/2$ contributions.

\section{Theoretical models for the decay rates}
\label{model}

We illustrate here the theoretical approaches which have been utilized 
for the formal derivation of $\Lambda$ decay rates in nuclei. 
We discuss first the general features 
of the approach used for direct finite nucleus calculations. 
It is usually called Wave Function Method (WFM), since it 
 makes use of shell model nuclear and hypernuclear wave functions 
(both at hadronic and quark level) as well as pion wave functions 
generated by pion--nucleus optical potentials. 
Then we consider the Polarization Propagator Method (PPM), which relies 
on a many--body description of the hyperon self--energy in nuclear matter.
The Local Density Approximation allows then one to implement the calculation
in finite nuclei. 
Finally, a microscopic approach, based again on the PPM, is shortly sketched: in 
this case the full $\Lambda$ self--energy is evaluated on the basis of 
Feynman diagrams, which are derived, within a functional integral approach,
in the framework of the so--called bosonic loop expansion.

\subsection{Wave Function Method: mesonic width}

The weak effective Hamiltonian for the ${\Lambda}\to \pi N$ decay can
be parameterized in the form:
\begin{equation}
\label{lagran}
{\mathcal H}^W_{{\Lambda}\pi N}=iG m_{\pi}^2\overline{\psi}_N(A+B\gamma_5)
{\vec \tau} \cdot {\vec \phi}_{\pi}{\psi}_{\Lambda} ,
\end{equation}
where the values of the weak coupling constants $G= 2.211\times 10^{-7}/m_{\pi}^2$, $A=1.06$ and
$B=-7.10$ are fixed on the free ${\Lambda}$ decay. The constants $A$ and $B$
determine the strengths of the parity
violating and parity conserving ${\Lambda}\rightarrow \pi N$ amplitudes, respectively.
In order to enforce the $\Delta I=1/2$ rule (which fixes
$\Gamma^{\rm free}_{\pi^-}/\Gamma^{\rm free}_{\pi^0}=2$), in Eq.~(\ref{lagran})
the hyperon is assumed to be an isospin spurion with $I=1/2$, $I_z=-1/2$.

In the non--relativistic approximation, the free $\Lambda$ decay width 
$\Gamma^{\rm free}_{\Lambda}=\Gamma^{\rm free}_{\pi^-}+\Gamma^{\rm free}_{\pi^0}$
is given by:
\begin{equation}
\Gamma^{\rm free}_{\alpha}=c_{\alpha}(G m^2_{\pi})^2\int
\frac{d\vec q}{(2\pi)^3\,2\omega({\vec q})}\,2\pi\,
\delta[m_{\Lambda}-\omega({\vec q})-E_N] \left(S^2+\frac{P^2}{m^2_{\pi}}{\vec q}\,^2\right) ,
\nonumber
\end{equation}
where $c_{\alpha}=1$ for $\Gamma_{\pi^0}$ and $c_{\alpha}=2$  for $\Gamma_{\pi^-}$
(expressing the $\Delta I=1/2$ rule),
$S=A$, $P=m_{\pi}B/(2m_N)$, whereas $E_N$ and $\omega(\vec q)$ 
are the total energies of nucleon and pion, respectively.
One finds the well known result:
\begin{equation}
\Gamma^{\rm free}_{\alpha}=c_{\alpha}(G m^2_{\pi})^2\frac{1}{2\pi}
\frac{m_Nq_{\rm c.m.}}{m_{\Lambda}}
\left(S^2+\frac{P^2}{m^2_{\pi}}q^2_{\rm c.m.}\right) ,\nonumber 
\end{equation}
which reproduces the observed rates. In the previous equation,
$q_{\rm c.m.}\simeq 100$ MeV is the pion momentum in the center--of--mass frame.

In a finite nucleus approach, the \emph{mesonic width}
$\Gamma_{\rm M}=\Gamma_{\pi^-}+\Gamma_{\pi^0}$ can be calculated by means
of the following formula \cite{Os93,Mo94}:
\begin{eqnarray}
\Gamma_{\alpha}&=&c_{\alpha}(G m^2_{\pi})^2\sum_{N\non F}\int
\frac{d\vec q}{(2\pi)^3\,2\omega({\vec q})}\,2\pi\,
\delta[E_{\Lambda}-\omega({\vec q})-E_N] \nonumber \\
&&\times \left\{S^2\left|\int d{\vec r} \phi_{\Lambda}(\vec r)
\phi_{\pi}({\vec q}, {\vec r})\phi^*_N(\vec r)\right|^2
+\frac{P^2}{m^2_{\pi}}
\left|\int d{\vec r} \phi_{\Lambda}(\vec r){\vec \bigtriangledown}
\phi_{\pi}({\vec q}, {\vec r}) \phi^*_N(\vec r)\right|^2 \right\} , \nonumber
\end{eqnarray}
where the sum runs over non--occupied nucleonic states and $E_{\Lambda}$ is the
hyperon total energy. The $\Lambda$ and nucleon wave functions $\phi_{\Lambda}$ and $\phi_N$
are obtainable within a shell model. The pion wave function $\phi_{\pi}$ 
corresponds to an outgoing wave, solution of the Klein--Gordon equation with the appropriate
pion--nucleus optical potential $V_{\rm opt}$:
\begin{equation}
\left\{{\vec \bigtriangledown}^2-m^2_{\pi}-2\omega V_{\rm opt}(\vec r)
+\left[\omega-V_C(\vec r)\right]^2\right\}
\phi_{\pi}({\vec q}, {\vec r})=0 , \nonumber
\end{equation}
where $V_C(\vec r)$ is the nuclear Coulomb potential and the energy eigenvalue
$\omega$ depends on $\vec q$.

Different calculations
\cite{Os93,Mo94} have shown how strongly the mesonic decay is sensitive to
the pion--nucleus optical potential, which can be parameterized in terms of the
nuclear density, as discussed in Refs.~\cite{Mo94}, or evaluated microscopically,
as in Ref.~\cite{Os93}.

\subsection{Wave Function Method: non--mesonic width}

Within the one--meson--exchange (OME) mechanism, the weak transition $\Lambda N\to nN$
is assumed to proceed via the mediation of virtual mesons of the pseudoscalar
($\pi$, $\eta$ and $K$) and vector ($\rho$, $\omega$ and $K^*$) octets
\cite{Du96,Pa97} (see Fig~\ref{nm12}).

The fundamental ingredients for the calculation of the $\Lambda N\to nN$ transition
within a OME model are the weak and strong hadronic vertices.
The ${\Lambda}\pi N$ weak Hamiltonian is given in Eq.~(\ref{lagran}).
For the strong $NN\pi$ Hamiltonian one has the usual pseudoscalar coupling:
\begin{equation}
{\mathcal H}^S_{NN\pi}=ig_{NN\pi}\overline{\psi}_N\gamma_5
{\vec \tau} \cdot {\vec \phi}_{\pi}{\psi}_N , \nonumber
\end{equation}
$g_{NN\pi}$ being the strong coupling constant for the $NN\pi$ vertex.
In momentum space, the non--relativistic transition potential in the OPE approximation
is then:
\begin{equation}
V_{\pi}({\vec q})=-G m_{\pi}^2\frac{g_{NN\pi}}{2m_N}\left(
A+\frac{B}{2\bar{m}}{\vec \sigma_1} \cdot {\vec q}\right)
\frac{{\vec \sigma_2} \cdot {\vec q}}{{\vec q}\,^2+m_{\pi}^2} 
{\vec \tau_1} \cdot {\vec \tau_2}, \nonumber
\end{equation}
where $\bar{m}=(m_{\Lambda}+m_N)/2$ and
${\vec q}$ is the momentum of the exchanged pion.

Due to the large momenta ($\simeq 420$ MeV) exchanged in the $\Lambda N\to nN$
transition, the OPE mechanism describes the long range part of the interaction,
and more massive mesons are expected to contribute at shorter distances.

Non--trivial difficulties arise with the heavier mesons, since their weak 
couplings in the $\Lam N$ vertex are not known experimentally. For example, if
one includes in the calculation the contribution of the $\rho$--meson,
the weak $\Lambda N\rho$ and strong $NN\rho$ Hamiltonians give rise to
the following $\rho$--meson transition potential:
\begin{eqnarray}
V_{\rho}({\vec q})=G m_{\pi}^2&&\left[g^V_{NN\rho}\alpha -
\frac{(\alpha+\beta)(g^V_{NN\rho}+g^T_{NN\rho})}{4m_n m}
({\vec \sigma}_1 \times {\vec q})\cdot ({\vec \sigma}_2 \times {\vec q}) \right. \nonumber \\
&&\left. +i\frac{\epsilon (g^V_{NN\rho}+g^T_{NN\rho})}{2m_m}({\vec \sigma}_1 \times {\vec \sigma}_2)
\cdot {\vec q}\right]
\frac{{\vec \tau}_1 \cdot {\vec \tau}_2}{{\vec q}\,^2+m^2_{\rho}}\,, 
\nonumber
\end{eqnarray}
where the weak coupling constants $\alpha$, $\beta$ and $\epsilon$ 
must be evaluated theoretically and turn out to be quite model--dependent.

The most general OME potential accounting for the exchange of pseudoscalar
and vector mesons can be expressed through the following decomposition:
\begin{equation}
\label{pot-ome}
V(\vec r)=\sum_m V_m(\vec r)=
\sum_m \sum_{\alpha}V^{\alpha}_m(r)\hat{O}^{\alpha}(\hat{{\vec r}})\hat{I}_m ,
\end{equation}
where $m=\pi$, $\rho$, $K$, $K^*$, $\omega$, $\eta$; the spin operators
$\hat{O}^{\alpha}$ are (PV stands for parity--violating):  
\begin{equation} 
\hat{O}^{\alpha}(\hat{{\vec r}})=
\begin{cases}
\hat{1} & \text{central spin--independent} , \\
{\vec \sigma}_1 \cdot {\vec \sigma}_2 & \text{central spin--dependent } , \\
S_{12}(\hat{{\vec r}})=3({\vec \sigma}_1 \cdot \hat{{\vec r}})
({\vec \sigma}_2 \cdot \hat{{\vec r}})-{\vec \sigma}_1 \cdot {\vec \sigma}_2 & \text{tensor} , \\
{\vec \sigma}_2 \cdot \hat{{\vec r}} & \text{PV for pseudoscalar mesons} , \\
({\vec \sigma}_1 \times {\vec \sigma}_2)\cdot \hat{{\vec r}} &
\text{PV for vector mesons},
\end{cases} \nonumber
\end{equation}
whereas the isospin operators $\hat{I}_m$ are:
\begin{equation}
\hat{I}_m=
\begin{cases}
\hat{1} & \text{isoscalars mesons ($\eta$, $\omega$)} , \\
{\vec \tau}_1 \cdot {\vec \tau}_2 & \text{isovector mesons ($\pi$, $\rho$)} ,\\
\text{linear combination of $\hat{1}$ and ${\vec \tau}_1 \cdot {\vec \tau}_2$} &
\text{isodoublet mesons ($K$, $K^*$)} .
\end{cases} \nonumber
\end{equation}  
For details concerning the potential (\ref{pot-ome}), see Ref.~\cite{Pa97,Pa01}.

Assuming the initial hypernucleus to be at rest,
the \emph{one--body induced non--mesonic decay rate} can then be written as:
\begin{equation}
\label{nm-wfm}
\Gamma_1=\int \frac{d{\vec p}_1}{(2\pi)^3}\int \frac{d{\vec p}_2}{(2\pi)^3}
\,2\pi\, \delta({\rm E.C.}) \overline{{\sum}}\left|{\mathcal M}({\vec p}_1,{\vec p}_2)\right|^2 ,
\end{equation} 
where $\delta({\rm E.C.})$ stands for the energy conserving delta function:
\begin{equation}
\delta({\rm E.C.})=\delta\left(m_H-E_{\it R}-2m_N-\frac{{\vec p}^{\,2}_1}{2m_N}-
\frac{{\vec p}^{\,2}_2}{2m_N}\right) . \nonumber
\end{equation}
Moreover:
\begin{equation}
{\mathcal M}({\vec p}_1,{\vec p}_2)
\equiv \langle \Psi_R; N({\vec p}_1)N({\vec p}_2)|
\hat{T}_{\Lambda N\to NN}|\Psi_H\rangle  \nonumber
\end{equation}
is the amplitude for the transition of the initial hypernuclear state $|\Psi_H\rangle$
of mass $m_H$ into a final state composed by a
residual nucleus $|\Psi_R\rangle$ with energy $E_R$ and an antisymmetrized
two nucleon state $|N({\vec p}_1)N({\vec p}_2)\rangle$, 
${\vec p}_1$ and ${\vec p}_2$ being the nucleon momenta. The 
sum $\overline{\sum}$
in Eq.~(\ref{nm-wfm}) indicates an average over the third component of the
hypernuclear total spin and a sum over the quantum numbers of the residual
system and over the spin and isospin third components of the outgoing
nucleons. Customarily, in shell model calculations
the weak--coupling scheme is used to describe the
hypernuclear wave function $|\Psi_H\rangle$, the nuclear core wave function
being obtained through the technique
of fractional parentage coefficients \cite{Pa97}. The many--body transition 
amplitude ${\mathcal M}({\vec p}_1,{\vec p}_2)$ is then expressed in terms of
two--body amplitudes $\langle NN|V|\Lambda N \rangle$ of the OME potential of
Eq.~(\ref{pot-ome}).

Two merits of the WFM must be remarked:
\begin{itemize}
\item
Since the $\Lambda$ decays from an orbital angular momentum $l=0$ state,
in the non--mesonic decay rate one can easily isolate 
the contributions of neutron-- and proton--induced transitions \cite{Pa97}, 
and the $\Gamma_n/\Gamma_p$ ratio can be directly evaluated. 
\item
The $nN$ final state interactions and the $\Lambda N$ correlations
(which are absent in an independent particle shell model)
can also be implemented in the calculation \cite{Pa97,Pa01}.
\end{itemize}

\subsection{Polarization Propagator Method and Local Density Approximation}
\label{pm}

The hypernuclear decay rates can be studied by using the 
Polarization Propagator Method~\cite{Os82} to 
evaluate the $\Lam$ self--energy inside the
nuclear medium. The polarization propagator is conveniently calculated
for a homogeneous system (nuclear matter), within the Random Phase 
Approximation (RPA) and eventually accounting for additional correlations.
The calculation can then be extended to finite nuclei via the Local Density 
Approximation (LDA).

This many--body technique provides a unified picture
of the different decay channels and it is equivalent to the
WFM \cite{Os94} (in the sense that it is a semiclassical
approximation of the exact quantum mechanical problem).
Obviously, for the mesonic rates the WFM is more reliable than the
PPM in LDA, since $\Gamma_M$ is rather sensitive to the 
shell structure of the hypernucleus. On the other
hand, the propagator method in LDA offers the possibility of calculating the
hypernuclear decay rates over a broad range of mass numbers, while the WFM 
is hardly exploitable for medium and heavy hypernuclei. 

To calculate the hypernuclear width one needs the 
imaginary part of the ${\Lambda}$ self--energy:
\begin{equation}
\label{Gamma}
{\Gamma}_{\Lambda}=-2\;{\rm Im}\,{\Sigma}_{\Lambda} ,
\end{equation}
which, in the non--relativistic limit, reads:
\begin{equation}
\label{Sigma1}
{\Sigma}_{\Lambda}(k)=3i(G m_{\pi}^2)^2\int \frac{d^4q}{(2\pi)^4}
\left(S^2+\frac{P^2}{m_{\pi}^2}\vec q\,^2\right)F_{\pi}^2(q)
G_N(k-q)G_{\pi}(q) .
\end{equation}
The nucleon and pion propagators in nuclear matter are, respectively:
\begin{equation}
G_N(p)=\frac{{\theta}(\mid \vec p \mid-k_F)}{p_0-E_N(\vec p)-V_N+i{\epsilon}}+
\frac{{\theta}(k_F-\mid \vec p \mid)}{p_0-E_N(\vec p)-V_N-i{\epsilon}} , 
\end{equation}
and:
\begin{equation}
\label{proppion}
G_{\pi}(q)=\frac{1}{q_0^2-\vec q\,^2-m_{\pi}^2-{\Sigma}_{\pi}^*(q)} .
\end{equation}
The above form of the non--relativistic nucleon propagator refers to a
non--interacting Fermi system but includes corrections due to Pauli principle
and an average binding. 
In the previous equations, $p=(p_0,\vec p)$ and $q=(q_0,\vec q)$ denote 
four--vectors, $k_F$ is the Fermi momentum, $E_N$ is the nucleon total free
energy, $V_N$ is the nucleon binding energy (which is density--dependent), 
and ${\Sigma}_{\pi}^*$ is the pion proper self--energy in nuclear matter. 
This quantity has been carefully evaluated within several many--body frameworks
and includes the (strong) coupling of the pion to particle--hole ({\sl p--h}) states,
collective RPA states and more complicated nuclear correlated states (e.g., the
two--particle two--hole states). A monopole form factor $F_\pi(q)$ describing
the hadronic structure of the $\pi\Lambda N$ vertex is also included
in Eq.~(\ref{Sigma1}).

We note here that the parity--conserving
term ($l=1$ term) in Eq.~(\ref{Sigma1}) contributes only about 12\% of the 
total free decay width. However, the $P$--wave interaction becomes dominant 
in the nuclear non--mesonic decay, because of the larger exchanged momenta. 

\begin{figure}
\begin{center}
\includegraphics[width=.8\textwidth]{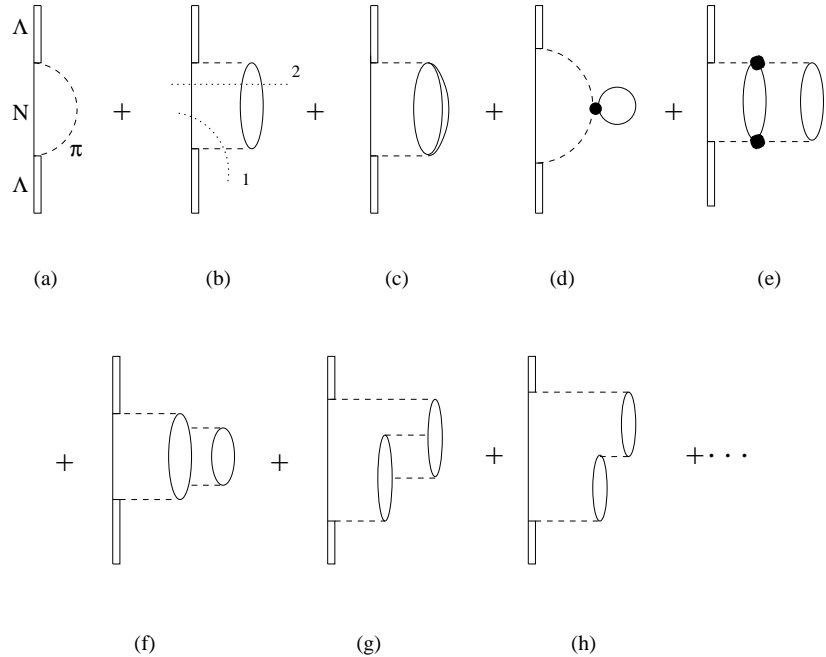}
\caption{Lowest order terms for the ${\Lambda}$ self--energy in 
nuclear matter. The meaning of the various diagrams is explained in the text.}
\label{self1}
\end{center}
\end{figure}

In Fig.~\ref{self1} we show the lowest order Feynman diagrams for the 
${\Lambda}$ self--energy in nuclear matter. Diagram (a) represents the bare
self--energy term, including the effects of the Pauli principle and of
binding on the intermediate nucleon.
In (b) and (c) the pion couples to a {\sl p--h} and a 
{\sl ${\Delta}$--h} pair, respectively. Diagram (d) is an insertion of 
$S$--wave pion self--energy at lowest order. In diagram (e) we show a 
{\sl 2p--2h} excitation coupled to the pion through $S$--wave $\pi N$ 
interactions. Other {\sl 2p--2h} excitations, coupled in $P$--wave, are 
shown in (f) and (g), while (h) is a RPA iteration of diagram (b). 

In Eq.~(\ref{Sigma1}) there are two different sources of imaginary part. 
The analytical structure of the integrand allows the integration over 
$q_0$ \cite{Os85}. After performing this integration, an imaginary part 
is obtained from the (renormalized) pion--nucleon pole and physically 
corresponds to the mesonic decay of the hyperon. Moreover, the
pion proper self--energy $\Sigma^*_{\pi}(q)$ has an imaginary part itself
for $(q_0, \vec q)$ values which correspond to the excitation of {\sl p--h},
{$\Delta$--h}, {\sl 2p--2h}, etc states on the mass shell. By expanding the 
pion propagator $G_{\pi}(q)$ as in Fig.~\ref{self1} and integrating 
Eq.~(\ref{Sigma1}) over $q_0$, the nuclear matter $\Lambda$ decay width of 
Eq.~(\ref{Gamma}) becomes \cite{Os85}:
\begin{eqnarray}
\label{Sigma2}
{\Gamma}_{\Lambda}(\vec k,\rho)&=&-6(G m_{\pi}^2)^2\int \frac{d\vec q}
{(2\pi)^3}{\theta}(\mid \vec k- \vec q \mid -k_F)
{\theta}(k_0-E_N(\vec k-\vec q)-V_N) \nonumber \\
& & \times {\rm Im}\left[{\alpha}(q)\right]_{q_0=k_0-E_N(\vec k-\vec q)-V_N} ,
\end{eqnarray}
where:
\begin{eqnarray}
\label{Alpha}
{\alpha}(q)&=&\left(S^2+\frac{P^2}{m_{\pi}^2}\vec q\,^2\right)F_{\pi}^2(q)
G_{\pi}^0(q)+\frac{\tilde{S}^2(q)U_L(q)}{1-V_L(q)U_L(q)} \nonumber \\
& & +\frac{\tilde{P}_L^2(q)U_L(q)}{1-V_L(q)U_L(q)}+
2\frac{\tilde{P}_T^2(q)U_T(q)}{1-V_T(q)U_T(q)} .
\end{eqnarray}
In Eq.~(\ref{Sigma2}) the first ${\theta}$ function forbids
intermediate nucleon momenta smaller than the Fermi momentum, while
the second one requires the pion energy $q_0$ to be 
positive. Moreover, the ${\Lambda}$ energy, 
$k_0=E_{\Lambda}(\vec k)+V_{\Lambda}$,
contains a phenomenological binding term. 
With the exception of diagram (a), the pion lines of Fig.~\ref{self1} 
have been replaced, in Eq.~(\ref{Alpha}), by the effective interactions 
$\tilde{S}$, $\tilde{P}_L$, $\tilde{P}_T$,$V_L$, $V_T$ 
($L$ and $T$ stand for spin--longitudinal and spin--transverse, respectively),
which include $\pi$-- and $\rho$--exchange
modulated by the effect of short range repulsive correlations.
The potentials $V_L$ and $V_T$ represent the (strong) {\sl p--h} 
interaction and include a Landau parameter $g^{\prime}$,
which accounts for the short range repulsion, 
while $\tilde{S}$, $\tilde{P}_L$ and
$\tilde{P}_T$ correspond to the lines connecting weak and strong 
hadronic vertices and contain another Landau parameter, $g^{\prime}_{\Lambda}$,
which is related to the strong $\Lambda N$
short range correlations. 

Furthermore, in Eq.~(\ref{Alpha}):
\begin{equation}
G_{\pi}^0(q)=\frac{1}{q_0^2-\vec q\,^2-m_{\pi}^2} , \nonumber
\end{equation}
is the free pion propagator,
while $U_L(q)$ and $U_T(q)$ contain the Lindhard functions for {\sl p--h} and 
{\sl ${\Delta}$--h} excitations \cite{Wa71}
and also account for the irreducible {\sl 2p--2h} polarization propagator:
\begin{equation}
\label{propU}
U_{L,T}(q)=U^{ph}(q)+U^{\Delta h}(q)+U^{2p2h}_{L,T}(q) .
\end{equation}
They appear in Eq.~(\ref{Alpha}) within the standard RPA expression:
\bea
\frac{U_{L(T)}(q)}{1-V_{L(T)}(q)U_{L(T)}(q)} &=& U_{L(T)}(q) +
U_{L(T)}(q)V_{L(T)}(q)U_{L(T)}(q) + \nonumber\\
&& + U_{L(T)}(q)V_{L(T)}(q)U_{L(T)}(q)V_{L(T)}(q)U_{L(T)}(q) + \dots
\label{propU-rpa}
\eea
The decay width (\ref{Sigma2}) depends both explicitly and through $U_{L,T}(q)$
on the nuclear matter density $\rho=2k_F^3/3{\pi}^2$.

The Lindhard function $U^{ph}$  ($U^{\Delta h}$) is given by:
\beq
U^{ph (\Delta h)}(q)=-4i\int\frac{d^4 p}{(2\pi)^3} 
G^0_N(p)G^0_{N(\Delta)}(p+q) ,
\label{Uph}
\eeq
with the \emph{free} nucleon and Delta propagators:
\beq
 G^0_N(p)= \frac{{\theta}(\mid\vec p\mid-k_F)}
{p_0-T_N(\vec p)+i{\epsilon}}+
\frac{{\theta}(k_F-\mid \vec p \mid)}{p_0-T_N(\vec p)-i{\epsilon}} ,
\label{G0N}
\eeq
\beq
G^0_{\Delta}(p)= \frac{4}{9}\,\frac{1}
{p_0-T_{\Delta}(\vec p)-\delta M_{\Delta N}+i\,\Gamma_\Delta/2} ,
\label{GDelta}
\eeq
where $T_{N(\Delta)}$ is the nucleon (Delta) kinetic energy, $\Gamma_\Delta$ the
free $\Delta$ width and $\delta M_{\Delta N}=m_\Delta-m_N$.
We remind that $U^{ph}(q)$ and $U^{\Delta h}(q)$ can be analytically 
evaluated in nuclear matter.

For the evaluation of $U^{2p2h}_{L,T}$ we discuss two different
approaches. In Refs.~\cite{Ra95,Al99} a phenomenological parameterization was 
adopted: this consists in relating 
$U^{2p2h}_{L,T}$ to the available phase space for on--shell {\sl 2p--2h} 
excitations in order
to extrapolate for off--mass shell pions the experimental data of
$P$--wave absorption of real pions in pionic atoms. 
In an alternative approach~\cite{Al99a}, $U^{2p2h}_{L,T}$ is
evaluated microscopically, starting from a classification of the relevant 
Feynman diagrams according to the so--called bosonic loop expansion, 
which is obtained within a functional approach. 

\begin{figure}
\begin{center}
\includegraphics[width=.8\textwidth]{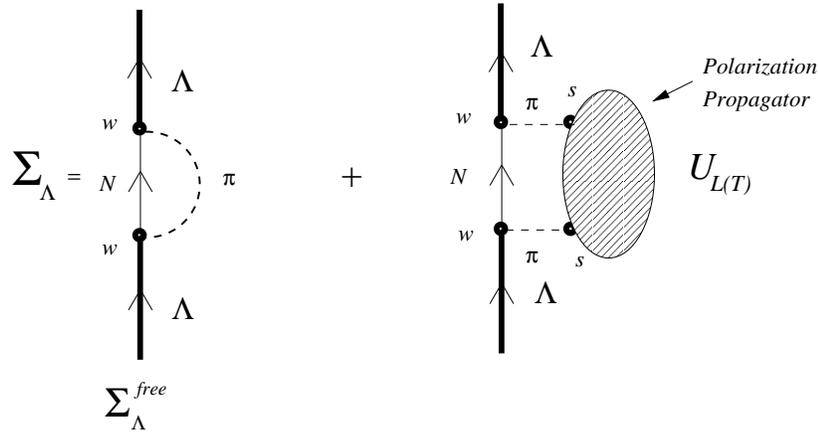}
\caption{Schematic representation of the total $\Lambda$ self-energy in 
terms of the $\pi$--self energy and polarization insertions; in the 
vertexes $w={\rm weak}$, $s={\rm strong}$.}
\label{prun1}
\end{center}
\end{figure}

We notice here, in connection with Eqs.~(\ref{Sigma1}) and (\ref{proppion}),
that the full (proper) pion self--energy 
\begin{equation}
{\Sigma}_{\pi}^*(q)={\Sigma}_{\pi}^{(S)\,*}(q)+{\Sigma}_{\pi}^{(P)\,*}(q) , 
\nonumber
\end{equation}
contains a $P$--wave term, which is related to the spin--longitudinal 
polarization propagator $U_L(q)$ according to:
\begin{equation}
{\Sigma}_{\pi}^{(P)\,*}(q)=\frac{\displaystyle \vec q\,^2
\frac{f_{\pi}^2}{m_{\pi}^2} F_{\pi}^2(q)U_L(q)}
{1-\displaystyle \frac{f_{\pi}^2}{m_{\pi}^2}g_L(q)U_L(q)} , 
\label{SigmaP-wave}
\end{equation}
the Landau function $g_L(q)$ being given in the appendix of Ref.~\cite{Al02}.
${\Sigma}_{\pi}^*(q)$ also contains an $S$--wave term, which, by using
the parameterization of Ref.~\cite{Se83}, can be written as: 
\begin{equation}
\Sigma_{\pi}^{(S)\,*}(q)=-4\pi \left(1+\frac{m_{\pi}}{m_N}\right)b_0\rho .
 \label{SigmaS-wave}
\end{equation}
The parameter $b_0$ is usually taken from the phenomenology of pionic 
atoms (see, for example, Ref.~\cite{Fried-Gal03}).
 The function ${\Sigma}_{\pi}^{(S)\,*}$ is real (constant and positive), 
therefore it contributes only to the mesonic decay 
[diagram (d) in Fig.~\ref{self1} is the relative lowest order]. 
On the contrary, the $P$--wave self--energy is complex and attractive: 
${\rm Re}\;{\Sigma}_{\pi}^{(P)\,*}(q)<0$. It contributes to all the decay 
channels, in particular to the two--body induced non--mesonic decay width.
A schematic picture of the relation between the $\Lam$ and $\pi$ 
self--energies is illustrated in Fig.~\ref{prun1}, where the polarization 
propagator insertion on the pionic line summarizes all contributions 
explicitly shown in Fig.~\ref{self1} and many others, up to infinite 
order.

The propagator method provides a unified picture of the decay widths. 
A non--vanishing imaginary part in a self--energy diagram requires placing 
simultaneously on--shell the particles of the considered intermediate state. 
For instance, diagram (b) in Fig.~\ref{self1} has two sources
of imaginary part.  One comes from cut 1, where the nucleon and the pion are 
placed on--shell. This term contributes to the mesonic channel: the final pion 
eventually interacts with the
medium through a {\sl p--h} excitation and then escapes from the nucleus. 
Diagram (b) and further iterations lead to a 
renormalization of the pion in the medium which may increase the mesonic rate 
even by one or two orders of magnitude in heavy nuclei \cite{Os85,Os93,Mo94}. 
The cut 2 in Fig.~\ref{self1}(b) places a nucleon and a {\sl p--h} pair on shell, 
so it is the lowest order contribution to the physical process 
$\Lambda N\to nN$. Analogous considerations apply to all the considered 
diagrams. 

\begin{figure}
\begin{center}
\includegraphics[width=.8\textwidth]{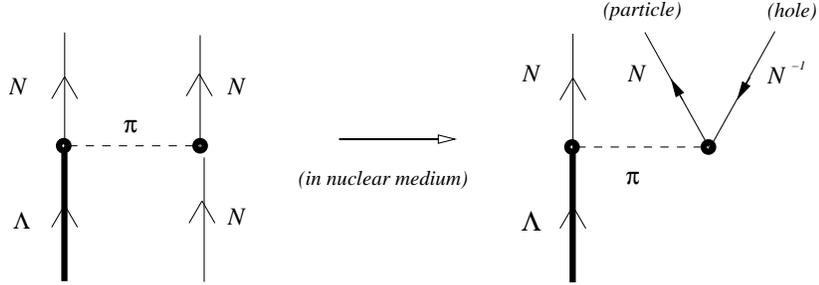}
\caption{One--body induced process in the language of particle--hole states.}
\label{prun2}
\end{center}
\end{figure}

\begin{figure}
\begin{center}
\includegraphics[width=.7\textwidth]{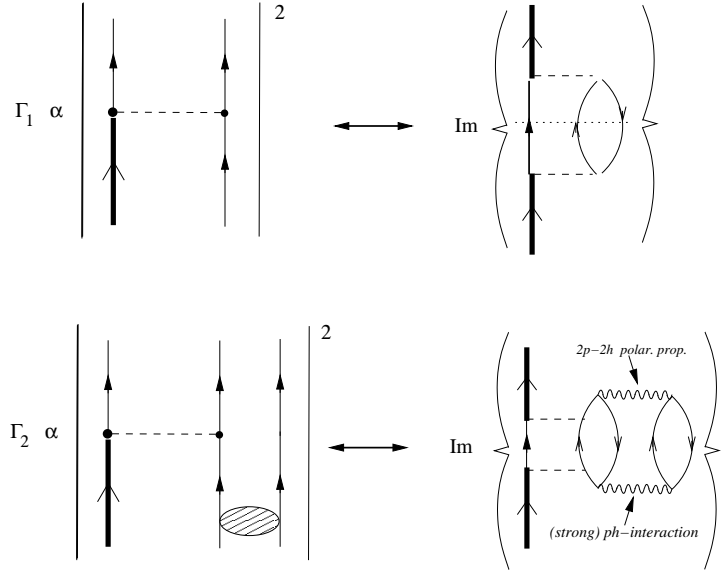}
\caption{Schematic relation between the one-body (upper panel) and
two--body (lower panel) $\Lambda$ decay amplitudes and the Imaginary
part of specific contributions to the $\Lam$ self-energy.}
\label{prun3}
\end{center}
\end{figure}

A couple of remarks may be useful, here, for those who are not familiar 
with the language of many--body theory.
\begin{enumerate}
\item
The one--body induced process $\Lam N\to nN$ (for example via the exchange 
of a pion) translates, inside the nuclear medium, into the creation of 
a particle--hole pair, the hole representing the state of the initial 
nucleon, which turns out to be vacant, in the medium, after the weak 
interaction with the $\Lam$. This is schematically illustrated in 
Fig.~\ref{prun2}.
\item
The relation between the one--body induced decay width, $\Gamma_1$, and the
imaginary part of the $\Lam$ self--energy, say, of diagram (b) in 
Fig.~\ref{self1}, is represented in the upper panel of Fig.~\ref{prun3}, 
where the modulus square of the decay amplitude is pictorially related to
the Imaginary part of the considered diagram, with a cut putting on--shell
the 2 nucleon, 1 hole intermediate state.
Analogously, the two--body induced decay width is related to the Imaginary
part of diagrams like, e.g., Fig.~\ref{self1}(f), the cut being placed on 
the 2p--2h intermediate state of the polarization propagator. This is 
represented in the lower panel of Fig.~\ref{prun3}.
\end{enumerate}

In order to evaluate the various contributions to the width stemming 
from Eq.~(\ref{Sigma2}), it is convenient to consider all the intervening 
free meson propagators as real. Then the
imaginary part of (\ref{Alpha}) will develop the following contributions:
\begin{equation}
\label{pres}
{\rm Im}\, \frac{U_{L,T}(q)}{1-V_{L,T}(q)U_{L,T}(q)}=
\frac{{\rm Im}\,U^{ph}(q)+{\rm Im}\,U^{\Delta h}(q)+{\rm Im}
\,U^{2p2h}_{L,T}(q)}
{\mid 1-V_{L,T}(q)U_{L,T}(q)\mid ^2} .
\end{equation}
The three terms in the numerator of Eq.~(\ref{pres}) can be interpreted as 
different decay mechanisms of the hypernucleus. The term proportional to 
${\rm Im}\,U^{ph}$ provides the one--nucleon induced non--mesonic rate, 
$\Gamma_1$. There is no overlap between ${\rm Im}\,U^{ph}(q)$ and the pole 
$q_0=\omega(\vec q)$ in the (dressed) pion propagator $G_{\pi}(q)$: 
thus the separation of the mesonic and one--body
stimulated non--mesonic channels is unambiguous. 

Further, ${\rm Im}\,U^{\Delta h}$
accounts for the $\Delta \to \pi N$ decay width, thus representing a 
(small) contribution to the mesonic decay. 

The third contribution of Eq.~(\ref{pres}), 
proportional to ${\rm Im}\,U^{2p2h}_{L,T}$, intervenes in a 
wide kinematic range, in which the above mentioned cuts put on the mass shell
not only the {\sl 2p--2h} lines, but possibly also the pionic lines.
Indeed, the renormalized pion pole in Eq.~(\ref{proppion}) is given by the
dispersion relation:
\begin{equation}
{\omega}^2(\vec q)-\vec q\,^2-m_{\pi}^2-{\rm Re}\,{\Sigma}_{\pi}^*
\left[{\omega}(\vec q),\vec q\,\right]=0 , \nonumber
\end{equation}
with the constraint:
\begin{equation}
\omega({\vec q})=k_0-E_N(\vec k -\vec q)-V_N . \nonumber
\end{equation}
At the pion pole, ${\rm Im}\,U^{2p2h}_{L,T}\neq 0$, thus the two--body induced
non--mesonic width, $\Gamma_2$, cannot be disentangled from the mesonic
width, $\Gamma_{\rm M}$. In other words, part of the decay rate calculated 
from ${\rm Im}\,U^{2p2h}_{L,T}$ is due to the excitations of the renormalized 
pion and gives in fact $\Gamma_{\rm M}$, with the exception of the mesonic 
contribution originating from ${\rm Im}\,U^{\Delta h}$, which is, however, 
only a small fraction of $\Gamma_{\rm M}$. 

In order to separate $\Gamma_{\rm M}$ from 
$\Gamma_2$, in the numerical calculation it is convenient to evaluate the
mesonic width by adopting the following prescription. We start from
Eq.~(\ref{Sigma2}), setting:
\begin{equation} 
\label{alphaprop} 
{\alpha}(q)={\alpha}_M(q)\equiv
\left(S^2+\frac{P^2}{m_{\pi}^2}\vec q\,^2\right)F^2_{\pi}(q)G_{\pi}(q) ,
\end{equation} 
and omitting ${\rm Im}\,{\Sigma}_{\pi}^*$ in $G_{\pi}$ (which corresponds to 
assume ${\rm Im}\, U^{ph}={\rm Im}\, U^{\Delta h}={\rm Im}\, 
U^{2p2h}_{L,T}=0$).
Then ${\rm Im}\,{\alpha}_M(q)$ only accounts for the (real) contribution of the
pion pole:
\begin{equation}
{\rm Im}\,G_{\pi}(q)= -\pi \delta \left[q_0^2-\vec q\,^2-m_{\pi}^2- 
{\rm Re}\,{\Sigma}_{\pi}^*(q)\right] . \nonumber 
\end{equation} 

Once the mesonic decay rate is known, one can calculate the three--body
non--mesonic rate by subtracting $\Gamma_{\rm M}$ and $\Gamma_1$ from the
total rate $\Gamma_{\rm T}$, which one gets via the full expression for
$\alpha(q)$ [Eq.~(\ref{Alpha})].

Using the PPM, the decay widths of finite nuclei can be obtained from the 
ones evaluated in nuclear matter via the LDA: in this approach 
the Fermi momentum is made $r$--dependent (namely 
a local Fermi sea of nucleons is introduced) and related to the 
nuclear density by the same relation which holds in nuclear matter: 
\begin{equation} 
\label{local} 
k_F(\vec r)=\left\{\frac{3}{2}{\pi}^2\rho 
(\vec r)\right\}^{1/3} . 
\end{equation} 
More specifically, one usually assumes the nuclear density to be a Fermi
distribution:
\beq
\rho_A(r)= \frac{\rho_0}
{\left\{\displaystyle 1+\exp\left[\frac{r-R(A)}{a}\right]\right\}}
\nonumber
\eeq
with $R(A)=1.12A^{1/3}-0.86A^{-1/3}$~fm, $a=0.52$~fm and
$\rho_0={A}\left\{\frac{4}{3}\pi R^3(A)
\left(1+\left[\frac{\pi a}{R(A)}\right]^2\right)\right\}^{-1}$.

Moreover, the nucleon binding potential $V_N$ also becomes $r$--dependent in 
LDA. In Thomas--Fermi approximation one assumes: 
\begin{equation} 
\epsilon_F(\vec r)+V_N(\vec r)\equiv \frac{k_F^2(\vec r)}{2m_N}+V_N(\vec r)=0 . \nonumber
\end{equation} 

With these prescriptions one can then evaluate the decay width in finite 
nuclei by using the semiclassical approximation, through the relation: 
\begin{equation} 
\label{local1} 
{\Gamma}_{\Lam}(\vec k)=\int d\vec r \,| {\psi}_{\Lambda}(\vec r)| ^2 
{\Gamma} _{\Lambda}\left[\vec k,\rho (\vec r)\right] , 
\end{equation} 
where ${\psi}_{\Lambda}$ is the appropriate ${\Lambda}$ wave function and 
${\Gamma} _{\Lambda}[\vec k,\rho (\vec r)]$ is given by 
Eqs.~(\ref{Sigma2}), (\ref{Alpha}) with a position dependent nuclear 
density. This decay rate can be regarded as the 
$\vec k$--component of the ${\Lambda}$ decay rate in the nucleus with density 
$\rho(\vec r)$. It can be used to estimate the decay rates 
by averaging over the ${\Lambda}$ momentum distribution 
$|\tilde{\psi}_{\Lambda}(\vec k)|^2$. One then obtains the
following total width: 
\begin{equation} 
\label{local2} 
{\Gamma}_{\rm T}=\int d\vec k \,| \tilde{\psi}_{\Lambda}(\vec k)|^2{\Gamma}_{\Lambda}
(\vec k) ,
\end{equation} 
which can be compared with the experimental results.

\subsubsection{{\sl Phenomenological {\sl 2p--2h} propagator}}
\label{p2p2hp}

Coming to the phenomenological evaluation of the {\sl 2p--2h}
contributions in the $\Lambda$ self--energy, the momentum dependence of 
the imaginary part of $U^{2p2h}_{L,T}$ can be obtained from the available 
phase space, through the following equation~\cite{Ra95}:
\begin{equation}
\label{pheim}
{\rm Im}\, U^{2p2h}_{L,T}(q_0,\vec q;\rho)=\frac{P(q_0,\vec q;\rho)}
{P(m_{\pi},\vec 0;\rho_{\rm eff})}
{\rm Im}\, U^{2p2h}_{L,T}(m_{\pi}, \vec 0; \rho_{\rm eff}) ,
\end{equation}
where $\rho_{\rm eff}=0.75\rho$. By neglecting the energy and 
momentum dependence of the {\sl p--h} interaction,
the phase space available for on--shell {\sl 2p--2h} excitations
[calculated, for simplicity, from
diagram \ref{self1}(e)] at energy--momentum $(q_0,\vec q)$ and density $\rho$
turns out to be:
\begin{eqnarray}
P(q_0,\vec q;\rho)&\propto &\int \frac{d^4k}{(2\pi)^4}\,{\rm Im}\, U^{ph}
\left(\frac{q}{2}+k;\rho\right){\rm Im}\,U^{ph}\left(\frac{q}{2}-k;
\rho\right) \nonumber \\
& &\times \theta \left(\frac{q_0}{2}+k_0\right)
\theta \left(\frac{q_0}{2}-k_0\right) . \nonumber
\end{eqnarray}

In the region of $(q_0,\vec q)$ where the {\sl p--h} and {\sl $\Delta$--h} excitations are
off--shell, the relation between $U^{2p2h}_L$ and the $P$--wave pion--nucleus 
optical potential $V_{\rm opt}$ is given by [see also Eq.~(\ref{SigmaP-wave}), 
in the language of pion self--energy]:
\begin{equation}
\label{phe1}
\frac{\displaystyle \vec q\,^2\frac{f^2_{\pi}}{m^2_{\pi}}F^2_{\pi}(q)
U^{2p2h}_L(q)}
{\displaystyle 1-\frac{f^2_{\pi}}{m^2_{\pi}}g_L(q)U_L(q)}=2q_0V_{\rm opt}(q) .
\end{equation}
At the pion threshold $V_{\rm opt}$ is usually parameterized as:
\begin{equation}
\label{phe2}
2q_0V_{\rm opt}(q_0\simeq m_{\pi},\vec q\simeq \vec 0; \rho)
=-4\pi \vec q\,^2 \rho^2 C_0 ,
\end{equation}
where $C_0$ is a complex number which can be extracted from experimental
data on pionic atoms. By combining Eqs.~(\ref{phe1}) and (\ref{phe2})
it is possible to parameterize the proper {\sl 2p--2h} excitations 
in the spin--longitudinal channel through Eq.~(\ref{pheim}), by setting:
\begin{equation}
\label{phebare}
\vec q\,^2\frac{f^2_{\pi}}{m^2_{\pi}}F^2_{\pi}(q_0\simeq m_{\pi},\vec q 
\simeq \vec 0) U^{2p2h}_L(q_0\simeq m_{\pi},\vec q \simeq \vec 0;\rho)=
-4\pi \vec q\,^2 \rho^2 C^*_0 .
\end{equation}
The relation between $C_0$ and $C^*_0$ is fixed on the basis of the 
same RPA expression for the polarization propagator contained in 
Eq.~(\ref{phe1}); hence the value of $C^*_0$ also depends 
on the correlation function $g_L$. From the analysis of pionic atoms data 
made in Ref.~\cite{Ga92} and taking $g'\equiv g_L(0)=0.615$, one obtains:
\begin{equation}
C^*_0=(0.105+i0.096)/m^6_{\pi} . \nonumber
\end{equation}

The spin--transverse component of $U^{2p2h}$ is assumed to be 
equal to the spin--longitudinal one, $U^{2p2h}_T=U^{2p2h}_L$, and
the real parts of $U^{2p2h}_{L}$ and $U^{2p2h}_{T}$ 
are considered constant [by using Eq.~(\ref{phebare})]
because they are not expected to be too sensitive to variations of $q_0$ and 
$\vec q$. The assumption $U^{2p2h}_{T}=U^{2p2h}_{L}$
is not \emph{a priori} a good approximation, but it is the only
one which can be employed in the present phenomenological description. Yet,
the differences between $U^{2p2h}_L$ and $U^{2p2h}_T$ 
can only mildly change the partial decay widths \cite{Al02}: 
in fact, $U^{2p2h}_{L,T}$ are summed to $U^{ph}$,
which gives the dominant contribution. Moreover, for $U^{2p2h}_L=U^{2p2h}_T$
the transverse contribution to $\Gamma_2$ [fourth term in the right hand side 
of Eq.~(\ref{Alpha})]  is only about 16\% of $\Gamma_2$ (namely
$2\div 3$\% of the total width) in medium--heavy hypernuclei.

The simplified form of the phenomenological {\sl 2p--2h} propagator, 
together with the availability of analytical expressions for $U^{ph}$ and 
$U^{\Delta-h}$, makes this approach particularly suitable for employing 
the above mentioned LDA.

\subsubsection{{\sl Functional approach to the $\Lambda$ self--energy}}
\label{func}

In alternative to the above mentioned phenomenological approach for the 
two--body induced  decay width, we briefly discuss here a truly microscopic 
approach. Indeed  the most relevant Feynman diagrams  
for the calculation of the $\Lambda$ self--energy can be obtained in 
the framework of a functional method: following Ref.~\cite{Al99a},
one can derive a classification of the diagrams according to the
prescription of the so--called bosonic loop expansion. 

The functional techniques can provide a theoretically founded derivation of 
new classes of Feynmann diagram expansions in terms of powers of suitably chosen parameters. 
For example, the already mentioned ring approximation (a subclass of RPA) 
automatically appears in this framework at the mean field level. 
This method has been extensively applied to the analysis of
different processes in nuclear physics \cite{Ne82,Al87,Ce97}. 
Here it is employed for the calculation of the $\Lambda$ self--energy 
in nuclear matter, which can be expressed through the nuclear responses 
to pseudoscalar--isovector and vector--isovector fields. 
The polarization propagators obtained in this framework 
include ring--dressed meson propagators and almost the whole spectrum of 
{\sl 2p--2h} excitations (expressed in terms of a 
one--loop expansion with respect to the ring--dressed meson propagators),
which are required for the evaluation of $\Gamma_2$. 

Let us first consider the polarization propagator in the
pionic (spin--longitudinal) channel. To illustrate the procedure,
 it is useful to start from a Lagrangian describing a system of 
nucleons interacting with pions through a pseudoscalar--isovector
coupling:
\begin{equation}
{\mathcal L}_{\pi N}=\overline \psi (i\dsla-m_N)\psi 
+\frac{1}{2}\partial_{\mu}{\vec \phi}\cdot{\partial^{\mu} {\vec \phi}}
-\frac{1}{2}m_{\pi}^2 \vec\phi\,^2-i\overline \psi \vec \Gamma\psi \cdot 
\vec \phi , 
\label{Lagrangian}
\end{equation}
where $\psi$ ($\vec \phi$) is the nucleonic (pionic) field and 
$\vec\Gamma=g\gamma_5\vec\tau$ ($g=2f_{\pi}m_N/m_{\pi}$)
is the spin--isospin matrix in the spin--longitudinal isovector channel. 
We remind the reader that in the calculation of the hypernuclear decay rates 
one also needs the polarization propagator in the transverse channel 
[see Eqs.~(\ref{Sigma2}) and (\ref{Alpha})]: hence, one has to include in the
model other mesonic degrees of freedom, like the $\rho$ meson. 
Since the bosonic loop expansion is characterized by the
topology of the diagrams, this is relatively straightforward 
and the same scheme can be easily 
applied to mesonic fields other than the pionic one.

The generating functional, expressed in terms of Feynman path integrals,
associated with the Lagrangian (\ref{Lagrangian}) has the form:
\begin{equation}
\label{Z}
Z[\vec \varphi\,]=\int {\mathcal D} \left[\overline \psi, \psi, \vec \phi \,\right]
\exp\left\{i \int dx \left[{\mathcal L}_{\pi N}(x)-i
\overline \psi(x) \vec\Gamma\psi(x) \cdot \vec \varphi(x)\right]\right\} ,
\end{equation}
where a \emph{classical} external field $\vec\varphi$ with the quantum numbers
of the pion has been introduced (here and in the following the coordinate 
integrals are 4--dimensional).
All the fields in the functional integrals have to be considered as classical
variables, but with the correct commuting properties
(hence the fermionic fields are Grassman variables). 

The physical quantities of interest for the problem are then deduced from 
the generating functional by means of functional differentiations. 
In particular, by introducing a new functional $Z_c$ such that:
\begin{equation}
\label{connect}
Z[\vec\varphi\,]=\exp{\left\{iZ_c[\vec\varphi\,]\right\}} ,
\end{equation}
the spin--longitudinal, isovector polarization propagator turns out to 
be the second functional derivative of $Z_c$ with respect to the source 
$\vec\varphi$ of the pionic field:
\begin{equation}
\label{proppol}
\Pi_{ij}(x,y)=-\left[\displaystyle \frac{\delta^2 Z_c[\vec\varphi\,]}{\delta \varphi_i(x) 
\delta \varphi_j(y)}\right]_{\vec \varphi=0}  .
\end{equation}
We note that the use of $Z_c$ instead of $Z$ in Eq.~(\ref{proppol}) amounts 
to cancel the disconnected diagrams of the corresponding perturbative 
expansion (linked cluster theorem). From the generating functional $Z$ one 
can obtain different approximation schemes according 
to the order in which the functional integrations are performed. 


For the present purpose, it is convenient to integrate Eq.~(\ref{Z}) 
over the nucleonic degrees of freedom \emph{first}.
Introducing the change of variable $\vec\phi\rightarrow \vec\phi-\vec\varphi$ 
one obtains:
\begin{eqnarray}
\label{Z1}
Z\left[\vec \varphi\,\right]&=&\exp\left\{\frac{i}{2}\int dx\,dy\, \vec\varphi(x) \cdot
G^{0^{-1}}_{\pi}(x-y)\vec\varphi(y)\right\} \\ 
& & \times \int {\mathcal D} \left[\overline \psi, \psi, \vec \phi\,\right]
\exp\left\{i \int dx\,dy\, \right.\left[\overline\psi(x)G^{-1}_N(x-y)\psi(y)\right. 
\nonumber \\
& & \left. \left. +\frac{1}{2}\vec\phi(x)\cdot G^{0^{-1}}_{\pi}(x-y)
\left(\vec\phi(y)+2\vec\varphi(y)\right)\right]\right\} , \nonumber
\end{eqnarray}
where the integral over $\left[\overline\psi, \psi\right]$ is gaussian:
\begin{equation}
\int {\mathcal D} \left[\overline \psi, \psi \right] 
\exp\left\{i \int dx\,dy\, \overline\psi(x)G^{-1}_N(x-y)\psi(y)\right\}=
\left({\rm det}\,G_N \right)^{-1} . \nonumber
\end{equation}
Hence, after multiplying Eq.~(\ref{Z1}) by the unessential factor 
${\rm det}\,G^0_N$ ($G^0_N$ being the free nucleon propagator), 
which only redefines the normalization constant of the generating functional, 
and using the property ${\rm det}\,X=\exp\left\{{\rm Tr}\ln X\right\}$, 
one obtains:
\begin{eqnarray}
\label{Z2}
Z[\vec \varphi\,]&=&\exp\left\{\frac{i}{2}\int dx\,dy\, \vec\varphi(x)\cdot
G^{0^{-1}}_{\pi}(x-y)\vec\varphi(y)\right\} \\
& &\quad\times
\int {\mathcal D} \left[\vec \phi \,\right]
\exp\left\{i S^B_{\rm eff}\left[\vec\phi \,\right]\right\} . \nonumber
\end{eqnarray}
In this expression we have introduced the {\sl bosonic effective action}:
\begin{equation}
\label{seffb}
S^B_{\rm eff}\left[\vec\phi\,\right]=\int dx\,dy\, 
\left\{\frac{1}{2}\vec\phi(x)\cdot G^{0^{-1}}_{\pi}(x-y)
\left[\vec\phi(y)+2\vec\varphi(y)\right]+V_{\pi}\left[\vec\phi\,\right]\right\} , 
\end{equation}
where\footnote{Eq.~(\ref{pionpot}) is a compact writing: for example,
the $n=2$ term must be interpreted as:
\begin{equation}
\frac{i}{2}{\rm Tr}\left(i\vec\Gamma\cdot\vec\phi G^0_N\right)^2=
\frac{i}{2}\int dx\,dy\,{\rm Tr}\sum_{i.j}i\Gamma_i G^0_N(x-y)\,
i\Gamma_j G^0_N(y-x)\phi_i(x)\phi_j(y)  , \nonumber
\end{equation}
where the trace in the right hand side acts on the vertices $\vec \Gamma$,
and so on.}:
\begin{eqnarray}
\label{pionpot}
V_{\pi}\left[\vec\phi\,\right]&=&i{\rm Tr}\sum_{n=1}^{\infty}\frac{1}{n}
\left(i\vec\Gamma\cdot\vec\phi G^0_N\right)^n \\
& =&\frac{1}{2}\sum_{i,j}{\rm Tr}\left(\Gamma_i\Gamma_j\right)\int dx\,dy\, \Pi^0(x,y)
\phi_i(x)\phi_j(y) \nonumber \\
& +&\frac{1}{3}\sum_{i,j,k}{\rm Tr}\left(\Gamma_i\Gamma_j\Gamma_k\right)
\int dx\,dy\,dz\, \Pi^0(x,y,z)\phi_i(x)\phi_j(y)\phi_k(z)+ 
{\mathcal O}(\vec\phi^4) . \nonumber
\end{eqnarray}
and: 
\begin{eqnarray}
\label{pizero}
-i\Pi^0(x,y)&=&iG^0_N(x-y)iG^0_N(y-x) , \\
-i\Pi^0(x,y,z)&=&iG^0_N(x-y)iG^0_N(y-z)iG^0_N(z-x) ,\;\; {\rm etc} .
\end{eqnarray}
The bosonic effective action (\ref{seffb}) contains a term for the
free pion field and also a highly non--local pion self--interaction 
$V_{\pi}$. This effective interaction 
is given by the sum of all diagrams containing one closed fermion loop 
and an arbitrary number of pionic legs. We note that the function in 
Eq.~(\ref{pizero}) is the free particle--hole polarization
propagator. Moreover, the functions $\Pi^0(x,y,\ldots,z)$ are 
symmetric for cyclic permutations of the arguments.


The next step is the evaluation of the functional integral over the
bosonic degrees of freedom in Eq.~(\ref{Z2}). A perturbative approach
to the bosonic effective action (\ref{seffb}) does not seem to provide any
valuable results within the capabilities of the present computing tools
and in Ref.~\cite{Al99a} another approximation scheme, the semiclassical 
method, was followed.


The lowest order of the semiclassical expansion is the stationary phase
approximation (also called saddle point approximation in the Euclidean space): 
the bosonic effective action is required to be stationary with respect to
arbitrary variations of the fields $\phi_i$:
\begin{equation}
\displaystyle\frac{\delta S^B_{\rm eff}\left[\vec\phi\,\right]}
{\delta \phi_i(x)}=0 . \nonumber
\end{equation}
From the partial derivative of Eq.~(\ref{seffb}) one obtains the following 
equation of motion for the classical field $\vec\phi$:
\begin{equation}
\label{eqmoto}
\left(\Box +m^2_{\pi}\right)\phi_i(x)=\int dy\, 
G^{0^{-1}}_{\pi}(x-y)\varphi_i(y)+
\displaystyle \frac{\delta V_{\pi}\left[\vec\phi\,\right]}{\delta \phi_i(x)} ,
\end{equation}
whose solutions are functional of the external source $\vec\varphi$. 

It can be shown  that in the saddle point approximation 
the  polarization propagator coincides with the well known ring expression:
\begin{eqnarray}
\Pi_{ij}(x,y)&=&\delta_{ij}\left[\Pi^0(x,y)+{\rm Tr}\left(\Gamma^2_i\right)
\int du\,dv\, \Pi^0(x,u)G^{\rm ring}_{\pi}(u-v)\Pi^0(v,y)\right] \nonumber \\
& \equiv & \delta_{ij}\Pi^{\rm ring}(x,y) , \nonumber
\end{eqnarray}
where:
\begin{equation}
G^{\rm ring}_{\pi}=\displaystyle \frac{G^0_{\pi}}
{1-{\rm Tr}\left(\Gamma_i^2\right)G^0_{\pi}\Pi^0}  \nonumber
\end{equation}
is the ring--dressed pion propagator.
Hence, the ring approximation corresponds to the mean field level
of the present effective theory.


In the next step of the semiclassical method, the bosonic 
effective action is expanded around the mean field solution:
\begin{eqnarray}
S^B_{\rm eff}\left[\vec\phi\,\right]&=&S^B_{\rm eff}\left[\vec\phi^0\right]+
\frac{1}{2} \sum_{ij} \int dx\,dy\, \displaystyle 
\left[\frac{\delta^2S^B_{\rm eff}\left[\vec\phi\,\right]}
{\delta\phi_i(x)\delta\phi_j(y)}
\right]_{\vec\phi=\vec\phi^0} \nonumber \\
& &\times \left[\phi_i(x)-\phi^0_i(x)\right]\left[\phi_j(y)-\phi^0_j(y)\right] . \nonumber
\end{eqnarray}
Then, after performing the gaussian integration over $\vec\phi$, the 
generating functional (\ref{Z2}) [up to an uninteresting normalization 
factor] reads:
\beq
\label{Z3}
Z\left[\vec\varphi \,\right]= \exp\left\{iS^B_{\rm eff}\left[\vec\phi^0\right]
-\frac{1}{2}{\rm Tr}\ln \left[\displaystyle
\frac{\delta^2S^B_{\rm eff}\left[\vec\phi\,\right]}
{\delta\phi_i(x)\delta\phi_j(y)}\right]_{\vec\phi=\vec\phi^0}\right\} . 
\eeq
Lengthy but straightforward calculations \cite{Al99a} lead then to
 the following total polarization propagator: 
\begin{equation}
\label{fluctua0}
\Pi_{ij}(x,y)=\delta_{ij}\left[\Pi^{\rm ring} (x,y)+\Pi^{\rm OBL} (x,y)\right],
\end{equation}
where:
\begin{eqnarray}
\nonumber
\Pi^{\rm OBL}(x,y)&=&\sum_{kl}{\rm Tr}\left(\Gamma_k\Gamma_l\right)
\int du\,dv\, G^{\rm ring}_{\pi}(u-v) \Pi^0(x,u,y,v) \\
& & +\sum_{kl}{\rm Tr}\left(\Gamma_k\Gamma_l\right)
\int du\,dv\,G^{\rm ring}_{\pi}(u-v) \left[\Pi^0(x,u,v,y)+\Pi^0(x,y,v,u)
\right] 
\nonumber \\
& &+\int du\,dv\,dw\,ds\, G^{\rm ring}_{\pi}(u-w) G^{\rm ring}_{\pi}(v-s)
\Pi^0(x,u,v)
\nonumber \\
& &\times 
\sum_{klmn}\left[{\rm Tr}\left(\Gamma_k\Gamma_l\Gamma_m\Gamma_n\right)
\Pi^0(y,w,s)+{\rm Tr}\left(\Gamma_k\Gamma_l\Gamma_n\Gamma_m\right)
\Pi^0(y,s,w)\right] .
\label{fluctua} 
\end{eqnarray}

\begin{figure}
\begin{center}
\includegraphics[width=.65\textwidth]{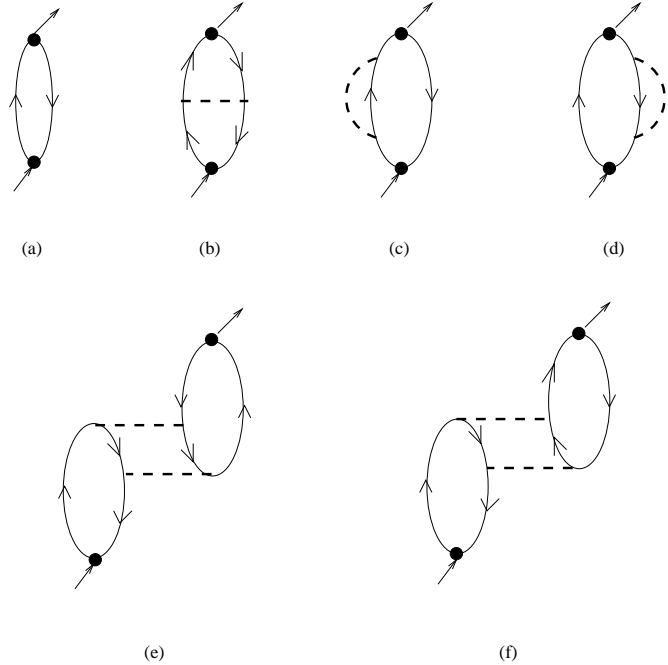}
\caption{Feynman diagrams for the polarization propagator of 
Eq.~\protect(\ref{fluctua0}): (a) particle--hole; (b) exchange; (c) and (d) 
self--energy--type; (e) and (f) correlation diagrams. 
Only the first contribution to the ring expansion has been drawn. The
dashed lines represent ring--dressed pion propagators.}
\label{onelooponeloop}
\end{center}
\end{figure}

The Feynman diagrams corresponding to Eqs.~(\ref{fluctua0}) and 
(\ref{fluctua}) are depicted in Fig.~\ref{onelooponeloop}.
Diagram (a) represents the Lindhard function $\Pi^0(x,y)$, 
which is just the first term of $\Pi^{\rm ring}(x,y)$. In (b) we have an 
\emph{exchange} diagram (the thick dashed lines representing 
\emph{ring--dressed
pion propagators}); (c) and (d) are \emph{self--energy} diagrams, while in (e) 
and (f) we show the \emph{correlation} diagrams of the present approach. 
The approximation scheme discussed here is also referred to as bosonic loop 
expansion (BLE). The practical rule to classify the Feynman diagrams according 
to their order in the BLE is to reduce to a point all its fermionic lines and
to count the number of bosonic loops left out. In this case the diagrams
(b)--(f) of Fig.~\ref{onelooponeloop}, which correspond to $\Pi^{\rm OBL}$
Eq.~(\ref{fluctua}), reduce to a one--boson--loop (OBL).

The polarization propagator of Eq.~(\ref{fluctua}) is the central result
of the present microscopic approach and it has been used \cite{Al99a} for 
the calculation of the $\Lambda$ decay width in nuclear matter. 
Note that the model can easily include the excitation of baryonic resonances, 
by replacing the fermionic field with multiplets. The topology
of the diagrams remains the same as in Fig.~\ref{onelooponeloop} but, 
introducing for example the $\Delta$ resonance, 
each fermionic line represents either a nucleon or a $\Delta$, taking care of
isospin conservation. 
Obviously this procedure substantially increases the number of diagrams.

Moreover, being the BLE characterized by the topology
of the diagrams, additional mesonic degrees of freedom together with 
phenomenological short range correlations can be included by changing 
the definition of the vertices $\Gamma_i$ in Eq.~(\ref{fluctua}). 
In particular, the formalism can be applied to evaluate the functions 
$U_{L,T}$ of Eq.~(\ref{propU}), which are required in Eqs.~(\ref{Sigma2}), 
(\ref{Alpha}). In the OBL approximation of Eq.~(\ref{fluctua0}) and 
Fig.~\ref{onelooponeloop} one has to replace Eq.~(\ref{Alpha}) with: 
\begin{eqnarray}
\label{Alpha1}
{\alpha}(q)&=&\left(S^2+\frac{P^2}{m_{\pi}^2}\vec q\,^2\right)F_{\pi}^2(q)
G_{\pi}^0(q)+\frac{\tilde{S}^2(q)U_1(q)}{1-V_L(q)U_1(q)} \\
& & +\frac{\tilde{P}_L^2(q)U_1(q)}{1-V_L(q)U_1(q)}+
2\frac{\tilde{P}_T^2(q)U_1(q)}{1-V_T(q)U_1(q)} \nonumber \\
& & +\left[\tilde S^2(q)+\tilde P_L^2(q)\right]U^{\rm OBL}_L(q)+2\tilde P_T^2(q)
U^{\rm OBL}_T(q) , \nonumber
\end{eqnarray}
where 
\begin{equation}
U_1=U^{ph}+U^{\Delta h} , \nonumber
\end{equation}
while $U^{\rm OBL}_{L,T}$ are evaluated from the diagrams 
\ref{onelooponeloop}(b)--\ref{onelooponeloop}(f)  
using the standard Feynman rules. The normalization of
these functions is such that $U^{ph}(x,y)=4\Pi^0(x,y)$, $\Pi^0$ being 
given by Eq.~(\ref{pizero}). One relevant difference between the
OBL formula (\ref{Alpha1}) and the RPA expression of
Eq.~(\ref{Alpha}) lies in the fact that in the former, to be consistent with 
 Eq.~(\ref{fluctua}), the {\sl 2p--2h} diagrams (which contribute to 
$U^{\rm OBL}_{L,T}$) are not RPA--iterated.

\section{Theory versus Experiment}
\label{exth}
In this Section the predictions of different theoretical models for the
total mesonic and non--mesonic weak decay rates are compared 
with experimental data. We mainly refer to recent
works. For a more specialized discussion
of theoretical and experimental results see Ref.~\cite{Al02}.

\subsection{Phenomenological approach} 
\label{pheres} 
We illustrate here the results which have
been obtained  for the decay rates by employing the 
PPM in the phenomenological approach.

In order to evaluate the widths from Eqs.~(\ref{local1}), (\ref{local2}), one 
needs to specify the wave function for the ${\Lambda}$. 
In Ref.~\cite{Al99} it has been obtained from a Woods--Saxon
$\Lambda$--nucleus potential
with fixed diffuseness $a=0.6$ fm and with radius and depth such that it
exactly reproduces the first two single particle eigenvalues ($s$ and $p$ $\Lambda$--levels)
of the hypernucleus under analysis. 

A crucial ingredient of the calculation
is the short range part of the strong $NN$ and ${\Lambda}N$ interactions,
$V_L$, $V_T$, $\tilde S$, $\tilde P_L$ and $\tilde P_T$ entering Eq.~(\ref{Alpha}). 
They are expressed by the functions $g_{L,T}(q)$ and $g_{L,T}^{\Lambda}(q)$
reported in the appendix of Ref.~\cite{Al02} and contain the Landau 
parameters $g^{\prime}$ and $g^{\prime}_{\Lambda}$, respectively. 
No experimental information is available on $g^{\prime}_{\Lambda}$, while 
many constraints have been set on $g^{\prime}$, for
example by the well known quenching of the Gamow--Teller resonance. Realistic
values of $g^{\prime}$ within the framework of the ring approximation are in 
the range $0.6\div 0.7$ \cite{Os82}. However, in the present context 
$g^{\prime}$ correlates not only {\sl p--h} pairs but also {\sl p--h} with 
{\sl 2p--2h} states. Accordingly, the correlation parameters have been fixed
 to $g^{\prime}=0.8$ and $g^{\prime}_{\Lambda}=0.4$~\cite{Al99} 
in order to reproduce the non--mesonic width measured for $^{12}_{\Lambda}$C.

Using these values for the Landau parameters, 
we illustrate in Table~\ref{wf sens} the sensitivity of the calculation
to the ${\Lambda}$ wave function in $^{12}_{\Lambda}$C. In addition
to the Woods--Saxon potential (New W--S) that reproduces the $s$ and $p$ 
$\Lambda$--levels, other choices have been introduced: an
harmonic oscillator wave function (H.O.) with an "empirical" frequency 
$\omega$, obtained from the $s-p$ energy shift, the 
Woods--Saxon wave function of Ref.~\cite{Do88} (Dover W--S) and
the microscopic wave function (Micr.) calculated, in Ref.~\cite{Po98},
from a non--local self--energy using a realistic $\Lambda N$ interaction.
The results (in units of the free $\Lambda$ width) are compared with the
experimental data from BNL and KEK.
\begin{table}
\begin{center}
\caption{\normalsize Sensitivity of the decay rates to the $\Lambda$ wave 
function for $^{12}_{\Lambda}$C.} 
\label{wf sens}
{\normalsize
\begin{tabular}{c c c c c c c c} \hline
\mc {1}{c}{} &
\mc {1}{c}{Micr.} &
\mc {1}{c}{Dover} &
\mc {1}{c}{H.O.} &
\mc {1}{c}{New} &
\mc {1}{c}{BNL} &
\mc {1}{c}{KEK} &
\mc {1}{c}{KEK New} \\
                &      & W--S  &      & W--S  & \cite{Sz91} & \cite{No95} & \cite{Bh98,Ou00,Sa04}
 \\ \hline
${\Gamma}_M$    & 0.25 & 0.25 & 0.26 & 0.25 & $0.11\pm 0.27$ & $0.36\pm 0.13$ & $0.31\pm 0.07$\\
${\Gamma}_1$    & 0.69 & 0.77 & 0.78 & 0.82 &                &                &    \\
${\Gamma}_2$    & 0.13 & 0.15 & 0.15 & 0.16 &                &                &\\
${\Gamma}_{NM}$ & 0.81 & 0.92 & 0.93 & 0.98 & $1.14\pm 0.20$ & $0.89\pm 0.18$ & $0.83\pm 0.09$ \\
${\Gamma}_{T}$  & 1.06 & 1.17 & 1.19 & 1.23 & $1.25\pm 0.18$ & $1.25\pm 0.18$ & $1.14\pm 0.08$ \\ \hline 
\end{tabular}}
\end{center}
\end{table}
By construction, the chosen $g^{\prime}$ and $g^{\prime}_{\Lambda}$ reproduce 
the experimental non--mesonic width using the
W--S wave function which gives the right $s$ and $p$ hyperon levels in
$^{12}_\Lambda$C. We note that it is possible to generate the microscopic wave function of
Ref.~\cite{Po98} for carbon via a local hyperon--nucleus W--S potential with 
radius $R=2.92$ fm and depth $V_0=-23$ MeV. 
Although this potential reproduces fairly well the experimental $s$--level for the 
${\Lambda}$ in $^{12}_{\Lambda}$C, it does not reproduce the $p$--level.
A completely phenomenological $\Lambda$--nucleus potential, that can easily
be extended to heavier nuclei and reproduces the experimental
$\Lambda$ single particle levels as well as possible, has been preferably
adopted in Ref.~\cite{Al99}: the potential parameters
obtained for carbon are $R=2.27$ fm and $V_0=-32$ MeV.

To analyze the results of Table~\ref{wf sens}, we note that the
microscopic wave function is substantially more extended than all the other 
wave functions. The Dover's parameters \cite{Do88}, namely $R=2.71$ fm and 
$V_0=-28$ MeV, give rise to a $\Lambda$ wave function that is somewhat more 
extended than the New W--S one but is very similar to the one obtained from 
a harmonic oscillator with an empirical frequency $\hbar \omega =10.9$ MeV. 
Consequently, the non--mesonic width from the Dover's
wave function is very similar to the one obtained from the harmonic oscillator
and slightly smaller than the new W--S one. 
The microscopic wave--function predicts the smallest non--mesonic widths due to
the more extended $\Lambda$ wave--function, which explores regions of lower 
density, where the probability of interacting 
with one or more nucleons is smaller. From Table~\ref{wf sens} one can also see
that, against intuition, the mesonic width is quite insensitive to 
the ${\Lambda}$ wave function: the different choices 
give rise to total decay widths which may differ at most by 15\%. 

Using the New W--S wave functions and the Landau parameters $g^{\prime}=0.8$
and $g^{\prime}_{\Lambda}=0.4$, in Refs.~\cite{Al99,Al02} the 
calculation has been extended to hypernuclei from $^5_{\Lambda}$He to 
$^{208}_{\Lambda}$Pb. Notice that, in order to reproduce 
the experimental $s$ and $p$ levels for the hyperon in the different 
nuclei one must use potentials with nearly 
constant depth, around $28\div 32$ MeV, in all but the lightest hypernucleus 
($^5_{\Lambda}$He).
For this hypernucleus  the $\Lambda$--nucleus mean potential
has a repulsive core and the most convenient
$\Lambda$ wave function turns out to be the one derived in
Ref.~\cite{St93} within a quark model description of $^5_{\Lambda}$He.

\begin{figure}
\begin{center}
\includegraphics[width=.7\textwidth]{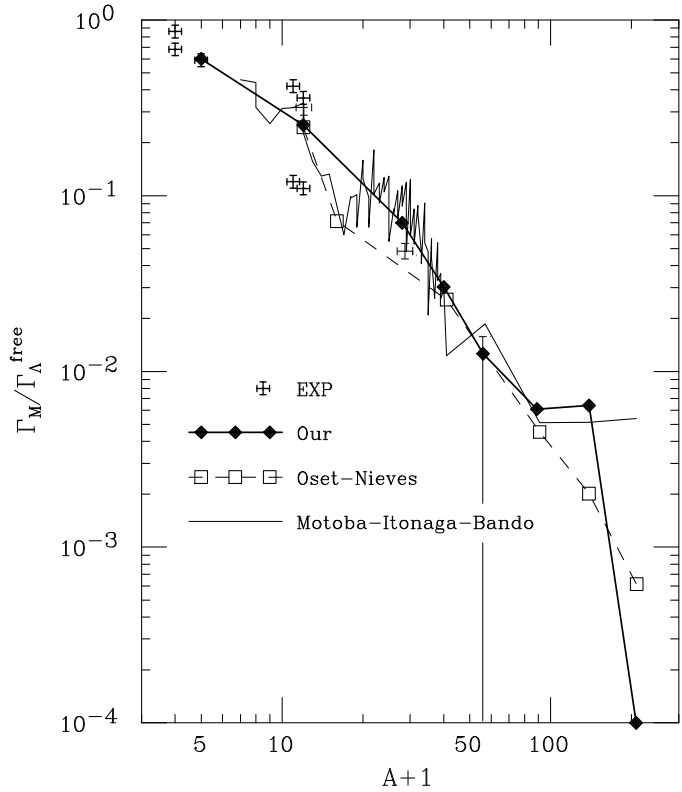}
\caption{Mesonic width as a function of the nuclear mass number $A$. The results
of Ref.~\cite{Al99,Al02} (thick solid line) are compared with the calculations of Nieves--Oset 
\protect\cite{Os93} (dashed line) and Motoba--Itonaga--Band$\overline{\rm o}$ 
\protect\cite{It88,Mo94} (solid line). Available experimental data 
\cite{Sz91,No95,Sa04} are also shown. See text for details on data (taken from Ref.~\cite{Al02}).}
\label{mesonic}
\end{center}
\end{figure}

 Table~\ref{sat} shows the resulting hypernuclear decay rates 
in units of the free $\Lambda$ width. 
\begin{table}
\begin{center}
\caption{\normalsize Mass dependence of the hypernuclear weak decay rates 
(taken from Ref.~\cite{Al02}).}
\label{sat}
{\normalsize
\begin{tabular}{c c c c c} \hline
\mc {1}{c}{$ ^{A+1}_{\Lambda}Z$} &
\mc {1}{c}{${\Gamma}_M$} &
\mc {1}{c}{${\Gamma}_1$} &
\mc {1}{c}{${\Gamma}_2$} &
\mc {1}{c}{${\Gamma}_{T}$} \\ \hline
$ ^{5}_{\Lambda}$He    & 0.60             & 0.27 & 0.04 & 0.91 \\
$ ^{12}_{\Lambda}$C    & 0.25             & 0.82 & 0.16 & 1.23 \\
$ ^{28}_{\Lambda}$Si   & 0.07             & 1.02 & 0.21 & 1.30 \\
$ ^{40}_{\Lambda}$Ca   & 0.03             & 1.05 & 0.21 & 1.29 \\
$ ^{56}_{\Lambda}$Fe   & 0.01             & 1.12 & 0.21 & 1.35 \\
$ ^{89}_{\Lambda}$Y    & $6\times 10^{-3}$ & 1.16 & 0.22 & 1.38 \\
$ ^{139}_{\Lambda}$La  & $6\times 10^{-3}$ & 1.14 & 0.18 & 1.33 \\
$ ^{208}_{\Lambda}$Pb  & $1\times 10^{-4}$ & 1.21 & 0.19 & 1.40 \\ \hline
\end{tabular}}
\end{center}
\end{table}
We observe that the mesonic rate rapidly vanishes 
by increasing the nuclear mass number $A$. 
This is well known and it is related to the decreasing phase space 
allowed for the mesonic channel, and to smaller overlaps between 
the ${\Lambda}$ wave function and the nuclear surface, as $A$ increases. 
In Fig.~\ref{mesonic} the results of Ref.~\cite{Al99,Al02}
for $\Gamma_{\rm M}$  are compared with the ones of Nieves--Oset \cite{Os93} 
and Motoba--Itonaga--Band$\overline{\rm o}$ \cite{It88,Mo94}, 
which were obtained within a shell model framework. Also the central values 
of the available experimental data are shown. Although the WFM is more 
reliable than the LDA for the evaluation of the mesonic rates (since  
the small energies involved in the decay amplify the effects of the nuclear 
shell structure), one sees that the LDA calculation of Ref.~\cite{Al99} fairly
agrees with the WFM ones  and with the data. 
In particular, Table~\ref{sat} shows that the results for $^{12}_{\Lambda}$C 
and $^{28}_{\Lambda}$Si are in agreement with the very recent KEK measurements 
\cite{Sa04}:
$\Gamma_{\rm M}(^{12}_{\Lambda}{\rm C})/\Gamma^{\rm free}_{\Lambda}= 
0.31\pm 0.07$, $\Gamma_{\pi^-}(^{28}_{\Lambda}{\rm Si})/\Gamma^{\rm free}_{\Lambda}=
0.046\pm 0.011$. The results for $^{40}_{\Lambda}$Ca, $^{56}_{\Lambda}$Fe 
and $^{89}_{\Lambda}$Y are in agreement 
with the old emulsion data (quoted in Ref.~\cite{Co90}), which indicates
$\Gamma_{\pi^-}/\Gamma_{\rm NM}\simeq (0.5\div 1)\times 10^{-2}$ 
in the region $40<A<100$. Moreover, the recent KEK experiments \cite{Sa04} 
obtained the limit: 
$\Gamma_{\pi^-}(^{56}_{\Lambda}{\rm Fe})/\Gamma^{\rm free}_{\Lambda} < 0.015$ 
(90\% CL). It is worth noticing, in figure \ref{mesonic}, the rather pronounced
oscillations of $\Gamma_{\rm M}$ in the calculation of Refs.~\cite{It88,Mo94}, 
which are caused by shell effects.

As a final comment to Table~\ref{sat}, we note that, with the exception of 
$^5_{\Lambda}$He, the two--body induced decay is rather independent of the 
hypernuclear dimension and it is about 15\% of the total width. Previous works
\cite{Al91,Ra95} gave more emphasis to this new channel,
without, however, reproducing the experimental non--mesonic
rates. The total width does not change much with $A$,
as it is also shown by the experiment.

\begin{figure}
\begin{center}                                                                                                            
\includegraphics[width=.68\textwidth]{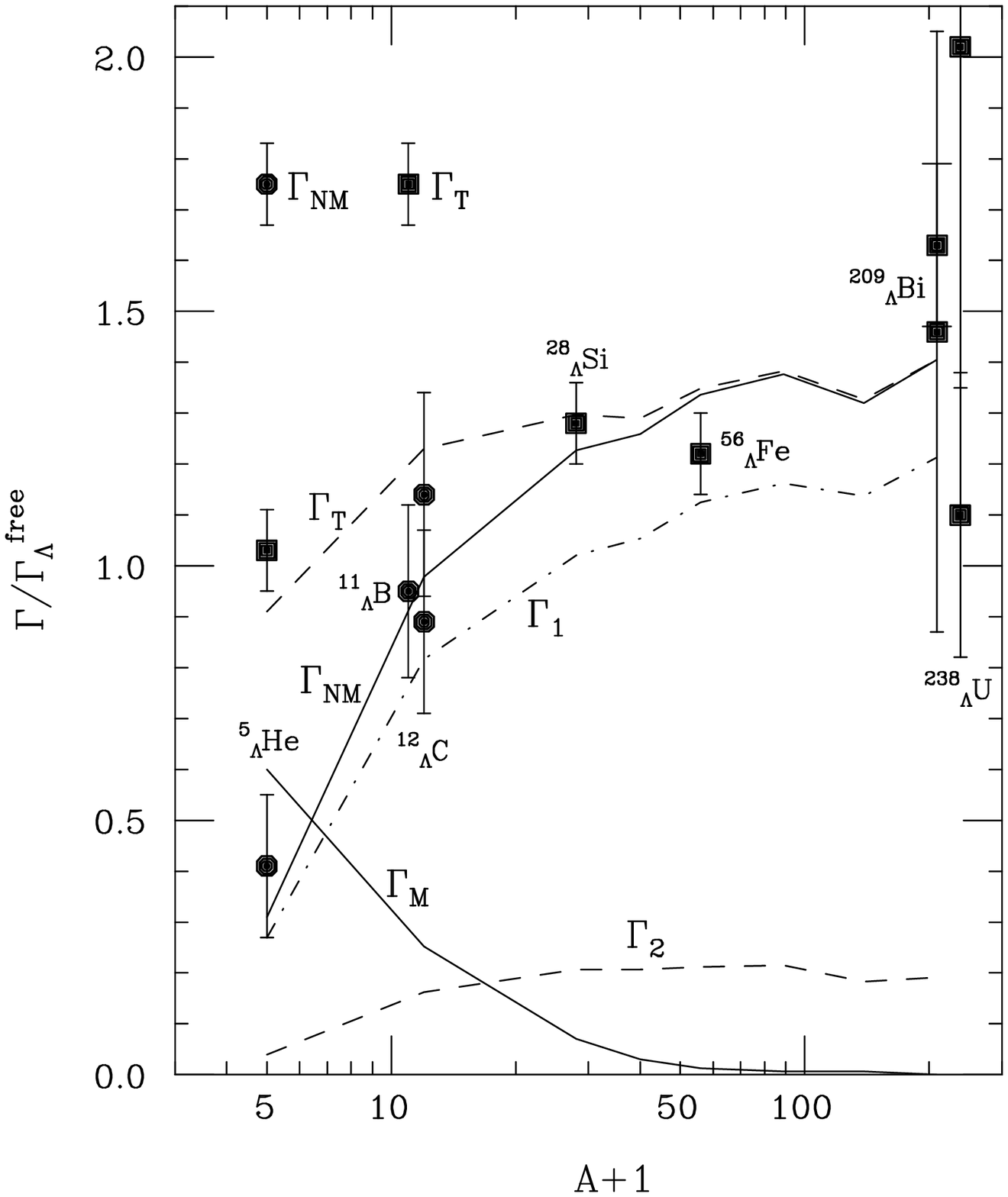}
\caption{Partial ${\Lambda}$ decay widths in finite nuclei as a function of
the nuclear mass number $A$. Experimental data are from 
Refs.~\cite{Sz91,Ar93,No95,Bh98,Ku98} (taken from Ref.~\cite{Al02}).}
\label{satu}
\end{center}
\end{figure}

In Fig.~\ref{satu} the results of Table~\ref{sat} are compared
with recent (after 1990) experimental data for $\Gamma_{\rm NM}$ and 
$\Gamma_{\rm T}$.
The theoretical results are in good agreement with the data 
over the whole hypernuclear mass range explored.
The saturation of the ${\Lambda}N\rightarrow nN$ interaction
in nuclei is well reproduced.

\subsection{Microscopic approach}
\label{micres}
The results discussed in this subsection have been obtained
by applying the one--boson--loop formalism developed in Section \ref{func}
for nuclear matter. 
Although, in principle, one could extend this calculation
to finite nuclei through the local density approximation, 
in practice this would require prohibitive computing times. 
Indeed, the latter are already quite conspicuous for the evaluation of the 
diagrams of Fig.~\ref{onelooponeloop} at \emph{fixed} Fermi momentum. 
Hence, in order to compare the results with the experimental data in finite 
nuclei, different Fermi momenta have been employed in the nuclear matter 
calculation.
The average, fixed Fermi momentum, which is appropriate for the present 
purposes, has  been obtained by weighting each local Fermi momentum $k^A_F(r)$ 
with the probability density of the hyperon in the considered nucleus:
\begin{equation}
\label{kf3}
\langle k_F\rangle_A=\int d\vec r \,k^A_F(\vec r)|\psi_{\Lambda}(\vec r)|^2 .
\end{equation}

It is then possible to classify the hypernuclei for which
experimental data on the non--mesonic decay rate 
are available into three mass regions
(medium--light: $A\simeq 10$; medium: $A\simeq 30\div 60$;
and heavy hypernuclei: $A\gsim 200$), as shown in Table~\ref{kmed}.
The experimental bands include values of the non--mesonic widths 
which are compatible with the quoted experiments. For medium and heavy 
hypernuclei the available experimental data actually refer to the \emph{total}
decay rate. However, the mesonic width for medium hypernuclei is at most 5\% 
of the total width and rapidly decreases as $A$ increases. 
Hence, one can safely approximate $\Gamma^{\rm exp}_{NM}$ with 
$\Gamma^{\rm exp}_{T}$ for medium and heavy systems.
In the third column of Table~\ref{kmed} we report the average Fermi momenta
obtained with Eq.~(\ref{kf3}).
\begin{table}
\begin{center}
\caption{\normalsize Average Fermi momenta for three representative mass regions. The experimental
data are in units of the free $\Lambda$ decay rate (taken from Ref.~\cite{Al99a}).}
\label{kmed}
{\normalsize
\begin{tabular}{c c c}
\hline
\mc {1}{c}{} &
\mc {1}{c}{$\Gamma^{\rm exp}_{NM}$} &
\mc {1}{c}{$\langle k_F\rangle$ (fm$^{-1}$)} \\ \hline
Medium--Light: $^{11}_{\Lambda}$B - $^{12}_{\Lambda}$C & $0.94\div 1.07$ \cite{Sz91,No95} &  1.08 \\
Medium : $^{28}_{\Lambda}$Si - $^{56}_{\Lambda}$Fe    & $1.20\div 1.30$ \cite{Bh98} &  $\simeq 1.2$\\
Heavy: $^{209}_{\Lambda}$Bi - $^{238}_{\Lambda}$U     & $1.45\div 1.70$ \cite{Ar93,Ku98} &  1.36 \\ \hline
\end{tabular}}
\end{center}
\end{table}
According to the Table, in Ref.~\cite{Al99a} the following average Fermi momenta 
have been employed: $k_F=1.1$~fm$^{-1}$ for medium--light, $k_F=1.2$~fm$^{-1}$ 
for medium and $k_F=1.36$~fm$^{-1}$ for heavy hypernuclei.

In addition to $\langle k_F\rangle$, other parameters enter into the 
microscopic calculation of hypernuclear decay widths, which are specifically
related to the baryon--meson vertices and to the short range correlations.
With the exception of the Landau parameters $g^{\prime}$ 
and  $g^{\prime}_{\Lambda}$, the values of these parameters have been kept 
fixed on the basis of the existing phenomenology (for example in the analysis 
of quasi--elastic electron--nucleus scattering, spin--isospin nuclear 
response functions, etc). For the complete list of these quantities we refer 
the reader to Ref.~\cite{Al99a}.

Instead, the zero energy and momentum limits of the strong $NN$ and 
$\Lambda N$ correlations, $g^{\prime}$ and $g^{\prime}_{\Lambda}$, 
are considered as \emph{free parameters}. We remind  the reader that 
the physical meaning of these Landau parameters is different 
in the present scheme (see Fig.~\ref{onelooponeloop}) with respect to the 
customary phenomenology based on the ring approximation. 

Fixing $g^{\prime}_{\Lambda}=0.4$, in ring approximation
the experimental decay rates can be reproduced
by using $g^{\prime}$ values which are compatible with the existing 
literature ($0.5\le g^{\prime}\le 0.7$). 
However, larger  $g^{\prime}$ values ($0.7\le g^{\prime}\le 0.9$)
appear to be more appropriate in the framework of the full OBL approximation. 
In addition, the OBL calculation  allows for a good description (keeping 
the same $g^{\prime}$ value) of the rates in the whole range of $k_F$ considered here. 

\begin{figure}
\begin{center}
\includegraphics[width=.75\textwidth]{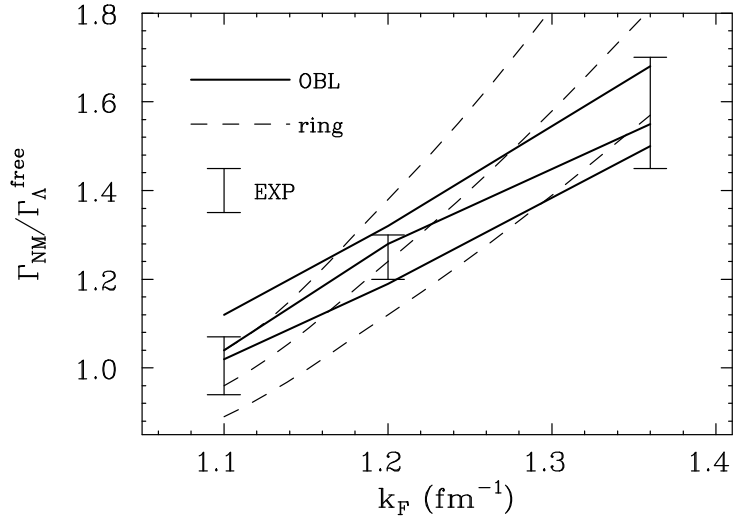}
\caption{Dependence of the non--mesonic width on the Fermi momentum of nuclear matter.
The solid curves refer to the one--boson--loop approximation
(with $g^{\prime}=0.7, 0.8, 0.9$ from the top to the bottom), 
while the dashed lines refer to the ring approximation ($g^{\prime}=0.5, 0.6, 0.7$). 
The experimental data are also shown (taken from Ref.~\cite{Al99a}).}
\label{gammakf}
\end{center}
\end{figure}

In Fig.~\ref{gammakf}
we see the dependence of the non--mesonic widths on the Fermi momentum.
The solid lines correspond to the OBL approximation, with 
$g^{\prime}=0.7, 0.8, 0.9$ from the top to the bottom, while the dashed
lines refer to the ring approximation, with $g^{\prime}=0.5, 0.6, 0.7$, 
again from the top to the bottom. One can then conclude that for the OBL
calculation the best choice for the Landau parameters is the following one:
$g^{\prime}=0.8, \hspace{0.2cm} g^{\prime}_{\Lambda}=0.4$. Accidentally,
this parameterization turns out to be the same  employed in the 
phenomenological evaluation of the {\sl 2p--2h} polarization propagator, 
although the role of strong short range correlations is different in the
two approaches.

\subsection{Other evaluations}
\label{exthothers}

In this subsection we briefly discuss how other calculations ---performed 
within the Polarization Propagator Model (PPM) and the
Wave Function Model (WFM) previously illustrated--- compare with existing data.

In Table~\ref{tab5} results for the mesonic decay width for $^{5}_{\Lambda}$He
are reported. All the theoretical evaluations agree with the BNL datum
but discrepancies appear if one compares them with the more accurate KEK observation.
A repulsive core in the $\Lambda$--$\alpha$ mean potential (used in all but the calculation of
Ref.~\cite{Os85}) is favoured. Moreover, it comes out naturally
in the quark model descriptions of Refs.~\cite{Mo91,St93}. The results of
Refs.~\cite{It88,Mo92} refer to the use of different pion--nucleus
optical potentials. On this point, we note that the more recent datum
is able to discriminate between different pion--nucleus optical potentials.
Thanks to such precise determinations of $\Gamma_{\rm M}$,
novel information on this potential will be possibly extracted in the next future.
The experimental and theoretical values of the ratio $\Gamma_{\pi^-}/\Gamma_{\pi^0}$,
not shown in the table, do not
deviate much from the $\Delta I=1/2$ value ($=2$) for free decays. We expect this
result, since $^5_{\Lambda}$He has a closed shell core with $N=Z$.
\begin{table}
\begin{center}
\caption{\normalsize Mesonic decay rate for $^{5}_{\Lambda}$He.}
\label{tab5}
{\normalsize
\begin{tabular}{c c c} \hline
\mc {1}{c}{Ref.} &
\mc {1}{c}{$\Gamma_{M}/\Gamma^{\rm free}_\Lambda$} &
\mc {1}{c}{Model} \\ \hline
\small Oset--Salcedo 1985 \cite{Os85}             & 0.65             & PPM \\ 
Oset--Salcedo--Usmani 1986 \cite{Os86a}             & 0.54             & PPM \\ 
Itonaga--Motoba--Band$\overline{\rm o}$ 1988 \cite{It88}            & $0.331\div0.472$ & WFM \\ 
Motoba {\em et al.}~1991 \cite{Mo91}            & 0.608            & WFM + Quark Model \\ 
Motoba        1992 \cite{Mo92}            & 0.61             & WFM   \\ 
Straub {\em et al.}~1993 \cite{St93}              & 0.670             & WFM + Quark Model   \\ 
Kumagai--Fuse {\em et al.}~1995  \cite{Ku95a}           & 0.60             & WFM  \\ \hline
Exp BNL 1991 \cite{Sz91}                  & $0.59^{+0.44}_{-0.31}$     &    \\ 
Exp KEK 2004 \cite{Ok04}                  & $0.541\pm 0.019$     &    \\ \hline
\end{tabular}}
\end{center}
\end{table}

The theoretical results for the mesonic rate for $^{12}_{\Lambda}$C, reported in 
Table~\ref{tab4}, are fairly compatible with data.
The most notable exception is the calculation of Ref.~\cite{Zh99}, supplying a
decay rate which strongly underestimates the KEK data.
Also in this case, accurate measurements such as the one
of Ref.~\cite{Ok04} would provide important information on the pion--nucleon optical potential.
The estimates obtained with the WFM of Refs.~\cite{It88,Os93,Mo94}
are consistent with the experimental ratio $(\Gamma_{\pi^0}/\Gamma_{\pi^-})^{\rm Exp}
\simeq 1\div 2>(\Gamma_{\pi^0}/\Gamma_{\pi^-})^{\rm free}=1/2$,
which reflects the particular nuclear shell structure of
$^{12}_{\Lambda}$C. 
\begin{table}
\begin{center}
\caption{\normalsize Mesonic decay rate for $^{12}_{\Lambda}$C.}
\label{tab4}
{\normalsize
\begin{tabular}{c c c} \hline
\mc {1}{c}{Ref.} &
\mc {1}{c}{$\Gamma_{M}/\Gamma^{\rm free}_\Lambda$} &
\mc {1}{c}{Model} \\ \hline
Oset--Salcedo 1985 \cite{Os85}             & 0.41             & PPM \\ 
Itonaga--Motoba--Band$\overline{\rm o}$ 1988 \cite{It88}            & $0.233\div0.303$ & WFM \\ 
Ericson--Band$\overline{\rm o}$ 1990 \cite{Er90}       & 0.229 & WFM \\ 
Nieves--Oset 1993 \cite{Os93}              & 0.245            & WFM \\ \
Itonaga--Motoba 1994 \cite{Mo94}            & 0.228            & WFM \\ 
Ramos {\em et al.}~1995 \cite{Ra95}             & 0.31             & PPM \\ 
Zhou--Piekarewicz 1999 \cite{Zh99}      & 0.112             & Relativistic PPM \\ 
Albertus {\em et al.}~2003 \cite{Al03}      &    0.25     & WFM \\ \hline
Exp BNL 1991 \cite{Sz91}                        & $0.11\pm0.27$    &   \\ 
Exp KEK 1995 \cite{No95}                        & $0.36\pm0.13$    &   \\ 
Exp KEK 2004 \cite{Sa04}                  & $0.313\pm 0.070$    & \\ 
Exp KEK 2004 \cite{Ok04}                  & $0.288\pm 0.017$    & \\ \hline
\end{tabular}}
\end{center}
\end{table}

Table~\ref{best-nmfull} refers to the non--mesonic rate for $^5_{\Lambda}$He,
$^{12}_{\Lambda}$C and nuclear matter. The different $\Lambda N\to nN$ potentials
used in the calculations reflect in a broad spectrum of predictions.
In particular, Dalitz \cite{Da73} (and more recently Jun {\em et al.}~\cite{Ju01}) 
used a phenomenological model in which the
OPE model at large distances is supplemented by a 4--baryon point interaction 
(4BPI) for the short range $\Lambda N\to nN$ interactions. 
Ref.~\cite{Ch83} used a hybrid model in which the long range interactions
are treated in terms of the OPE, while the short range interactions
are described by a 6--quark cluster model which includes both
$\Delta I=1/2$ and $\Delta I=3/2$ components.
The PPM of Ref.~\cite{Os85} overestimate the data, although the decay rate
is reduced when a more realistic $\Lambda$--wave function (less superimposed with 
the nuclear core) is used~\cite{Os86a,Al99,Al99a}, and particularly if different short range
correlations are considered \cite{Al99,Al99a}
(see results discussed in subsections \ref{pheres} and \ref{micres}).
Antisymmetrization of the final nucleons, as in Ref.~\cite{Os01}, would
also moderately decrease the non--mesonic rate. 
Sasaki {\em et al.}~\cite{Ok99} treated the non--mesonic decay
within a direct quark (DQ) model combined with a OME 
potential containing pion and kaon exchange.
In their model the $NN$ and $\Lambda N$ repulsion at short
distance originates from quark exchange between baryons (induced by the quark anti--symmetrization) and 
gluon exchange between quarks. The intervals shown for the 
OME calculation of Ref.~\cite{Pa01} correspond to the use of different Nijmegen models for
the hadronic coupling constants. Itonaga {\em et al.}~\cite{It02} considered
the correlated two--pion--exchange.

Large part of the predictions of Table~\ref{best-nmfull} agree with the available data. 
The non--mesonic width in $^{12}_{\Lambda}$C seems to be
reduced by a factor of about $2$ with respect to the nuclear matter value.
From inspection of data on heavy hypernuclei one concludes that realistic
values of the $\Lambda$ decay rate in nuclear matter lie in the range $1.5\div 2$. 
From both experiment and theory, it also follow that 
$\Gamma_{\rm NM}(^{12}_{\Lambda}{\rm C})\simeq 2\,\Gamma_{\rm NM}(^5_{\Lambda}{\rm He})$.
\begin{table}
\caption{\normalsize Non--mesonic decay rate for $^5_\Lambda$He, $^{12}_\Lambda$C and nuclear matter
in units of the free $\Lambda$ decay width.}
\label{best-nmfull}
\begin{center}
{\normalsize
\begin{tabular}{c c c c} \hline
\mc {1}{c}{Ref. and Model} &
\mc {1}{c}{$^5_{\Lambda}$He} &
\mc {1}{c}{$^{12}_{\Lambda}$C} &
\mc {1}{c}{Nuclear Matter} \\ \hline
Dalitz~1973 \cite{Da73}  & 0.5 & & 2  \\
(WFM: OPE + 4BPI) & &      & \\
Cheung {\em et al.}~1983 \cite{Ch83}   & & 1.28 & 3  \\
(WFM: hybrid) & &      & \\
Oset--Salcedo~1985 \cite{Os85}  & 1.15 & 1.5 & 2.2 \\
(PPM: Correlated OPE) & &      & \\
Oset--Salcedo--Usmani~1986 \cite{Os86a}  & 0.54 &  &  \\
(PPM: Correlated OPE) & &      & \\
Sasaki {\em et al.}~2000 \cite{Ok99} &0.519 & &2.456 \\
(WFM: $\pi +K+$ DQ)               & & & \\ 
Jun {\em et al.}~2001 \cite{Ju01} &0.426 &1.174 & \\
(WFM: OPE + 4BPI)                  & & & \\ 
Jido {\em et al.}~2001 \cite{Os01} & &0.769 & \\
(PPM: $\pi +K+2\pi +\omega$)      & & & \\ 
Parre\~{n}o--Ramos~2001 \cite{Pa01} &$0.317\div 0.425$ & $0.554\div 0.726$ & \\
(WFM: $\pi +\rho +K + K^* + \omega +\eta$)                        & & & \\ 
Itonaga {\em et al.}~2002 \cite{It02}    & 0.422 & 1.060 & \\
(WFM: $\pi + 2\pi/\rho + 2\pi/\sigma+\omega$) & &      & \\ \hline
Exp BNL 1991 \cite{Sz91} &$0.41\pm0.14$    &$1.14\pm0.20$ & \\ 
Exp CERN 1993 \cite{Ar93}   & & & $\bar{p}+$Bi: $1.46^{+1.83}_{-0.52}$ \\
                            & & & $\bar{p}+$U:  $2.02^{+1.74}_{-0.63}$ \\ 
Exp KEK 1995 \cite{No95} & &$0.89\pm0.18$ & \\ 
Exp KEK 1995 \cite{No95a} &$0.50\pm0.07$ & & \\ 
Exp COSY 1998 \cite{Ku98}    & & & $p+$Bi: $1.63^{+0.19}_{-0.14}$  \\ 
Exp KEK 2000 \cite{Bh98,Ou00} & &$0.83\pm 0.11$ & $^{56}_{\Lambda}$Fe: $1.22\pm 0.08$ \\ 
Exp COSY 2001 \cite{Ka01}   & & & $p+$Au: $2.02^{+0.56}_{-0.35}$ \\ 
Exp COSY 2001 \cite{Ku01}   & & & $p+$U:  $1.91^{+0.28}_{-0.22}$ \\ 
Exp KEK 2004 \cite{Ok04}    & $0.406\pm 0.020$ & $0.953\pm 0.032$ & \\ \hline
\end{tabular}}
\end{center}
\end{table}

\section{The ratio $\Gamma_n/\Gamma_p$}
\label{ratio}

Up to very recent times, the main challenge of hypernuclear weak decay studies
has been to provide a theoretical explanation of the large experimental values for the 
ratio ${\Gamma}_n/{\Gamma}_p$ between the neutron-- and the proton--induced decay widths.
Until recently, the large uncertainties involved in the extraction of
the ratio from data did not allow to reach any definitive
conclusion. The data were quite limited and not precise due to the difficulty of
detecting the products of the non--mesonic decays, especially the
neutrons. Moreover, up to now it has not been possible to distinguish between
nucleons produced by the one--body induced and the
(non--negligible) two--body induced decay mechanism.
However, due to recent theoretical \cite{Ok99,Os01,Pa01,It02,Ga03} and experimental 
\cite{outa-coinc,kim,Ok04l} progress, we are now towards a solution of the
$\Gamma_n/\Gamma_p$ puzzle. In this section we review these important developments.

The one--pion--exchange approximation with the $\Delta I=1/2$ rule
provides small ratios:
\begin{equation}
\label{ope-new}
\left[ \frac{\Gamma_n}{\Gamma_p} \right]^{\rm OPE} \simeq 0.05\div 0.20 ,
\end{equation}
for all the considered systems.
This is due to the $\Delta I=1/2$ rule, which fixes the vertex ratio 
$V_{\Lambda \pi^- p}/V_{\Lambda \pi^0 n}=-\sqrt{2}$ (both in $S$-- and $P$--wave
interactions), and to the particular form of
the OPE potential, which has a strong tensor 
component requiring isospin $0$ $np$ pairs in the antisymmetric final state.
In $p$--shell and heavier hypernuclei the relative $\Lambda N$ $L=1$ state is found
to give only a small contribution to tensor transitions for the neutron--induced
decay, so it cannot improve the OPE ratio.
The contribution of the $\Lambda N$ $L=1$ relative state to $\Gamma_{\rm NM}$ 
seems to be of about $5\div 15$\% in $p$--shell hypernuclei \cite{Pa01}. 
For these systems we expect the dominance of the $S$--wave interaction in the initial state,
due to the small $\Lambda N$ relative momentum. By using a 
simple argument about the isospin structure of the transition $\Lambda N\rightarrow nN$
in OPE, it is possible to estimate that for pure $\Delta I=3/2$ transitions
($V_{\Lambda \pi^-p}/V_{\Lambda \pi^0n}=1/\sqrt{2}$) the OPE ratio
can increase up to about $0.5$. However, the OPE model with $\Delta I=1/2$ couplings
has been able to reproduce the one--body stimulated non--mesonic rates 
$\Gamma_{1}=\Gamma_n+\Gamma_p$ for light and medium hypernuclei
\cite{Ok99,Al99,Al99a,Pa01,Os01,It02}. 

Other interaction mechanisms beyond the OPE might then be responsible for the 
overestimation of $\Gamma_p$ and the underestimation of $\Gamma_n$.
Many attempts have been made up to now in order to solve the $\Gamma_n/\Gamma_p$ puzzle. 
We recall here the inclusion in the ${\Lambda}N\rightarrow nN$
transition potential of mesons heavier than the pion 
\cite{Du96,Pa97,Pa01,Os01,It02}, the inclusion of interaction terms that
explicitly violate the ${\Delta}I=1/2$ rule 
\cite{Pa98} and the description of the 
short range baryon--baryon interaction in terms of quark degrees of freedom
\cite{Ch83,Ok99}, which automatically introduces $\Delta I=3/2$ contributions.
It is important to note that a few calculations with $\Lambda N \rightarrow nN$ transition 
potentials including heavy--meson--exchange and/or direct quark contributions
\cite{Ok99,Os01,Pa01,It02} have improved the situation, without providing, nevertheless, a
satisfactory explanation of the origin of the puzzle.
The tensor component of $K$--exchange has opposite sign with respect to the one
for $\pi$--exchange, resulting in a reduction of $\Gamma_p$. The parity violating
$\Lambda N(^3S_1)\to nN(^3P_1)$ transition, which contributes to both the $n$--
and $p$--induced processes, is considerably enhanced by $K$--exchange
and direct quark mechanism and tends to increase $\Gamma_n/\Gamma_p$.

In table~\ref{best-ratio} we summarize the calculations
that predicted ratios considerably enhanced with respect to the 
OPE values. Experimental data are given for comparison.
Almost all calculations reproduce the observed non--mesonic widths
$\Gamma_n+\Gamma_p$, as one can see in Table~\ref{best-nm}
[we remind the reader that the experimental data should also include (at least a part)
of the two--nucleon induced decay rate]. Although no calculation
is able to explain the old data on $\Gamma_n/\Gamma_p$, some
predictions are in agreement with the recent determinations \cite{Ga03,GaDAFNE} from KEK
nucleon coincidence data \cite{outa-coinc}. We discuss in detail these analyses in \ref{coin}.
\begin{table}
\caption{\normalsize $\Gamma_n/\Gamma_p$ ratio.}
\label{best-ratio}
\begin{center}
{\normalsize
\begin{tabular}{c c c} \hline
\mc {1}{c}{Ref. and Model} &
\mc {1}{c}{$^5_{\Lambda}$He} &
\mc {1}{c}{$^{12}_{\Lambda}$C} \\ \hline
Sasaki {\em et al.}~2000 \cite{Ok99} &0.701 &  \\
$\pi +K+$ DQ               & & \\ 
Jido {\em et al.}~2001 \cite{Os01} & &0.53  \\
$\pi +K+2\pi +\omega$      & & \\ 
Parre\~{n}o--Ramos~2001 \cite{Pa01} &$0.343\div 0.457$ & $0.288\div 0.341$  \\
$\pi +\rho +K + K^* + \omega +\eta$                        & &  \\
Itonaga {\em et al.}~2002 \cite{It02}    & 0.386 & 0.368  \\
$\pi + 2\pi/\rho + 2\pi/\sigma+\omega$ & &      \\ \hline 
BNL 1991 \cite{Sz91} &$0.93\pm0.55$ &$1.33^{+1.12}_{-0.81}$  \\ 
KEK 1995 \cite{No95} & &$1.87^{+0.67}_{-1.16}$  \\ 
KEK 1995 \cite{No95a} &$1.97\pm0.67$ &  \\ 
KEK 2004 \cite{Sa04} & &$0.87\pm 0.23$  \\ 
KEK 2004 (coincidence) \cite{outa-coinc} & $0.40\pm 0.11$ & $0.38\pm 0.14$ \\
(with Garbarino et al.~\cite{Ga03,GaDAFNE} analysis) & & \\ \hline
\end{tabular}}
\end{center}
\end{table}
\begin{table}
\caption{\normalsize 
Non--mesonic width $\Gamma_n+\Gamma_p$ (in units of $\Gamma^{\rm free}_{\Lambda}$).}
\label{best-nm}
\begin{center}
{\normalsize
\begin{tabular}{ccc} \hline
\mc {1}{c}{Ref. and Model} &
\mc {1}{c}{$^5_{\Lambda}$He} &
\mc {1}{c}{$^{12}_{\Lambda}$C} \\ \hline
Sasaki {\em et al.}~2000 \cite{Ok99} &0.519 &  \\
$\pi +K+$ DQ               & &  \\ 
Jido {\em et al.}~2001 \cite{Os01} & &0.769 \\
$\pi +K+2\pi +\omega$      & &  \\ 
Parre\~{n}o--Ramos~2001 \cite{Pa01} &$0.317\div 0.425$ & $0.554\div 0.726$  \\
$\pi +\rho +K + K^* + \omega +\eta$                        & &  \\ 
Itonaga {\em et al.}~2002 \cite{It02}    & 0.422 & 1.060  \\
$\pi + 2\pi/\rho + 2\pi/\sigma+\omega$ & &       \\ \hline
BNL 1991 \cite{Sz91} &$0.41\pm0.14$    &$1.14\pm0.20$ \\ 
KEK 1995 \cite{No95} & &$0.89\pm0.18$ \\ 
KEK 1995 \cite{No95a} &$0.50\pm0.07$ &  \\ 
KEK 2000 \cite{Ou00} & &$0.83\pm 0.11$ \\ 
KEK 2004 \cite{Ok04} & $0.406\pm 0.020$ &$0.953\pm 0.032$ \\ \hline
\end{tabular}}
\end{center}
\end{table}

\subsection{Two--body induced decay and nucleon final state interactions}
\label{FSI2b}
The analysis of the ratio ${\Gamma}_n/{\Gamma}_p$ is influenced by the  
two--nucleon induced process ${\Lambda}NN\rightarrow nNN$, whose experimental 
identification is rather difficult and it is a challenge for the future.  
By assuming that the meson produced in the weak  
vertex is mainly absorbed by an isoscalar $np$ correlated pair 
(quasi--deuteron approximation), the three--body process turns out to be
${\Lambda}np\rightarrow nnp$, so that    
a considerable fraction of the measured neutrons 
could come from this channel and not only from $\Lambda n \rightarrow nn$ 
and $\Lambda p \rightarrow np$. In this way it might be 
possible to explain the large experimental ${\Gamma}_n/{\Gamma}_p$ ratios, 
which originally have been analyzed without taking into account  
the two--body stimulated process. 
Nevertheless, the situation is far from being clear and simple, both from the  
theoretical and experimental viewpoints. 
The new non--mesonic mode was introduced in Ref.~\cite{Al91} and its calculation 
was improved in Ref.~\cite{Ra95}, where the authors found that the inclusion of the 
new channel would bring to extract from the experiment
even larger values for the ${\Gamma}_n/{\Gamma}_p$ ratios, thus worsening the 
disagreement with the theoretical estimates. However, in the 
hypothesis that only two out of the three nucleons coming from $\Lambda np \to nnp$
are detected \cite{Ga95}, the reanalysis of the experimental data would lead back to smaller 
ratios.

These observations show that ${\Gamma}_n/{\Gamma}_p$ is sensitive to  
the detailed kinematics of the non--mesonic processes and to the experimental threshold for
nucleon detection. In Ref.~\cite{Ra97} the nucleon energy distributions have been  
calculated by using a Monte Carlo simulation to describe the nucleon
rescattering inside the nucleus: the ratio 
$\Gamma_n/\Gamma_p$ has been taken as a free parameter and extracted  
by comparing the simulated spectra with data. The momentum distributions of 
the primary nucleons were determined within the polarization propagator scheme.
In their way out of the nucleus, the nucleons, due to the collisions 
with other nucleons, continuously change energy, direction, charge, and secondary  
nucleons are emitted as well. Then, the energy distribution of the observable nucleons,  
which also loose their energy by the interactions with 
the experimental set--up, is quite different from the one at the level of the primary nucleons.
The nucleons from two--body stimulated decays appear mainly at low energies, while
those from the one--nucleon stimulated
process peak around 75 MeV only for light hypernuclei such as $^5_{\Lambda}$He. 

Quite recently, at KEK--E307 \cite{Ha01}, the proton spectra 
for $^{12}_{\Lambda}$C, $^{28}_{\Lambda}$Si and $^{56}_{\Lambda}$Fe were 
measured and compared with theoretical simulations of the intranuclear 
cascades after the weak processes, obtained with the MC code of Ref.~\cite{Ra97}. 
Corrections for the detector geometry and the
nucleon interactions inside the target and detector materials
were necessary: they were implemented through a GEANT MC code. After fitting the
KEK--E307 spectra using $\Gamma_n/\Gamma_p$ as a free parameter,
the following results were obtained \cite{Sa04} by neglecting the two--nucleon 
induced decay channel:
\begin{eqnarray}
\label{newratio}
\frac{\Gamma_n}{\Gamma_p}(^{12}_{\Lambda}{\rm C})&=&0.87\pm 0.23 ,  \\
\frac{\Gamma_n}{\Gamma_p}(^{28}_{\Lambda}{\rm Si})&=&0.79^{+0.28}_{-0.26} , \nonumber \\
\frac{\Gamma_n}{\Gamma_p}(^{56}_{\Lambda}{\rm Fe})&=&1.13^{+0.29}_{-0.27} . \nonumber
\end{eqnarray}
All the existing calculations underestimate the $^{12}_{\Lambda}$C result
(see Table~\ref{best-ratio}).
The $\Gamma_n/\Gamma_p$ ratios of Eq.~(\ref{newratio}) confirm
the results of previous experiments: the neutron--
and proton--induced decay rates appear to be of the same order of magnitude over a large
hypernuclear mass number range. However, since the new experiments have
significantly improved the quality of the data, small values of $\Gamma_n/\Gamma_p$
(say smaller than $0.5$ for $^{12}_{\Lambda}$C, as predicted by theory) seem to be excluded.

\subsection{Towards a solution of the $\Gamma_n/\Gamma_p$ puzzle}
\label{improvements}

Fortunately, recent important developments have contributed to approach the solution
of the $\Gamma_n/\Gamma_p$ puzzle. This progress has been based on the following
main ideas:
\begin{itemize}
\item [1)] The proton spectra originating from neutron-- and proton--induced
processes (and, eventually, from two--nucleon stimulated decays)
are added \emph{incoherently} in the Monte Carlo intranuclear cascade
code used to determine $\Gamma_n/\Gamma_p$ from data. In this way a possible 
quantum--mechanical \emph{interference} effect between the two channels 
is lost. Therefore, extracting $\Gamma_n/\Gamma_p$ from data with the help of
a classical intranuclear cascade calculation may not be a clean task. 
The consequences of this idea have been explored in Ref.~\cite{Ga03}.

\item [2)] In order to perform, as desirable, a direct measurement of $\Gamma_n/\Gamma_p$ one needs to 
detect the outgoing \emph{neutrons}. In principle, neutron spectra 
can be measured down to about 10 MeV kinetic
energy since they are less affected than the proton ones by 
energy losses in the target and detector materials. Besides,
the joint observation of proton and neutron spectra could help to disentangle 
the set--up material effects from the nucleon FSI 
occurring inside the residual nucleus.
A recent experiment, KEK--E369, measured neutron spectra from $^{12}_{\Lambda}$C
and $^{89}_{\Lambda}$Y non--mesonic decays \cite{kim}. 
With other experiments, KEK--E462 and KEK--E408 \cite{Ok04l}, proton and neutron spectra could be
simultaneously measured for $^5_{\Lambda}$He and $^{12}_{\Lambda}$C.
An analysis of these data has been done in Ref.~\cite{Ga03} and is discussed in 
\ref{pnsing}.

\item [3)] One could avoid the possible deficiencies of the \emph{single} nucleon spectra measurements 
discussed in point 1) by employing nucleon--nucleon \emph{coincidence} measurements, which 
should indeed permit a cleaner extraction of $\Gamma_n/\Gamma_p$.
Coincidence observations are also expected to be less affected by
FSI effects and to help in the direct observation 
of two--nucleon induced decay events. In the experiments KEK--E462 
and KEK--E508 \cite{outa-coinc}, $nn$ and $np$ angular and energy correlation
measurements have been performed for the decay of $^5_{\Lambda}$He and $^{12}_{\Lambda}$C.
Other experiments, at BNL \cite{Gi01} and J--PARC \cite{jparc}, 
will determine $\Gamma_n/\Gamma_p$ for $s$--shell hypernuclei,
again by $nn$ and $np$ coincidence measurements. More data are expected 
in the near future from Da$\Phi$ne \cite{FI}.
\end{itemize}

\subsubsection{{\sl Analyses of single nucleon spectra}}
\label{pnsing}
In Ref.~\cite{Ga03} a OME model for the $\Lambda N\to nN$ transition
in finite nuclei was incorporated in the intranuclear cascade code of 
Ref.~\cite{Ra97} for the calculation of single and coincidence nucleon
distributions from hypernuclear non--mesonic weak decay. 
The OME weak transition potential \cite{Pa01} contained the exchange
of $\pi$, $\rho$, $K$, $K^*$, $\omega$ and $\eta$ mesons. Strong couplings and
strong FSI acting between the weak decay nucleons were taken into
account by means of a $nN$ wave function from the Lippmann--Schwinger equation
obtained with NSC97 (versions ``a" and ``f") potentials \cite{Ri99a}.
The two--nucleon stimulated channel is also taken into account, using the
polarization propagator method and treating the
nuclear finite size effects by means of a local density approximation.

\begin{figure}
\begin{center}
\includegraphics[width=.8\textwidth]{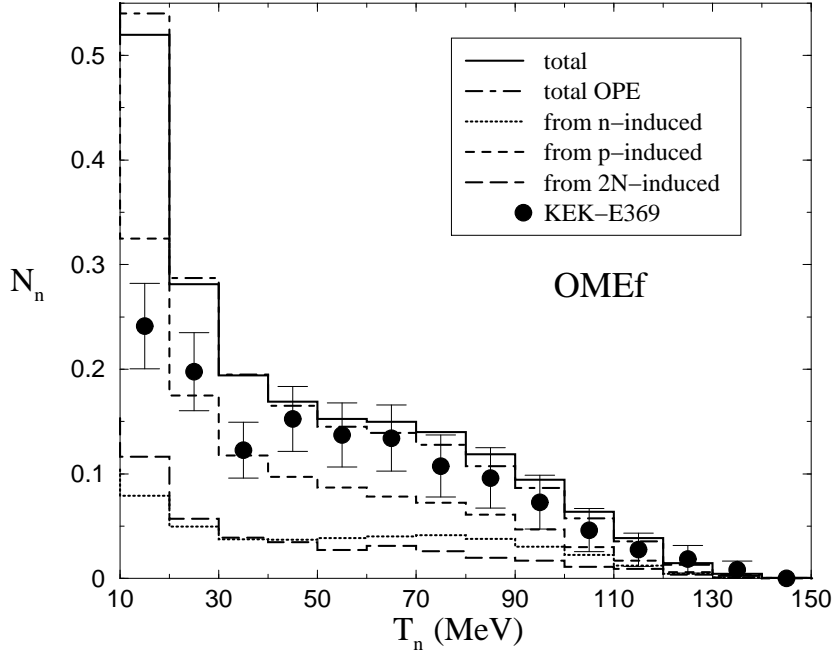}
\caption{Single neutron kinetic energy spectra for the non--mesonic weak decay of
$^{12}_\Lambda$C. The total spectrum $N_n$ (normalized per non--mesonic weak decay)
has been decomposed in its components $N^{\Lambda n\to nn}_{np}$,
$N^{\Lambda p\to np}_{np}$ and $N^{\Lambda np\to nnp}_{np}$ according to Eq.~(\ref{single-n})
(taken from Ref.~\cite{Ga03}).}
\label{sing-n}
\end{center}
\end{figure}

Single neutron and proton spectra have been calculated and compared
with data for $^5_\Lambda$He and $^{12}_\Lambda$C. 
In Fig.~\ref{sing-n} we show results of Ref.~\cite{Ga03} for the kinetic energy neutron distribution
from $^{12}_\Lambda$C based on two models [OPE and OMEf (using the NSC97f potential)], 
which predict quite different $\Gamma_n/\Gamma_p$ ratios.
The KEK--E369 spectrum \cite{kim} is well reproduced by the calculations. Unfortunately,
the dependence of the neutron spectrum $N_n$ on variations of $\Gamma_n/\Gamma_p$ is very weak
(the same is true also for the proton spectrum)
and a precise extraction of the ratio from the KEK--E369 distribution is not possible.
The little sensitivity of $N_n$ and $N_p$ to $\Gamma_n/\Gamma_p$ is mainly due to
the fact that these numbers are normalized to the same total non--mesonic decay rate
(i.e., per non--mesonic weak decay).
The non--normalized nucleon spectra, $S_n\equiv N_n \, \Gamma_{\rm NM}$
and $S_p\equiv N_p \, \Gamma_{\rm NM}$ [see Eq.~(\ref{single-n})],
have indeed a stronger dependence on $\Gamma_n/\Gamma_p$.
As a consequence, in order to discriminate between different weak decay models, one should
separately compare the complementary observable, $\Gamma_{\rm NM}$, with experiment.
For $^{12}_\Lambda$C, the calculations of Ref.~\cite{Ga03} supply
$\Gamma_{\rm NM}\equiv \Gamma_n+\Gamma_p+\Gamma_2=0.91$ or $0.69$ 
when model OMEa or OMEf is
used, which agree quite well with the experimental determinations (see Table \ref{best-nm}).

Let us now introduce the number of nucleons of the type $N$ ($N=n,p$) produced in $n$--induced
($N^{\rm 1Bn}_N$), $p$--induced ($N^{\rm 1Bp}_N$) and two--nucleon induced
($N^{\rm 2B}_N$) decays. If we normalize these quantities per
$n$--induced, $p$--induced, and two--nucleon induced decay, respectively,
the \emph{total} number of nucleons of the type $N$
(normalized per non--mesonic weak decay) is given by:
\begin{eqnarray}
\label{single-n}
N_N&=&\frac{N^{\rm 1Bn}_N\, \Gamma_n+N^{\rm 1Bp}_N\,
\Gamma_p+N^{\rm 2B}_N\, \Gamma_2}
{\Gamma_n+\Gamma_p+\Gamma_2} \\
&\equiv&N^{\Lambda n\to nn}_N+N^{\Lambda p\to np}_N
+N^{\Lambda np\to nnp}_N, \nonumber
\end{eqnarray}
where $N^{\Lambda n\to nn}_N$, $N^{\Lambda p\to np}_N$ and
$N^{\Lambda np\to nnp}_N$ have obvious meaning and are shown in Fig.~\ref{sing-n}
for the neutron spectrum from $^{12}_\Lambda$C.
All these nucleon numbers can be considered either as being functions of the nucleon
kinetic energy, $N_N(T_N)$, as it is in Fig.~\ref{sing-n},
or as the corresponding integrated quantities,
$N_N=\int dT_N N_N(T_N)$.
By construction, $N^{\rm 1Bn}_N$, $N^{\rm 1Bp}_N$ and $N^{\rm 2B}_N$
($N^{\Lambda n\to nn}_N$, $N^{\Lambda p\to np}_N$ and $N^{\Lambda np\to nnp}_N$)
\emph{do not} depend (\emph{do} depend) on the interaction model
employed to describe the weak decay.

The problem of the small sensitivity of $N_n$ and $N_p$
to variations of $\Gamma_n/\Gamma_p$ can be overcome if one concentrates
on another single nucleon observable, the ratio $N_n/N_p$.
The number of \emph{weak decay} neutrons and protons produced per non--mesonic weak 
decay are given by $N^{\rm wd}_n= (2\,\Gamma_n+\Gamma_p)/\Gamma_{\rm NM}$
and $N^{\rm wd}_p= \Gamma_p/\Gamma_{\rm NM}$, 
respectively. Thus, the ratio $\Gamma_n/\Gamma_p$ can be obtained as:
\begin{equation}
\label{ratio-np1}
\frac{\Gamma_n}{\Gamma_p}\equiv \frac{1}{2}\left(\frac{N^{\rm wd}_n}{N^{\rm wd}_p}-1\right) .
\end{equation}
Due to two--body induced decays and (especially) nucleon FSI, one expects the inequality:
\begin{equation}
\label{ratio-np2}
\frac{\Gamma_n}{\Gamma_p}
\neq \frac{1}{2}\left(\frac{N_n}{N_p}-1\right)\equiv
R_1\left[\Delta T_n, \Delta T_p\right]
\end{equation}
to be valid in a situation, such as the experimental one, in which particular intervals
of variability of the neutron and proton kinetic energy,
$\Delta T_n$ and $\Delta T_p$, are employed in the determination of 
the \emph{observable} numbers $N_n$ and $N_p$.

The results of Ref.~\cite{Ga03} clearly show a pronounced dependence of $R_1$ on
$\Delta T_n$ and $\Delta T_p$ (see Table~\ref{sing-r}); $N_n/N_p$ turns out to be much less sensitive to
FSI effects and variations of the energy cuts than $N_n$ and $N_p$ separately.
The OMEf prediction of Table~\ref{sing-r}
corresponding to $T^{\rm th}_N=60$ MeV is compatible with 
the very recent determination by KEK--E462 \cite{Ok04l}. 
\begin{table}
\begin{center}
\caption{\normalsize Predictions of Ref.~\cite{Ga03} for the quantity $R_1$ of Eq.~(\ref{ratio-np2}) for
$^5_\Lambda$He corresponding to different nucleon kinetic energy thresholds 
$T^{\rm th}_N$ and to the OPE, OMEa and OMEf models.}
\label{sing-r}
{\normalsize
\begin{tabular}{c c c c c} \hline
\mc {1}{c}{} &
\mc {1}{c}{} &
\mc {1}{c}{$T^{\rm th}_N$ (MeV)} &
\mc {1}{c}{} &
\mc {1}{c}{} \\
               & $0$     & $30$    & $60$   & $\Gamma_n/\Gamma_p$ \\ \hline
   OPE      & $0.04$  & $0.13$  & $0.16$ & $0.09$  \\
   OMEa     & $0.15$  & $0.32$  & $0.39$ & $0.34$  \\
   OMEf   & $0.19$  & $0.40$  & $0.49$ & $0.46$ \\ \hline
KEK--E462 \cite{Ok04l} & & & $0.59 \pm 0.11$ & \\ \hline
\end{tabular}}
\end{center}
\end{table}

\subsubsection{{\sl Analyses of coincidence spectra}}
\label{coin}
Due to the reduction of interferences and FSI effects, nucleon correlation
analyses are expected to provide a cleaner determination of $\Gamma_n/\Gamma_p$
than single nucleon observables \cite{Al02}. Ref.~\cite{Ga03}
evaluated double--nucleon energy and angular correlations and analyzed the
data obtained by the experiments KEK--E462 and KEK--E508 \cite{outa-coinc}
for $^5_\Lambda$He and $^{12}_\Lambda$C.

\begin{figure}
\begin{center}
\includegraphics[width=.8\textwidth]{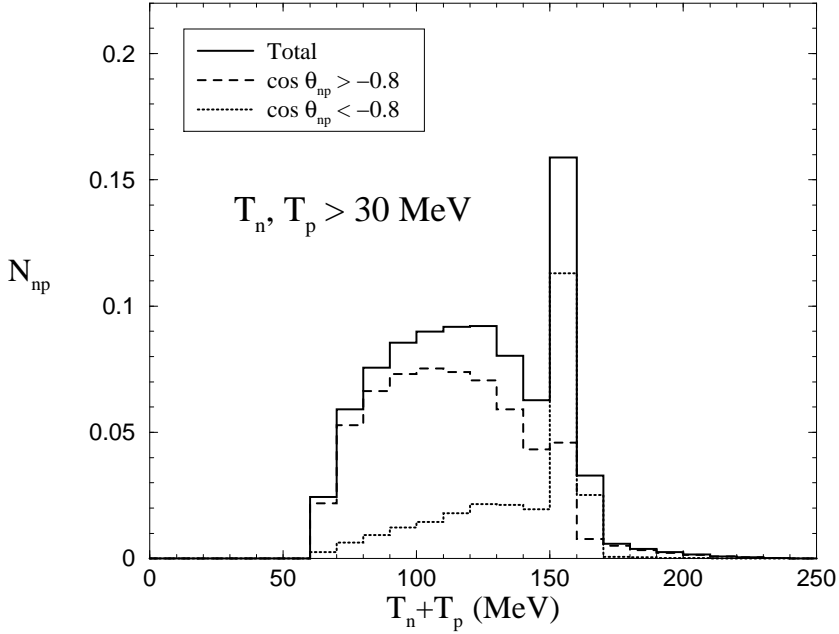}
\caption{Kinetic energy correlations of $np$ pairs emitted
per non--mesonic decay of $^{12}_\Lambda$C (taken from Ref.~\cite{Ga03}).}
\label{c-ene}
\end{center}
\end{figure}

In Fig.~\ref{c-ene} we report the prediction of Ref.~\cite{Ga03} for the kinetic energy 
correlation of $np$ pairs emitted in the non--mesonic decay of $^{12}_\Lambda$C. 
To facilitate a comparison with experiments,
whose kinetic energy threshold for proton (neutron) detection
is typically of about $30$ MeV ($10$ MeV), and to avoid a possible
non--realistic behaviour of the intranuclear cascade simulation
for low nucleon energies, the distributions correspond to $T_n$, $T_p\geq 30$ MeV.
A narrow peak, mainly originated by the back--to--back kinematics
(cos$\, \theta_{np}<-0.8$) of the one--nucleon induced decay, 
is predicted around $155$ MeV, i.e., close to the
$Q$--value expected for the proton--induced three--body process
$^{12}_\Lambda{\rm C}\to ^{10}\hspace{-1mm}{\rm B}+n+p$. 
A broad peak, predominantly due to $\Lambda p \to np$ or
$\Lambda n \to nn$ weak transitions
followed by the emission of secondary (less energetic) nucleons,
has been found around $110$ MeV for cos$\, \theta_{np}>-0.8$.

\begin{figure}
\begin{center}
\includegraphics[width=.8\textwidth]{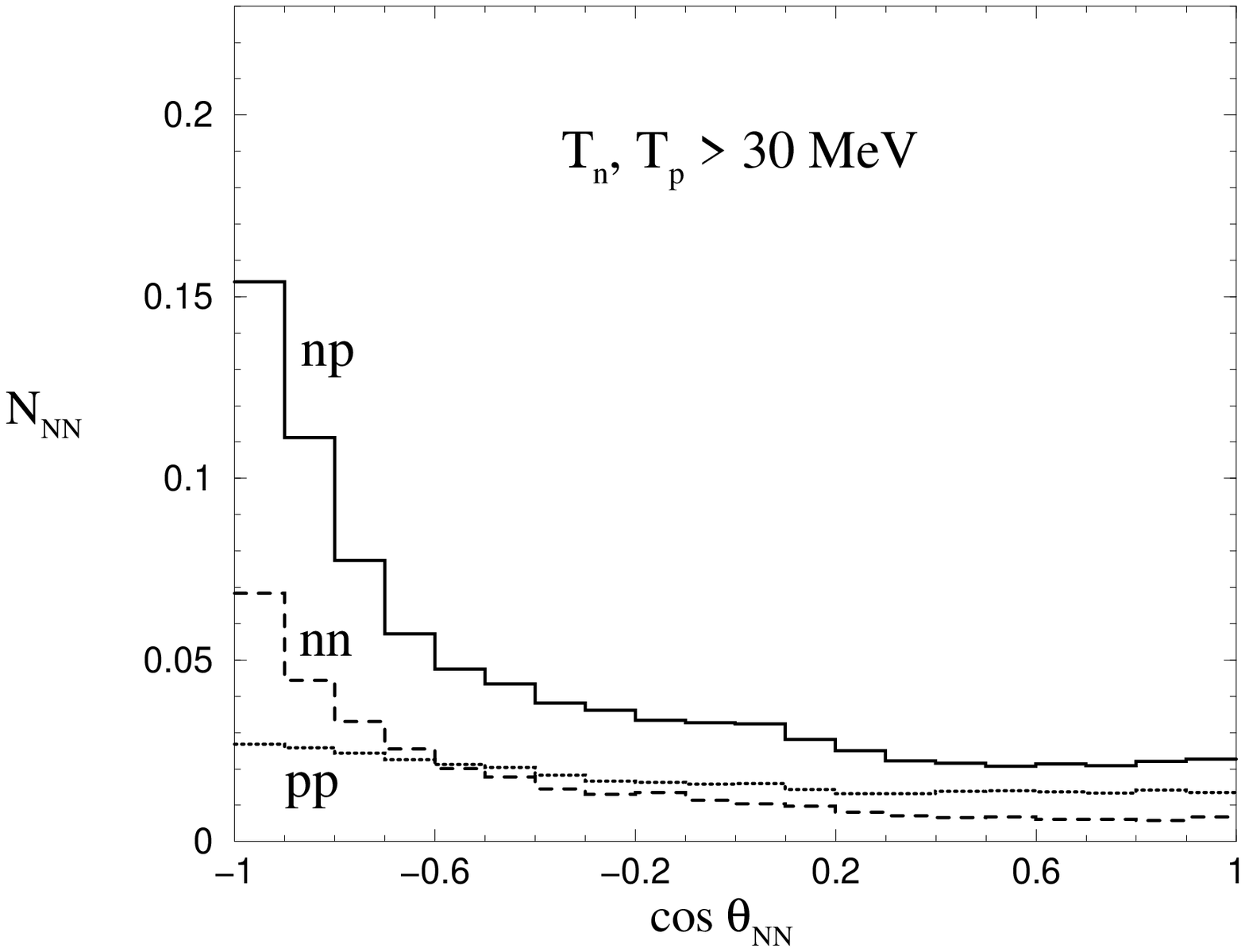}
\caption{
Opening angle correlations of $nn$, $np$ and $pp$ pairs
emitted per non--mesonic decay of $^{12}_\Lambda$C
(taken from Ref.~\cite{Ga03}).}
\label{c-ang}
\end{center}
\end{figure}

Figure \ref{c-ang} shows the opening angle correlations of $nn, np$ and
$pp$ pairs from the non--mesonic weak decay of $^{12}_\Lambda\rm{C}$
and for a $30$ MeV energy threshold. The peaking structure at $\cos\, \theta\simeq -1$ 
for the $nn$ ($np$) distribution is a clear signal of the back--to--back
kinematics of the neutron (proton) induced decay. For a discussion
of the different effects of FSI in the energy and angular coincidence distributions 
for $^5_\Lambda\rm{He}$ and $^{12}_\Lambda\rm{C}$ see Ref.~\cite{Ga03}.

The ratio $\Gamma_n/\Gamma_p$ is defined as the ratio
between the number of \emph{weak decay} $nn$ and $np$ pairs,
$N^{\rm wd}_{nn}$ and $N^{\rm wd}_{np}$. However, due to
two--body induced decays and (especially) nucleon FSI effects, one has:
\begin{equation}
\label{ratio-nn}
\frac{\Gamma_n}{\Gamma_p}\equiv \frac{N^{\rm wd}_{nn}}{N^{\rm wd}_{np}}
\neq \frac{N_{nn}}{N_{np}}
\equiv R_2\left[\Delta \theta_{12}, \Delta T_n, \Delta T_p\right]
\end{equation}
when the \emph{observable} numbers $N_{nn}$ and $N_{np}$ 
are determined by employing particular intervals
of variability of the pair opening angle, $\Delta \theta_{12}$, and the nucleon
kinetic energies, $\Delta T_n$ and $\Delta T_p$. The discussion
of Ref.~\cite{Ga03} proves that $N_{nn}/N_{np}$ is much less sensitive
to FSI effects and variations of the energy
cuts and angular restrictions than $N_{nn}$ and $N_{np}$ separately.

The numbers of nucleon pairs $N_{NN}$ discussed up to now and
normalized per non--mesonic weak decay
are related to the corresponding quantities for the one--nucleon ($N^{\rm 1B}_{NN}$)
and two--nucleon ($N^{\rm 2B}_{NN}$) induced processes
[the former (latter) being normalized per one--body (two--body) stimulated
non--mesonic weak decay] via the following equation:
\begin{equation}
\label{1-2}
N_{NN}=\frac{N^{\rm 1B}_{NN}\, \Gamma_1+N^{\rm 2B}_{NN}\, \Gamma_2}{\Gamma_1+\Gamma_2} 
\equiv N^{\Lambda n\to nn}_{NN}+
N^{\Lambda p\to np}_{NN}+N^{\Lambda np\to nnp}_{NN},
\end{equation}
where:
\begin{eqnarray}
\label{1-22}
N^{\rm 1B}_{NN}&=&\frac{N^{\rm 1Bn}_{NN}\, \Gamma_n+N^{\rm 1Bp}_{NN}\,
\Gamma_p} {\Gamma_1} ,
\end{eqnarray}
and the remaining $N$'s have obvious meaning. Therefore, the quantities
$N^{\rm 1Bn}_{NN}$, $N^{\rm 1Bp}_{NN}$ and $N^{\rm 2B}_{NN}$
($N^{\Lambda n\to nn}_{NN}$, $N^{\Lambda p\to np}_{NN}$ and $N^{\Lambda np\to nnp}_{NN}$)
\emph{do not} depend (\emph{do} depend) on the interaction model
employed to describe the weak decay.

In Table \ref{ome-che} the ratio $N_{nn}/N_{np}$ predicted by the
OPE, OMEa and OMEf models of Refs.~\cite{Ga03,GaDAFNE} for $^5_\Lambda$He
and $^{12}_\Lambda$C is given for a nucleon energy threshold of $30$ MeV and 
for the back--to--back kinematics (cos $\theta_{NN}\leq -0.8$). 
The predictions of the different weak decay models 
for $\Gamma_n/\Gamma_p$ are also quoted. The results of the OMEa and OMEf
models are in agreement with the preliminary KEK--E462 and KEK--E508 data
\cite{outa-coinc}: this comparison provides an indication for a ratio
$\Gamma_n/\Gamma_p\simeq 0.3$ in both hypernuclei.
\begin{table}
\begin{center}
\caption{\normalsize 
Predictions of Refs.~\cite{Ga03,GaDAFNE} for the ratio $N_{nn}/N_{np}$
for $^5_\Lambda$He and $^{12}_\Lambda$C. They correspond to
an energy thresholds $T^{\rm th}_N$ of $30$ MeV 
and to the back--to--back kinematics ($\cos\, \theta_{NN}\leq -0.8$). 
The (preliminary) data are from KEK--E462 
and KEK--E508 \protect\cite{outa-coinc}.}
\label{ome-che}
{\normalsize
\begin{tabular}{c c c c c} \hline
\mc {1}{c}{} &
\mc {1}{c}{$^5_\Lambda$He} &
\mc {1}{c}{} &
\mc {1}{c}{$^{12}_\Lambda$C} &
\mc {1}{c}{} \\
       & $N_{nn}/N_{np}$ & $\Gamma_n/\Gamma_p$ 
       & $N_{nn}/N_{np}$ & $\Gamma_n/\Gamma_p$ \\ \hline
OPE     & $0.25$  & $0.09$  & $0.24$ & $0.08$  \\
OMEa    & $0.51$  & $0.34$  & $0.39$ & $0.29$  \\
OMEf    & $0.61$  & $0.46$  & $0.43$ & $0.34$ \\ \hline
EXP  & $0.45\pm 0.11$  &     &  $0.40\pm 0.09$   &  \\ \hline
\end{tabular}}
\end{center}
\end{table}

A weak--decay--model independent analysis of KEK coincidence data has been
performed in Ref.~\cite{Ga03}. The six weak--interaction--model independent quantities
$N^{\rm 1Bn}_{nn}$, $N^{\rm 1Bp}_{nn}$, $N^{\rm 2B}_{nn}$,
$N^{\rm 1Bn}_{np}$, $N^{\rm 1Bp}_{np}$ and $N^{\rm 2B}_{np}$ of Eqs.~(\ref{1-2}) and (\ref{1-22})
are used to evaluate $\Gamma_n/\Gamma_p$ as:
\begin{equation}
\label{fit}
\frac{\Gamma_n}{\Gamma_p}=
\frac{\displaystyle N^{\rm 1Bp}_{nn}+N^{\rm 2B}_{nn} \frac{\Gamma_2}{\Gamma_1}
-\left(N^{\rm 1Bp}_{np}+N^{\rm 2B}_{np} \frac{\Gamma_2}{\Gamma_1}
\right)\frac{N_{nn}}{N_{np}}}
{\displaystyle \left(N^{\rm 1Bn}_{np}+N^{\rm 2B}_{np}
\frac{\Gamma_2}{\Gamma_1} \right) \frac{N_{nn}}{N_{np}}
-N^{\rm 1Bn}_{nn}-N^{\rm 2B}_{nn}
\frac{\Gamma_2}{\Gamma_1}} 
\end{equation}
[which can be easily obtained from Eqs.~(\ref{1-2}) and (\ref{1-22})],
using appropriate values for $\Gamma_2/\Gamma_1$ and data for $N_{nn}/N_{np}$. 
For the KEK values of Table \ref{ome-che} the following ratios have been determined:
\begin{equation}
\label{fit1}
\frac{\Gamma_n}{\Gamma_p}\left(^5_\Lambda {\rm He}\right)=0.40\pm0.11\,\,\,
{\rm if}\,\,\, \Gamma_2=0\,\, , \,\,
\frac{\Gamma_n}{\Gamma_p}\left(^5_\Lambda {\rm He}\right)=0.27\pm0.11\,\,\,
{\rm if}\,\,\, \frac{\Gamma_2}{\Gamma_1}=0.2 ,
\end{equation}
\begin{equation}
\label{fit2}
\frac{\Gamma_n}{\Gamma_p}\left(^{12}_\Lambda {\rm C}\right)=0.38\pm 0.14\,\,\,
{\rm if}\,\,\, \Gamma_2=0\,\, , \,\,
\frac{\Gamma_n}{\Gamma_p}\left(^{12}_\Lambda {\rm C}\right)=0.29\pm 0.14\,\,\,
{\rm if}\,\,\, \frac{\Gamma_2}{\Gamma_1}=0.25 ,
\end{equation}
where the non--vanishing values adopted for $\Gamma_2/\Gamma_1$ are predictions obtained 
in Ref.~\cite{Al99} within the polarization propagator method in local density approximation.
The $\Gamma_n/\Gamma_p$ values of Eqs.~(\ref{fit1}) and (\ref{fit2}) are in agreement 
with the pure theoretical predictions of Refs.~\cite{Os01,Pa01,It02} but
are substantially smaller than those obtained experimentally from \emph{single} nucleon spectra analyses 
(see Table \ref{best-ratio}). Actually, all the previous experimental analyses of single nucleon
spectra, supplemented in some cases by
intranuclear cascade calculations, derived $\Gamma_n/\Gamma_p$ values in
disagreement with all existing theoretical predictions. 
The fact that the calculations of Refs.~\cite{Ga03,GaDAFNE} reproduce coincidence data
for values of $\Gamma_n/\Gamma_p$ as small as $0.3\div 0.4$
could signal the existence of non--negligible interference effects between
the $n$-- and $p$--induced channels in those old single nucleon data.

In our opinion, the achievements discussed in this subsection clearly exhibit the 
interest of analyses of correlation observables and represent an important
progress towards the solution of the $\Gamma_n/\Gamma_p$ puzzle.
Forthcoming coincidence data from KEK, BNL \cite{Gi01}, J--PARC \cite{jparc} 
and FINUDA \cite{FI} could be directly compared with the results discussed here 
and in Refs.~\cite{Ga03,GaDAFNE}.
This will permit to achieve better determinations of $\Gamma_n/\Gamma_p$
and to establish the first constraints on $\Gamma_2/\Gamma_1$.

\section{Hypernuclei of the $s$--shell and $\Delta I=1/2$ rule}
\label{passh}
The analysis of the non--mesonic decays for $s$--shell hypernuclei offers an important 
and complementary tool for the solution of the $\Gamma_n/\Gamma_p$ puzzle.
In addition, it is very suitable for testing the 
validity of the related $\Delta I= 1/2$ rule. 

For $s$--shell hypernuclei the $\Lambda N$   
pair is necessarily in the $L=0$ relative state, thus the only possible 
$\Lambda N\rightarrow nN$ transitions are the following ones 
(we use the spectroscopic notation $^{2S+1}L_J$):  
\begin{eqnarray}
\label{partial}  
{^1S_0} &\rightarrow &{^1S_0} \hspace{0.3cm}(I_f=1)  \\  
        &\rightarrow &{^3P_0} \hspace{0.3cm}(I_f=1) \nonumber \\
{^3S_1} &\rightarrow &{^3S_1} \hspace{0.3cm}(I_f=0) \nonumber \\ 
        &\rightarrow &{^1P_1} \hspace{0.3cm}(I_f=0) \nonumber \\ 
        &\rightarrow &{^3P_1} \hspace{0.3cm}(I_f=1) \nonumber \\
        &\rightarrow &{^3D_1} \hspace{0.3cm}(I_f=0) . \nonumber   
\end{eqnarray} 
The $\Lambda n\rightarrow nn$ process has final states with isospin  
$I_f=1$ only, while for $\Lambda p\rightarrow np$ both $I_f=1$ and  
$I_f=0$ are allowed. 

\subsection{Phenomenological model of Block and Dalitz}
In the following we discuss an analysis performed by 
the authors \cite{Al99b} in order to explore the validity of 
the $\Delta I= 1/2$ rule in the one--nucleon induced $\Lambda$--decay.
This analysis is based on a phenomenological model due to Block and Dalitz
\cite{Bl63}, which we briefly outline now.

The interaction probability of a particle which crosses an infinite
homogeneous system of thickness $ds$ is, classically, $dP=ds/\lambda$, where 
$\lambda=1/(\sigma \rho)$ is the mean free path of the projectile, 
$\sigma$ is the relevant cross section and $\rho$ is the
density of the system. Then, if we refer to the process $\Lambda N\to nN$, the width
$\Gamma_{\rm NM}=dP_{\Lambda N\to nN}/dt$ can be written as:  
\begin{equation} 
\Gamma_{\rm NM}=v\sigma \rho , \nonumber
\end{equation}
$v=ds/dt$ being the $\Lambda$ velocity in the rest frame of the homogeneous system.  For a
finite nucleus of density $\rho(\vec r)$, after introducing a local Fermi sea of nucleons
one can write, within the semiclassical approximation:  
\begin{equation} 
\Gamma_{\rm NM}=\langle v\sigma \rangle \int d{\vec r} \rho({\vec r}) \mid 
\psi_{\Lambda}({\vec r})\mid^2 , \nonumber
\end{equation}
where $\psi_{\Lambda}({\vec r})$ is the $\Lambda$ wave function in the  
hypernucleus and $\langle \rangle$ denotes an average over spin and isospin
states. In the above equation the nuclear density is normalized to the 
mass number $A=N+Z$, hence the integral gives the average nucleon density $\rho_A$ 
at the position of the $\Lambda$ hyperon. In this scheme, the non--mesonic width 
$\Gamma_{\rm NM}=\Gamma_n+\Gamma_p$ of the hypernucleus $^{A+1}_{\Lambda}Z$ turns
out to be:  
\begin{equation}  
\Gamma_{\rm NM}(^{A+1}_{\Lambda}Z)= 
\frac{N\overline{R}_n(^{A+1}_{\Lambda}Z)+Z\overline{R}_p(^{A+1}_{\Lambda}Z)}{A}\rho_A , \nonumber
\end{equation}
where $\overline{R}_n$ ($\overline{R}_p$) denotes the spin--averaged rate for   
the neutron--induced (proton--induced) process appropriate for the 
considered hypernucleus. 

Furthermore, by introducing the rates $R_{NJ}$ for the spin--singlet ($R_{n0}$, $R_{p0}$) and
spin--triplet ($R_{n1}$, $R_{p1}$) elementary $\Lambda N\to nN$ 
interactions, the non--mesonic decay widths of
$s$--shell hypernuclei are \cite{Bl63}:  
\begin{eqnarray} 
\label{phen}
\Gamma_{\rm NM}(^3_{\Lambda}{\rm H})&=&
\left(3R_{n0}+R_{n1}+3R_{p0}+R_{p1}\right)\frac{\rho_2}{8} ,\\
\Gamma_{\rm NM}(^4_{\Lambda}{\rm H})&=& 
\left(R_{n0}+3R_{n1}+2R_{p0}\right)\frac{\rho_3}{6},\nonumber \\ 
\Gamma_{\rm NM}(^4_{\Lambda}{\rm He})&=&
\left(2R_{n0}+R_{p0}+3R_{p1}\right)\frac{\rho_3}{6} ,\nonumber \\
\Gamma_{\rm NM}(^5_{\Lambda}{\rm He})&=&
\left(R_{n0}+3R_{n1}+R_{p0}+3R_{p1}\right)\frac{\rho_4}{8} .
\nonumber 
\end{eqnarray}
These relations take into account that the total hypernuclear angular momentum is 0 for
$^4_{\Lambda}$H and $^4_{\Lambda}$He and 1/2 for $^3_{\Lambda}$H and $^5_{\Lambda}$He.
In terms of the rates associated to the partial--wave transitions (\ref{partial}), the
$R_{NJ}$'s of Eqs.~(\ref{phen}) read:  
\begin{eqnarray} 
R_{n0}&=&R_n(^1S_0)+R_n(^3P_0) , \nonumber \\ 
R_{p0}&=&R_p(^1S_0)+R_p(^3P_0) , \nonumber \\
R_{n1}&=&R_n(^3P_1) , \nonumber \\ 
R_{p1}&=&R_p(^3S_1)+R_p(^1P_1)+R_p(^3P_1)+R_p(^3D_1), \nonumber 
\end{eqnarray} 
the quantum numbers of the $nN$ final state being reported in brackets.  

If one assumes that the $\Lambda N\to nN$ weak interaction occurs with a
change $\Delta I=1/2$ of the isospin, the following relations (simply derived by
angular momentum coupling coefficients)  hold among the rates for transitions to $I_f=1$
final states:  
\begin{equation} 
\label{uno1} 
R_n(^1S_0)=2R_p(^1S_0) , \hspace{0.2cm}
R_n(^3P_0)=2R_p(^3P_0) , \hspace{0.2cm} R_n(^3P_1)=2R_p(^3P_1) .
\end{equation}
Hence: 
\begin{equation}
\label{uno2} 
\frac{R_{n1}}{R_{p1}}\leq \frac{R_{n0}}{R_{p0}}=2 .  
\end{equation}
For pure $\Delta I=3/2$ transitions, the factors 2 in 
Eqs.~(\ref{uno1}) are replaced by 1/2. Then, by further introducing the ratio:  
\begin{equation}
r=\frac{\langle I_f=1||A_{1/2}||I_i=1/2\rangle} {\langle
I_f=1||A_{3/2}||I_i=1/2\rangle} \nonumber
\end{equation}
between the $\Delta I=1/2$ and $\Delta I=3/2$
$\Lambda N\to nN$ transition amplitudes for isospin 1 final states ($r$ being real, as
required by time reversal invariance), for a general $\Delta I=1/2$--$3/2$
isospin mixture one gets:  
\begin{equation} 
\label{mixt1} 
\frac{R_{n1}}{R_{p1}}=
\frac{4r^2-4r+1}{2r^2+4r+2+6\lambda^2}\leq
\frac{R_{n0}}{R_{p0}}=\frac{4r^2-4r+1}{2r^2+4r+2} , 
\end{equation}
where:  
\begin{equation} 
\label{lam}
\lambda=\frac{\langle I_f=0||A_{1/2}||I_i=1/2\rangle} 
{\langle I_f=1||A_{3/2}||I_i=1/2\rangle} .  
\end{equation}
The partial rates of Eq.~(\ref{mixt1}) supply
$\Gamma_n/\Gamma_p$ for $s$--shell hypernuclei through Eqs.~(\ref{phen}).

By using Eqs.~(\ref{phen}) and (\ref{mixt1}) together with available
data it is possible to determine the spin--isospin structure
of the $\Lambda N\to nN$ interaction without resorting to a detailed knowledge 
of the interaction mechanism. 

\subsection{Experimental data and $\Delta I=1/2$ rule} 
\label{blda} 
In Ref.~\cite{Al99b} a phenomenological analysis
of the data summarized in Table~\ref{data} 
is employed to study the possible violation of the $\Delta I=1/2$ rule 
in the process $\Lambda N\to nN$. 
\begin{table}  
\begin{center} 
\caption{\normalsize Experimental data for the non--mesonic weak decay of $s$--shell hypernuclei.} 
\label{data} 
{\normalsize
\begin{tabular}{c c c c c c} 
\hline
\mc {1}{c}{} & 
\mc {1}{c}{$\Gamma_n/\Gamma^{\rm free}_\Lambda$} &  
\mc {1}{c}{$\Gamma_p/\Gamma^{\rm free}_\Lambda$} & 
\mc {1}{c}{$\Gamma_{\rm NM}/\Gamma^{\rm free}_\Lambda$} &
\mc {1}{c}{$\Gamma_n/\Gamma_p$} & 
\mc {1}{c}{Ref.} \\ \hline  
$^4_{\Lambda}$H      &                       &                & $0.22\pm 0.09$ & &  
reference value \\
                     &                       &                & $0.17\pm 0.11$ & 
& 
KEK \cite{Ou98}\\ 
                     &                       &                & $0.29\pm 0.14$ & 
& 
\cite{Bl63}\\  
$^4_{\Lambda}$He     &$0.04\pm 0.02$         & $0.16\pm 0.02$ & $0.20\pm 0.03$ & $0.25\pm  
0.13$         &   
BNL \cite{Ze98} \\
$^5_{\Lambda}$He     &$0.20\pm 0.11$         & $0.21\pm 0.07$ & $0.41\pm 0.14$ & $0.93\pm  
0.55$         & 
BNL \cite{Sz91} \\ \hline  
\end{tabular}} 
\end{center} 
\end{table} 

Unfortunately, no data are available on the non--mesonic decay of hypertriton
($^3_\Lambda$H) and on $\Gamma_n/\Gamma_p$ for $^4_{\Lambda}$H. Indeed, we shall see in the 
following that future measurements of $\Gamma_n/\Gamma_p$ for $^4_{\Lambda}$H at BNL 
\cite{Gi01} and J--PARC \cite{jparc} will be of great importance for testing the $\Delta I=1/2$ rule.
The BNL data \cite{Ze98,Sz91} for $^4_{\Lambda}$He and $^5_{\Lambda}$He of Table~\ref{data} 
together with the \emph{reference value} for $^4_{\Lambda}$H have been used
in the analysis of Ref.~\cite{Al99b}. This last number is the weighted
average of two previous estimates \cite{Ou98,Bl63},
which have not been obtained from direct measurements but rather by using 
theoretical constraints. One has then 5 independent data which allow to fix,  
from Eqs.~(\ref{phen}), the 4 rates $R_{N,J}$ and $\rho_3$. Indeed, the 
average nucleon density at the $\Lambda$ position for $^5_{\Lambda}$He, 
also entering Eqs.~(\ref{phen}), has been estimated to be 
$\rho_4=0.045$ fm$^{-3}$ by employing the 
$\Lambda$ wave function of Ref.~\cite{St93} (which was obtained through a quark model
description of the $\Lambda N$ interaction) and the gaussian density for $^4$He that  
reproduces the experimental mean square radius of the nucleus.    
For $^4_{\Lambda}$H and $^4_{\Lambda}$He, instead, no realistic hyperon wave 
function is available and we can obtain the value  
$\rho_3=0.026$ fm$^{-3}$ from the data of Table~\ref{data},  
by imposing that [see Eqs.~(\ref{phen})]:  
\begin{equation} 
\frac{\Gamma_p(^5_{\Lambda}{\rm He})}{\Gamma_p(^4_{\Lambda}{\rm He})}= 
\frac{3}{4} \frac{\rho_4}{\rho_3} .  \nonumber
\end{equation}

The best choice to determine the rates $R_{N,J}$ by fitting measured values
corresponds to use the relations for the observables:  
\begin{equation}
\Gamma_{\rm NM}(^4_{\Lambda}{\rm H}) , \hspace{0.3cm} \Gamma_{\rm NM}(^4_{\Lambda}{\rm He}) ,
\hspace{0.3cm} \Gamma_{\rm NM}(^5_{\Lambda}{\rm He}) , \hspace{0.3cm} 
\frac{\Gamma_n}{\Gamma_p}(^4_{\Lambda}{\rm He}) ,  \nonumber
\end{equation}
which have the smallest experimental uncertainties. 
After solving these equations we obtained the following partial rates 
(as usual, the decay widths of Eqs.~(\ref{phen}) are considered in units of the 
free $\Lambda$ decay width):
\begin{eqnarray}
\label{results} 
R_{n0}&=&(4.7\pm 2.1)\:{\rm fm}^3 , \\ 
R_{p0}&=&(7.9^{+16.5}_{-7.9})\:{\rm fm}^3 , \\
R_{n1}&=&(10.3\pm 8.6)\:{\rm fm}^3 , \\   
\label{results2}  
R_{p1}&=&(9.8\pm 5.5)\:{\rm fm}^3 , \\  
\overline{R}_n(^5_{\Lambda}{\rm He})\equiv   
\frac{1}{4}\left(R_{n0}+3R_{n1}\right)&=&(8.9\pm 6.5)\:{\rm fm}^3,  \nonumber \\  
\overline{R}_p(^5_{\Lambda}{\rm He})\equiv 
\frac{1}{4}\left(R_{p0}+3R_{p1}\right)&=&(9.3\pm 5.8)\:{\rm fm}^3 . \nonumber
\end{eqnarray} 
The errors have been obtained with the standard method, i.e.,  
by treating the data as independent and uncorrelated.

For the ratios of Eq.~(\ref{mixt1}) we have then:  
\begin{eqnarray}  
\label{gen1} 
\frac{R_{n0}}{R_{p0}}&=&0.6^{+1.3}_{-0.6} , \\
\label{gen2}
\frac{R_{n1}}{R_{p1}}&=&1.0^{+1.1}_{-1.0} . 
\end{eqnarray} 
while the ratios of the spin--triplet to the spin--singlet interaction 
rates are:  
\begin{eqnarray} 
\frac{R_{n1}}{R_{n0}}&=&2.2\pm 2.1 , \nonumber \\
\frac{R_{p1}}{R_{p0}}&=&1.2^{+2.7}_{-1.2} . \nonumber
\end{eqnarray}
The large uncertainties do not allow 
to draw definite conclusions about the possible violation of the 
$\Delta I=1/2$ rule and the spin--dependence of the transition rates. 
Eqs.~(\ref{gen1}) and (\ref{gen2}) are still compatible 
with Eq.~(\ref{uno2}), namely with the $\Delta I=1/2$ rule, although 
the central value in Eq.~({\ref{gen1}}) is more in 
agreement either with a pure $\Delta I=3/2$ transition ($r\simeq 0$) or
with $r\simeq 2$ [see Eq.~(\ref{mixt1})].
Actually, Eq.~(\ref{gen1}) is compatible with $r$   
in the range $-1/4\div 40$, while the ratio $\lambda$ of 
Eqs.~(\ref{mixt1}) and (\ref{lam}) is completely undetermined. 

By using the results of Eqs.~(\ref{results})--(\ref{results2}) one predicts:
\begin{eqnarray} 
\frac{\Gamma_n}{\Gamma_p}(^3_{\Lambda}{\rm H})&=&0.7^{+1.1}_{-0.7} , \nonumber \\
\frac{\Gamma_n}{\Gamma_p}(^4_{\Lambda}{\rm H})&=&2.3^{+5.0}_{-2.3} , \nonumber \\
\frac{\Gamma_n}{\Gamma_p}(^5_{\Lambda}{\rm He})&=&0.95\pm 0.92 , \nonumber
\end{eqnarray}
and, by using $\rho_2=0.001$ fm$^{-3}$ \cite{Bl63},
\begin{equation} 
\Gamma_{\rm NM}(^3_{\Lambda}{\rm H})=0.007\pm 0.006 . \nonumber
\end{equation}
The ratio obtained for $^5_{\Lambda}$He is in agreement with the data of Table \ref{data}
and with the recent determinations discussed in \ref{coin}. 
An accurate measurement of $\Gamma_{\rm NM}(^3_{\Lambda}{\rm H})$ and  
$\Gamma_n/\Gamma_p$ for $^3_{\Lambda}{\rm H}$
and $^4_{\Lambda}{\rm H}$ would then provide a test of the weak 
decay model of Eqs.~(\ref{phen}) if the rates of Eqs.~(\ref{results})--(\ref{results2})
could be extracted with less uncertainty from data.

The compatibility of the data with the $\Delta I=1/2$ 
rule can be discussed in a different way: by \emph{assuming} this rule, 
we fix $R_{n0}/R_{p0}=2$. Then, by using the observables:  
\begin{equation}  
\Gamma_{\rm NM}(^4_{\Lambda}{\rm He}) , 
\hspace{0.3cm} \Gamma_{\rm NM}(^5_{\Lambda}{\rm He}) , \hspace{0.3cm}
\frac{\Gamma_n}{\Gamma_p}(^4_{\Lambda}{\rm He}) , \nonumber
\end{equation}
the extracted partial rates are ($R_{n0}$, $R_{n1}$, $\overline{R}_n$ and $\overline{R}_p$ 
are unchanged with respect to the above derivation): 
\begin{eqnarray} 
R_{n0}&=&(4.7\pm 2.1)\:{\rm fm}^3 , \nonumber \\  
R_{p0}\equiv R_{n0}/2&=&(2.3\pm 1.0)\:{\rm fm}^3 , \nonumber \\ 
R_{n1}&=&(10.3\pm 8.6)\:{\rm fm}^3 , \nonumber \\ 
R_{p1}&=&(11.7\pm 2.4)\:{\rm fm}^3 . \nonumber
\end{eqnarray} 
These values are compatible with the ones in Eqs.~(\ref{results})--(\ref{results2}). 
For pure $\Delta I=1/2$ transitions the spin--triplet interactions 
seem to dominate over the spin--singlet ones:   
\begin{eqnarray}
\frac{R_{n1}}{R_{n0}}&=&2.2\pm 2.1 , \nonumber \\
\frac{R_{p1}}{R_{p0}}&=&5.0\pm 2.4 . \nonumber
\end{eqnarray}
Moreover, since: 
\begin{equation} 
\frac{R_{n1}}{R_{p1}}=0.9\pm 0.8 , \nonumber
\end{equation}
from Eq.~(\ref{mixt1}) one obtains the following estimate   
for the ratio between the $\Delta I=1/2$ amplitudes: 
\begin{equation} 
\left|\frac{\langle I_f=0||A_{1/2}||I_i=1/2\rangle} 
{\langle I_f=1||A_{1/2}||I_i=1/2\rangle}\right| \simeq \frac{1}{3.7}\div 2.3 . \nonumber
\end{equation}
The other independent observables which have not been utilized are then predicted to be:   
\begin{equation}
\Gamma_{\rm NM}(^4_{\Lambda}{\rm H})=0.17\pm 0.11 , \nonumber
\end{equation}
\begin{equation}
\label{gngphe}
\frac{\Gamma_n}{\Gamma_p}(^5_{\Lambda}{\rm He})=0.95\pm 0.72 ,
\end{equation} 
in agreement with the data of Table~\ref{data}, with a $\chi^2$ for 
one degree of freedom of 0.31 (corresponding to a $0.56\, \sigma$ deviation).
This means that these data are consistent with the hypothesis of validity  
of the $\Delta I=1/2$ rule at the level of 60\%. In other words, 
the $\Delta I=1/2$ rule is excluded at the 40\% confidence level. 
On the contrary, if, according to the analysis of subsection \ref{coin},
one compares with the datum $\Gamma_n/\Gamma_p=0.40\pm 0.11$ 
for $^5_\Lambda$He, the $\Delta I=1/2$ rule turns out to be completely excluded by 
the central value of Eq~(\ref{gngphe}).

The observables for which experimental data are not available at present
are predicted to be:
\begin{eqnarray}
\frac{\Gamma_n}{\Gamma_p}(^3_{\Lambda}{\rm H})=1.3\pm 0.6 , \nonumber  \\
\frac{\Gamma_n}{\Gamma_p}(^4_{\Lambda}{\rm H})=7.6\pm 6.2 , \nonumber 
\end{eqnarray}
and, for $\rho_2=0.001$ fm$^{-3}$,
\begin{equation}
\Gamma_{\rm NM}(^3_{\Lambda}{\rm H})=0.005\pm 0.003 . \nonumber
\end{equation}
We note that the central value of $\Gamma_n/\Gamma_p$ for $^4_{\Lambda}{\rm H}$
in the analysis which enforces the $\Delta I=1/2$ rule is considerably 
larger than the central value obtained in the general analysis previously
discussed. Thus, the future measurements \cite{Gi01,jparc} of this quantity
will represent an important test of the $\Delta I=1/2$ rule.

\section{Non--mesonic decay of polarized $\Lambda$--hypernuclei: the asymmetry puzzle}
\label{polarized}

Lambda hypernuclear states can be produced with a sizeable amount of polarization 
\cite{Ba89}. The development of angular distribution measurements of decay 
particles (photons, pions and protons) from polarized hypernuclei is of crucial importance 
in order to extract new information on hypernuclear production, structure and decay.

Despite the recent progress discussed in section \ref{ratio}, the reaction mechanism for 
the non--mesonic weak decay does not seem to be fully understood. Indeed,
a new intriguing problem, of more recent origin, is open: it
concerns a strong disagreement between theory and experiment 
on the asymmetry of the angular emission of non--mesonic decay protons from polarized 
hypernuclei. This asymmetry is due to the interference between parity--violating and
parity--conserving $\vec \Lambda p \to np$ transition amplitudes \cite{Ba90}.
The non--mesonic rates $\Gamma_n$ and $\Gamma_p$ are dominated by parity--conserving amplitudes.
The study of the asymmetric emission of protons from polarized hypernuclei, with
its information on the spin--parity structure of the $\Lambda p \to np$ process, is thus
supposed to provide new constraints on the dynamics of the non--mesonic decay.

Thanks to the large momentum transfer involved, the $n(\pi^+,K^+)\Lambda$ reaction has been used
\cite{Aj92,Aj00}, at $p_{\pi}=1.05$ GeV and small $K^+$ laboratory scattering angles 
($2^\circ\lsim \theta_{K}\lsim 15^\circ$), 
to produce hypernuclear states with a substantial amount of spin--polarization
preferentially aligned along the axis normal to the reaction plane.
The origin of hypernuclear polarization is twofold \cite{Ba89}. 
It is known that the distortions (absorptions) of the initial ($\pi^+$) 
and final ($K^+$) meson--waves produce a small polarization of the
hypernuclear orbital angular momentum up to laboratory scattering angles $\theta_{K}\simeq 15^\circ$
(at larger scattering angles, the orbital polarization increases with a negative sign). 
At small but non--zero angles, the main source of polarization is due to 
an appreciable spin--flip term in the elementary reaction $\pi^+ n\to \Lambda K^+$,
which interferes with the spin--nonflip amplitude.
In a typical experimental situation with $p_{\pi}=1.05$ GeV and $\theta_{K}\simeq 15^\circ$,
the polarization of the hyperon spin in the free $\pi^+ n\to \Lambda K^+$ process is about 0.75. 

\subsection{Spin--polarization observables}
\label{pol-obs}
The intensity of protons emitted in $\vec \Lambda p \to np$ decays
along a direction forming an angle $\Theta$ 
with the polarization axis is given by (see Ref.~\cite{Ra92} for more details):
\begin{equation}
\label{wdint}
I(\Theta, J)=I_0(J)\left[1+\mathcal{A}(\Theta, J)\right] , 
\end{equation}
where
\begin{equation}
\label{iso}
I_0(J)=\frac{\sum_M \sigma(J,M)}{2J+1} 
\end{equation} 
is the (isotropic) intensity for an unpolarized hypernucleus. In Eq.~(\ref{iso}):
\begin{equation}
\label{inten-prot}
\sigma(J, M)=\sum_{F} \left|\langle F|\mathcal{M}|I; J, M\rangle \right| ^2 \nonumber
\end{equation}
is the intensity of protons emitted along the quantization axis
for a projection $M$ of the hypernuclear total spin $J$.
The proton asymmetry parameter, $\mathcal{A}$, can be written in the following form \cite{Ra92}:
\begin{equation}
\label{asymm-a-vec}
\mathcal{A}(\Theta, J)=P_y(J)\, A_y(J)\, {\rm cos}\, \Theta .
\end{equation}
The quantity:
\begin{equation}
\label{asymm-y} 
A_y(J) = \frac{3}{J+1} \, \frac{\sum_{M} M\, \sigma(J, M)}{\sum_{M} \sigma(J, M)} , \nonumber
\end{equation}
which is a property of the hypernuclear non--mesonic decay only, is usually referred to
as the hypernuclear asymmetry parameter. The hypernuclear polarization $P_y$
depends both on the kinematics ($p_{\pi}$ and $\theta_{K}$) and dynamics of the production
reaction. 

In the shell model weak--coupling scheme, $P_y$ is directly related to the
polarization $p_{\Lambda}$ of the $\Lambda$ spin in the hypernucleus as follows:
\begin{equation} 
\label{p-lambda}
p_{\Lambda}(J)=
\begin{cases}
\displaystyle -\frac{J}{J+1}P_y(J) & \text{if}\, \, J=J_C-\frac{1}{2}  \\
P_y(J) & \text{if}\, \, J=J_C+\frac{1}{2} , 
\end{cases} 
\end{equation} 
$J_C$ being the total spin of the nuclear core.
It is useful to introduce an \emph{intrinsic lambda asymmetry parameter}
$a_{\Lambda}$, which is characteristic of the elementary process $\vec{\Lambda} p\to np$
and should be independent of the hypernucleus, such that:
\begin{equation}
\label{asymm-a-vec1} 
\mathcal{A}(\Theta, J)=p_{\Lambda}(J)\, a_{\Lambda}\, {\rm cos}\, \Theta . 
\end{equation} 
From Eqs.~(\ref{asymm-a-vec}) and (\ref{p-lambda}) it follows then:
\begin{equation}
\label{a-lambda}
a_{\Lambda}= 
\begin{cases} 
\displaystyle -\frac{J+1}{J}A_y(J) & \text{if}\, \, J=J_C-\frac{1}{2}  \\ 
A_y(J) & \text{if}\, \, J=J_C+\frac{1}{2} . \nonumber
\end{cases}
\end{equation}

\subsection{Experiment versus theory}
\label{expts}

Nucleon FSI acting after the non--mesonic weak decay are expected to modify the weak decay intensity
of Eq.~(\ref{wdint}). Experimentally one has access to a proton intensity 
$I^{\rm M}(\Theta, J)$ which is generally assumed to have the same $\Theta$--dependence
as $I(\Theta, J)$:
\begin{equation}
\label{wdint2}
I^{\rm M}(\Theta, J)=I^{\rm M}_0(J)\left[1+p_\Lambda(J)\, a^{\rm M}_\Lambda(J)
\cos \Theta\right] .
\end{equation}
The observable asymmetry, $a^{\rm M}_\Lambda(J)$, which is expected to depend on the 
hypernucleus, is then determined as:
\begin{equation}
\label{asym-exp}
a^{\rm M}_\Lambda(J)=\frac{1}{p_\Lambda(J)} \frac{I^{\rm M}(0^{\circ},J)-
I^{\rm M}(180^{\circ},J)}{I^{\rm M}(0^{\circ},J)+I^{\rm M}(180^{\circ},J)} .
\end{equation}

Until now, four KEK experiments measured the proton asymmetric emission
from polarized $\Lambda$--hypernuclei.
The experiment KEK--E160 \cite{Aj92}, which studied $p$--shell hypernuclei,
suffered from large uncertainties: only
poor statistics and energy resolution could be used; moreover, the
values of the $\Lambda$ polarization $p_\Lambda$
needed to determine the asymmetry $a^{\rm M}_\Lambda$,
had to be evaluated theoretically. More recently, $a^{\rm M}_\Lambda$
was measured by KEK--E278 \cite{Aj00} from the
decay of $^5_\Lambda{\vec {\rm H}{\rm e}}$. The values of $p_\Lambda$
used to obtain $a^{\rm M}_\Lambda$ were
determined by observing the asymmetry, ${\mathcal{A}}^{\pi^-}=p_\Lambda\, a^{\pi^-}_\Lambda$,
in the emission of negative pions in the $^5_\Lambda{\vec {\rm H}{\rm e}}$ mesonic decay,
after assuming $a^{\pi^-}_\Lambda$
to be equal to the value for the free $\Lambda \to \pi^- p$
decay, $\alpha_{\pi^-}=-0.642\pm 0.013$. Unfortunately, the small branching ratio
and expected asymmetry ${\mathcal{A}}^{\pi^-}$
for the mesonic decay of $p$--shell hypernuclei makes a similar
measurement of $p_\Lambda$ very difficult
for these systems; even the recent and more accurate experiment
KEK--E508 \cite{Ma04} had to resort to theoretical estimates 
for the $\Lambda$ polarization in 
$^{12}_\Lambda{\vec {\rm C}}$ and $^{11}_\Lambda{\vec {\rm B}}$. 
In the other recent experiment KEK--E462 \cite{Ma04}, $a^{\rm M}_\Lambda$ was measured again for
$^5_\Lambda {\vec {\rm H}{\rm e}}$, but with improved statistics.
\begin{table}
\begin{center}
\caption{\normalsize Theoretical and experimental determinations of
the asymmetry parameters ($a_\Lambda$ and $a^{\rm M}_\Lambda$, respectively).
The predictions for $a_\Lambda$ have been obtained
with different OME weak transition potentials
and with the direct quark mechanism (DQ).}
\label{other-res}
{\normalsize
\begin{tabular}{c c c} \hline
\mc {1}{c}{Ref. and Model} &
\mc {1}{c}{$^5_\Lambda {\vec {\rm H}}{\rm e}$} &
\mc {1}{c}{$^{12}_\Lambda {\vec {\rm C}}$} \\ \hline
Sasaki et al. \cite{Ok99} &   & \\
$\pi+K+{\rm DQ}$                & $-0.68$     \\ 
Parre\~no et al. \cite{Pa01} &    & \\
$\pi+\rho+K+K^*+\omega+\eta$    & $-0.68$ & $-0.73$    \\ 
Itonaga et al. \cite{It03}   &    & \\
$\pi+K+2\pi/\rho+2\pi/\sigma+\omega$    & $-0.33$ &   \\ 
Barbero et al. \cite{Ba03}           &    & \\
$\pi+\rho+K+K^*+\omega+\eta$            & $-0.54$ &    \\ \hline
  KEK--E160 \cite{Aj92}
 &                &  $-0.9\pm0.3\, ^*$  \\
  KEK--E278 \cite{Aj00}    & $0.24\pm0.22$  &                  \\
  KEK--E462 (preliminary) \cite{Ma04}   & $0.07\pm0.08$  & \\ 
  KEK--E508 (preliminary) \cite{Ma04}    &                & $-0.44\pm0.32$  \\ \hline
\end{tabular}}
\end{center}
\vskip -4mm
{\footnotesize * This result correspond to the weighted average (discussed on pag.~95 of
Ref.~\cite{Al02}) among different $p$--shell hypernuclear data.}
\end{table}

In Table \ref{other-res} we report the results for $a^{\rm M}_\Lambda$ obtained by
the above mentioned experiments together with available theoretical estimates
for $a_\Lambda$. While theory predicts a negative intrinsic $\Lambda$
asymmetry, with a moderate dependence on the hypernucleus,
the measurements seem to favor positive
values for $a^{\rm M}_\Lambda(^5_\Lambda{\vec {\rm H}{\rm e}})$ and negative values
for $a^{\rm M}_\Lambda(^{12}_\Lambda{\vec {\rm C}})$.

\subsection{Recent developments}
\label{rec-dev}
Concerning the above comparison between theory and experiment,
it is important to stress that, while one predicts
$a_\Lambda(^5_\Lambda{\vec {\rm H}{\rm e}})\simeq a_\Lambda(^{12}_\Lambda{\vec {\rm C}})$,
there is no known reason to expect this approximate equality to be valid
for $a^{\rm M}_\Lambda$.
Indeed, the relationship between $I(\Theta,J)$ of Eq.~(\ref{wdint}) and
$I^{\rm M}(\Theta,J)$ of Eq.~(\ref{wdint2}) can be strongly affected by FSI
of the emitted protons: this fact prevents establishing a direct
relation between $a_\Lambda$ and $a^{\rm M}_\Lambda$ and to make
a direct comparison among results for these quantities.
In order to overcome this obstacle, an evaluation of the effects of the
nucleon FSI on the non--mesonic weak decay of $^5_\Lambda{\vec {\rm H}}{\rm e}$,
$^{11}_\Lambda{\vec {\rm B}}$ and $^{12}_\Lambda{\vec {\rm C}}$
has been performed very recently \cite{GaDAFNE}. 
We summarize here some results of this investigation, which is the first one
evaluating $a^{\rm M}_\Lambda$.

The simulated proton intensities turned out to be well fitted by Eq.~(\ref{wdint2}),
then one can actually evaluate $a^{\rm M}_\Lambda$ through Eq.~(\ref{asym-exp}).
In Table~\ref{results-asy} we show predictions of the OMEf model
(a OME model built with the NSC97f potential)
for the weak decay and observable proton intensities,
$I(\Theta,J)$ and $I^{\rm M}(\Theta, J)$, respectively.
As a result of the nucleon rescattering in the nucleus, 
$|a_\Lambda|\gsim |a^{\rm M}_\Lambda|$ for any
value of the proton kinetic energy threshold: when $T^{\rm th}_p=0$,
$a_\Lambda/a^{\rm M}_\Lambda\simeq 2$ for $^5_\Lambda {\vec {\rm H}}{\rm e}$
and $a_\Lambda/a^{\rm M}_\Lambda\simeq 4$ for
$^{12}_\Lambda {\vec {\rm C}}$; $|a^{\rm M}_\Lambda|$ increases
with $T^{\rm th}_p$ and $a_\Lambda/a^{\rm M}_\Lambda\simeq 1$
for $T^{\rm th}_p=70$ MeV in both cases. 
Asymmetries $a^{\rm M}_\Lambda$ rather independent of the hypernucleus
are obtained for $T^{\rm th}_p=30$, $50$ and $70$ MeV.
\begin{table}
\begin{center}
\caption{\normalsize Results of Ref.~\cite{GaDAFNE} for the proton intensities
[Eqs.~(\ref{wdint}) and (\ref{wdint2})] from the non--mesonic 
weak decay of $^5_\Lambda{\vec {\rm H}{\rm e}}$
and $^{12}_\Lambda{\vec {\rm C}}$.}
\label{results-asy}
{\normalsize
\begin{tabular}{l c c c c} \hline
\mc {1}{c}{} &
\mc {1}{c}{$^5_\Lambda{\vec {\rm H}}{\rm e}$} &
\mc {1}{c}{} &
\mc {1}{c}{$^{12}_\Lambda{\vec {\rm C}}$} &
\mc {1}{c}{} \\
                  & $I^{\rm M}_0$    & $a^{\rm M}_\Lambda$   &
$I^{\rm M}_0$    &     $a^{\rm M}_\Lambda$ \\ \hline
{\small Without FSI}                    & $I_0=0.69$  & $a_\Lambda=-0.68$  & 
$I_0=0.75$ & $a_\Lambda=-0.73$   \\
{\small FSI and $T^{\rm th}_p=0$}       & $1.27$  & $-0.30$  & $2.78$ & $-0.16$  \\
{\small FSI and $T^{\rm th}_p=30$ MeV}  & $0.77$  & $-0.46$  & $1.05$ & $-0.37$  \\
{\small FSI and $T^{\rm th}_p=50$ MeV}  & $0.59$  & $-0.52$  & $0.65$ & $-0.51$  \\
{\small FSI and $T^{\rm th}_p=70$ MeV}  & $0.39$  & $-0.55$  & $0.38$ & $-0.65$  \\ \hline
  KEK--E462 (preliminary) \cite{Ma04}  &   & $0.07\pm 0.08$  &  &   \\
  KEK--E508 (preliminary) \cite{Ma04}  &     &  & & $-0.44\pm 0.32$  \\ \hline
\end{tabular}}
\end{center}
\end{table}
The KEK data quoted in the table refer to a $T^{\rm th}_p$ varying between
$30$ and $50$ MeV: the corresponding predictions of Ref.~\cite{GaDAFNE} agree with the
$^{12}_\Lambda{\vec {\rm C}}$ datum but are inconsistent with the observation for
$^5_\Lambda{\vec {\rm H}}{\rm e}$.

In conclusion, nucleon FSI turn out to be an important ingredient also when studying the non--mesonic
weak decay of polarized hypernuclei, but they cannot explain the present asymmetry data.
Further investigations are then required to clarify the issue.
On the theoretical side there seems to be no 
reaction mechanism which may be responsible for positive or vanishing asymmetry values.
On the experimental side the present anomalous
discrepancy between different data needs to be resolved.
Future experimental studies of the
inverse reaction $\vec{p} n\to p\Lambda $ should also be encouraged, since they
could help in disentangling the puzzling 
situation. Indeed, the weak production of the $\Lambda$--hyperon
could give a richer and cleaner 
(with respect to the non--mesonic hypernuclear decay) piece of
information on the lambda--nucleon weak interaction and especially on 
the $\Lambda$ polarization observables \cite{Na99}.

\section{Summary and perspectives}
\label{concl}

In these Lectures we have discussed several aspects of 
the weak decay of $\Lambda$--hypernuclei.
Beyond the mesonic channel, which is observed also for a free $\Lambda$,
the hypernuclear decay proceeds through non--mesonic processes, mainly induced
by the interaction of the $\Lam$ with one nucleon or with a pair of correlated 
nucleons. This channel is the dominant one in medium--heavy hypernuclei, 
where the Pauli principle strongly suppresses the mesonic decay.

Various models have been proposed to describe the
mesonic and non--mesonic decay rates as well as the asymmetry parameters
in the decay of the $\Lambda$--hyperon in nuclei. 
The results obtained within these models  have been thoroughly discussed.
The mesonic rates have been reproduced quite well by calculations performed in
different frameworks. Also the non--mesonic rates have been considered within a 
variety of phenomenological and microscopic models. 
In this context, particular interest has been devoted to the ratio 
$\Gamma_n/\Gamma_p$. Indeed, in spite of the fact that several calculations 
were able to reproduce, already at the OPE level, the total non--mesonic width, 
$\Gamma_{NM}=\Gamma_n+\Gamma_p (+\Gamma_2)$,
the values therewith obtained for $\Gamma_n/\Gamma_p$ revealed a strong 
disagreement with the experimental data. Although some of these calculations 
represented an improvement of the situation, further efforts were required 
in order to approach a solution the $\Gamma_n/\Gamma_p$ puzzle.
From the experimental side, recent experiments measured 
nucleon--nucleon coincidence observables with good statistics.
Recent  analyses of these data, complemented with the theoretical estimate of
final state interactions,  allowed the determination of 
$\Gamma_n/\Gamma_p$ values in agreement with the theoretical expectations.
Yet, good statistics coincidence measurements of $nn$ and $np$ emitted pairs 
are further required. These correlation measurements will also allow to establish
the first constraints on the two--nucleon 
induced decay width, thus testing the various models which have been proposed
for the evaluation of the different decay channels.

The analysis of the $\Delta I=1/2$ rule in the non--mesonic decay of light 
hypernuclei appears to be feasible, also within a relatively simple 
phenomenological model: again, the measurements of a few 
delicate, missing decay rates, would help in understanding the role of this
empirical selection rule.

As far as the decay of polarized hypernuclei is concerned, the situation is even more 
puzzling. While theory predicts negative values for both the intrinsic asymmetry 
$a_\Lambda$ and the observable asymmetry $a^{\rm M}_\Lambda$, with a moderate dependence
on the hypernucleus, experiments seem to favour negative values for 
$a^{\rm M}_\Lambda(^{12}_\Lambda {\vec {\rm C}})$ 
but small, positive values for $a^{\rm M}_\Lambda(^5_\Lambda {\vec {\rm He}})$.
Further investigations are then required to clarify the issue. Theoretically, there
seems to be no reaction mechanism which may be responsible for positive or vanishing
$a^{\rm M}_\Lambda$ values.
Improved experiments, establishing with certainty the sign and magnitude of 
$a^{\rm M}_\Lambda$ for $s$-- and $p$--shell hypernuclei, are strongly awaited.
                                                                
We conclude this work by reminding the reader that hypernuclear physics is 
52 years old, yet a lot of efforts remain to be done, both experimentally 
and theoretically, in order to fully understand the hyperon dynamics and 
decay inside the nuclear medium.
The impressive progress experienced in the last few years is promising and 
we hope that it deserves a definite answer to the intriguing open questions 
which we have illustrated here.

\acknowledgments
We warmly acknowledge our colleagues
A. De Pace, R. Cenni, A. Parre\~{n}o and A. Ramos, who collaborated with
us in obtaining some of the results discussed in these Lectures.


\begin{thebibliography}{99}
\label{biblio}

\bibitem{Da53} \BY{M. Danysz \atque J. Pniewski} 
\IN{Philos. Mag.}{44}{1953}{348}.

\bibitem{Gal-Varenna} \BY{A. Gal} These Proceedings.

\bibitem{Nagae-Varenna} \BY{T. Nagae} These Proceedings.

\bibitem{Outa-Varenna} \BY{H. Outa} These Proceedings.

\bibitem{Al02} \BY{W. M. Alberico \atque G. Garbarino}
\IN{Phys. Rep.} {369} {2002} {1}.

\bibitem{Gi79} \BY{F. J. Gillman \atque M. B. Wise} 
\IN{Phys. Rev. D} {20} {1979} {2392}.

\bibitem{Faccioli} \BY{M. Cristoforetti, P. Faccioli, E. V. Shuryak 
\atque M. Traini} \IN{Phys. Rev. D} {70} {2004} {054016}.

\bibitem{Os93} \BY{J. Nieves \atque E. Oset} 
\IN{Phys. Rev. C} {47} {1993} {1478}.

\bibitem{Mo94} \BY{T. Motoba \atque K. Itonaga} 
\IN{Prog. Theor Phys. Suppl.} {117}{1994} {477}.

\bibitem{Ak97} \BY{Y. Akaishi \atque T. Yamazaki} 
\IN{Prog. Part. Nucl. Phys.} {39} {1997} {565}.

\bibitem{Ou98} \BY{H. Outa {\sl et al.}} 
\IN{Nucl. Phys. A} {639} {1998} {251c}.

\bibitem{Co90} \BY{J. Cohen} \IN{Prog. Part. Nucl. Phys.} {25} {1990} {139}.
                                                                                                                 
\bibitem{Os98} \BY{E. Oset \atque A. Ramos} 
\IN{Prog. Part. Nucl. Phys.} {41} {1998} {191}.

\bibitem{Du96} \BY{J. F. Dubach, G. B. Feldman \atque B. R. Holstein}
\IN{Ann. Phys.} {249} {1996} {146}.

\bibitem{Pa97} \BY{A. Parre\~{n}o, A. Ramos \atque C. Bennhold}
\IN{Phys. Rev. C} {56} {1997} {339}.

\bibitem{It02} \BY{K. Itonaga, T. Ueda \atque T. Motoba} 
\IN{Phys. Rev. C} {65} {2002} {034617}.

\bibitem{Pa01} \BY{A. Parre\~{n}o \atque A. Ramos} 
\IN{Phys. Rev. C} {65} {2002} {015204}.

\bibitem{Os01} \BY{D. Jido, E. Oset \atque J. E. Palomar} 
\IN{Nucl. Phys. A} {694} {2001} {525}.

\bibitem{Pa98} \BY{A. Parre\~{n}o, A. Ramos, C. Bennhold \atque K. Maltman} 
\IN{Phys. Lett. B} {435} {1998} {1}.

\bibitem{Ch83} \BY{C.-Y. Cheung, D. P. Heddle \atque L. S. Kisslinger}
\IN{Phys. Rev. C} {27} {1983} {335};
\BY{D. P. Heddle \atque L. S. Kisslinger} \IN{Phys. Rev. C} {33} {1986} {608}.

\bibitem{Ok99} \BY{K. Sasaki, T. Inoue \atque M. Oka}
\IN{Nucl. Phys. A} {669} {2000} {331}; \SAME{678} {2000} {455(E)}; 
\SAME{707} {2002} {477}.

\bibitem{Os82} \BY{E. Oset, H. Toki \atque W. Weise} 
\IN{Phys. Rep.} {83} {1982} {281}.

\bibitem{Os94} \BY{E. Oset, P. Fern\'andez de C\'ordoba, J. Nieves, A. Ramos 
\atque L. L. Salcedo} \IN{Prog. Theor. Phys. Suppl.} {117} {1994} {461}.

\bibitem{Os85} \BY{E. Oset \atque L. L. Salcedo} 
\IN{Nucl. Phys. A} {443} {1985} {704}.

\bibitem{Wa71} \BY{A. L. Fetter \atque J. D. Walecka} 
\TITLE{Quantum Theory of Many Particle Systems} 
(McGraw--Hill, New York) 1971.

\bibitem{Ra95} \BY{A. Ramos, E. Oset \atque L. L. Salcedo}
\IN{Phys. Rev. C} {50} {1994} {2314}.

\bibitem{Al99} \BY{W. M. Alberico, A De Pace, G. Garbarino \atque A. Ramos}
\IN{Phys. Rev. C} {61} {2000} {044314}.

\bibitem{Al99a} \BY{W. M. Alberico, A. De Pace, G. Garbarino \atque R. Cenni}
\IN{Nucl. Phys. A} {668} {2000} {113}.

\bibitem{Se83} \BY{R. Seki \atque K. Masutani}
\IN{Phys. Rev. C} {27} {1983} {2799}.

\bibitem{Fried-Gal03} \BY{E. Friedman \atque A. Gal} 
\IN{Nucl. Phys. A} {724} {2003} {143}

\bibitem{Ga92} \BY{C. Garcia-Recio, J. Nieves \atque E. Oset}
\IN{Nucl. Phys. A} {547} {1992} {473}.

\bibitem{Ne82} \BY{J. W. Negele} \IN{Rev. Mod. Phys.} {54} {1982} {813}.

\bibitem{Al87} \BY{W. M. Alberico, R. Cenni, A. Molinari \atque P. Saracco}
\IN{Ann. Phys.} {174} {1987} {131}.

\bibitem{Ce97} \BY{R. Cenni, F. Conte \atque P. Saracco}
\IN{Nucl. Phys. A} {623} {1997} {391}.

\bibitem{Do88} \BY{C. B. Dover, A. Gal \atque D. J. Millener}
\IN{Phys. Rev. C}{38} {1988} {2700}.

\bibitem{Po98} \BY{J. Vida\~na, A. Polls, A. Ramos \atque M. Hjorth-Jensen} 
\IN{Nucl. Phys. A} {644} {1998} {201}.

\bibitem{Sz91} \BY{J. J. Szymanski {\sl et al.}} 
\IN{Phys. Rev. C} {43} {1991} {849}.

\bibitem{No95} \BY{H. Noumi {\sl et al.}} 
\IN{Phys. Rev. C} {52} {1995} {2936}.

\bibitem{Bh98} \BY{H. C. Bhang {\sl et al.}} 
\IN{Phys. Rev. Lett.} {81} {1998} {4321};
\BY{H. Park {\sl et al.}} \IN{Phys. Rev. C} {61} {2000} {054004}.

\bibitem{Ou00} \BY{H. Outa {\sl et al.}} 
\IN{Nucl. Phys. A} {670} {2000} {281c}.

\bibitem{Sa04} \BY{Y. Sato {\em et al.}} 
{\bf nucl--ex/0409007} (submitted to {\em Phys. Rev. C}).

\bibitem{St93} \BY{U. Straub, J. Nieves, A. Faessler \atque E. Oset}
\IN{Nucl. Phys. A} {556} {1993} {531}.

\bibitem{It88} \BY{K. Itonaga, T. Motoba \atque H. Band$\overline{\rm o}$} 
\IN{Z. Phys. A} {330} {1988} {209}; 
\IN{Nucl. Phys. A} {489} {1988} {683}.

\bibitem{Al91} \BY{W. M. Alberico, A. De Pace, M. Ericson \atque A. Molinari}
\IN{Phys. Lett. B} {256} {1991} {134}.

\bibitem{Ar93} \BY{T. A. Armstrong {\sl et al.}} 
\IN{Phys. Rev. C} {47} {1993} {1957}.

\bibitem{Ku98} \BY{H. Ohm {\sl et al.}} \IN{Phys. Rev. C} {55} {1997} {3062};
\BY{P. Kulessa {\sl et al.}} \IN{Phys. Lett. B} {427} {1998} {403}.

\bibitem{Mo91} \BY{T. Motoba, H. Band$\overline{\rm o}$, T. Fukuda 
\atque J. \v{Z}ofka}
\IN{Nucl. Phys. A} {534} {1991} {597}.

\bibitem{Mo92} \BY{T. Motoba} 
\IN{Nucl. Phys. A} {547} {1992} {115c}.

\bibitem{Os86a} \BY{E. Oset, L. L. Salcedo \atque Q. N. Usmani} 
\IN{Nucl. Phys. A} {450} {1986} {67c}.

\bibitem{Ku95a} \BY{I. Kumagai-Fuse, S. Okabe \atque Y. Akaishi} 
\IN{Phys. Lett. B} {345} {1995} {386}.

\bibitem{Ok04} \BY{S. Okada {\em et al.}}
\TITLE{VIII International Conference on Hypernuclear and Strange Particle
Physics} (HYP2003), JLab, Newport News, Virginia, {\bf nucl--ex/0402022} 
[{\em Nucl. Phys. A} (to be published)].

\bibitem{Zh99} \BY{L. Zhou \atque J. Piekarewicz} 
\IN{Phys. Rev. C} {60} {1999} {024306}.

\bibitem{Er90} \BY{M. Ericson \atque H. Band$\overline{\rm o}$} 
\IN{Phys. Lett. B} {237} {1990} {169}.

\bibitem{Al03} \BY{C. Albertus, J. E. Amaro, \atque J. Nieves} 
\IN{Phys. Rev. C} {67} {2003} {034604}.

\bibitem{Da73} \BY{R. H. Dalitz} 
Proceedings of the \TITLE{Summer Study Meeting on Nuclear
and Hypernuclear Physics with Kaon Beams}, BNL Report No. 18335 (1973) p. 41.

\bibitem{Ju01} \BY{J.-H. Jun \atque H. C. Bhang} 
\IN{Nuovo Cim.} {112 A} {1999} {649};
\BY{J.-H. Jun} \IN{Phys. Rev. C} {63} {2001} {044012}.

\bibitem{No95a} \BY{H. Noumi {\sl et al.}} 
Proceedings of the \TITLE{IV International
Symposium on Weak and Electromagnetic Interactions in Nuclei},
H. Ejiri, T. Kishimoto and T. Sato eds, World Scientific (1995) p. 550.

\bibitem{Ka01} \BY{B. Kamys {\sl et al.}}
\IN{Eur. Phys. J. A} {11} {2001} {1}.

\bibitem{Ku01} \BY{P. Kulessa {\sl et al.}} 
\IN{Acta Phys. Polon. B} {33} {2002} {603}.

\bibitem{Ga03} \BY{G. Garbarino, A. Parre\~{n}o \atque A. Ramos}
\IN{Phys. Rev. Lett.} {91} {2003} {112501};
\IN{Phys. Rev. C} {69} {2004} {054603}.

\bibitem{outa-coinc} \BY{H. Outa} 
these proceedings and in
\TITLE {VIII International Conference on Hypernuclear and Strange Particle 
Physics} (HYP2003), JLab, Newport News, Virginia [{\em Nucl. Phys. A} 
(to be published)].

\bibitem{kim} \BY{J. H. Kim {\sl et al.}}
\IN{Phys. Rev. C} {68} {2003} {065201}.

\bibitem{Ok04l} \BY{S. Okada {\em et al.}} 
\IN{Phys. Lett. B} {597} {2004} {249}.

\bibitem{GaDAFNE} \BY{W. M. Alberico, G. Garbarino, A. Parre\~{n}o \atque 
A. Ramos} \TITLE {DAPHNE2004: Physics at meson factories}, Laboratori 
Nazionali di Frascati, Frascati, Italy, {\bf nucl--th/0407046} 
[Frascati Physics Series (to be published)].

\bibitem{Ga95} \BY{A. Gal} 
in \TITLE{Weak and electromagnetic interactions in nuclei}, H. Ejiri, T.
Kishimoto and T. Sato eds, World Scientific (1995) p. 573.

\bibitem{Ra97} \BY{A. Ramos, M. J. Vicente--Vacas \atque E. Oset}
\IN{Phys. Rev. C} {55} {1997} {735}; \SAME{66} {2002} {039903E}.

\bibitem{Ha01} \BY{O. Hashimoto {\sl et al.}} 
\IN{Phys. Rev. Lett.} {88} {2002} {042503}.

\bibitem{Gi01} \BY{R. L. Gill} 
\IN{Nucl. Phys. A} {691} {2001} {180c}.

\bibitem{jparc} \BY{S. Ajimura} 
\TITLE{Precise measurement of the non--mesonic weak decay of $A=4,5$
$\Lambda$--hypernuclei}, Letter of intent (LOI21) for experiments at J--PARC 
(2003).

\bibitem{FI} \BY{A. Zenoni} these proceedings;
\BY{A. Feliciello} \IN{Nucl. Phys. A} {691} {2001} {170c};
\BY{P. Gianotti} \IN{Nucl. Phys. A} {691} {2001} {483c}.

\bibitem{Ri99a} \BY{V. G. J. Stoks \atque Th. A. Rijken} 
\IN{Phys. Rev. C} {59} {1999} {3009}.

\bibitem{Al99b} \BY{W. M. Alberico \atque G. Garbarino}
\IN{Phys. Lett. B} {486} {2000} {362}.

\bibitem{Bl63} \BY{M. M. Block \atque R. H. Dalitz} 
\IN{Phys. Rev. Lett.} {11} {1963} {96}.

\bibitem{Ze98} \BY{V. J. Zeps} 
\IN{Nucl. Phys. A} {639} {1998} {261c}.

\bibitem{Ba89} \BY{H. Band$\overline{\rm o}$, T. Motoba M. Sotona 
\atque J. \v{Z}ofka} \IN{Phys. Rev. C} {39} {1989} {587};
\BY{H. Ejiri, T. Fukuda, T. Shibata, H. Band$\overline{\rm o}$ 
\atque K.-I. Kubo} \IN{Phys. Rev. C} {36} {1987} {1435}.

\bibitem{Ba90} \BY{H. Band$\overline{\rm o}$, T. Motoba \atque J. \v{Z}ofka}
\IN{Int. J. Mod. Phys. A} {5} {1990} {4021}.

\bibitem{Aj92} \BY{S. Ajimura {\sl et al.}} 
\IN{Phys. Lett. B} {282} {1992} {293}.

\bibitem{Aj00} \BY{S. Ajimura {\sl et al.}}
\IN{Phys. Rev. Lett.} {84} {2000} {4052}.

\bibitem{Ra92} \BY{A. Ramos, E. van Meijgaard, C. Bennhold \atque B. K. 
Jennings} \IN{Nucl. Phys. A} {544} {1992} {703}.

\bibitem{Ma04} \BY{T. Maruta {\sl et al.}} 
\TITLE{VIII International Conference on Hypernuclear and Strange
Particle Physics} (HYP2003), JLAB, Newport News, Virginia, 
{\bf nucl--ex/0402017} [{\em Nucl. Phys.} {\bf A} (to be published)].

\bibitem{It03} \BY{K. Itonaga, T. Motoba \atque T. Ueda} in
\TITLE{Electrophoto--production of Strangeness on Nucleons and Nuclei} 
(Sendai03), K. Maeda, H. Tamura, S. N. Nakamura and O. Hashimoto
eds, World Scientific (2004) p. 397.

\bibitem{Ba03} \BY{C. Barbero, C. De Conti, A. P. Gale\~ao \atque 
F. Krmpoti\'c} \IN{Nucl. Phys. A} {726} {2003} {267}.

\bibitem{Na99} \BY{H. Nabetani, T. Ogaito, T. Sato \atque T. Kishimoto}
\IN{Phys. Rev. C} {60} {1999} {017001}.





\end{thebibliography}
\end{document}